\documentstyle[aps,axodraw,eqsecnum]{revtex}
\begin{document}

\title{Nucleon effects on
the photon dispersion relations in matter}
\author{Juan Carlos D'Olivo}
\address{Instituto de Ciencias Nucleares\\
Universidad Nacional Aut\'{o}noma de M\'{e}xico\\
Apartado Postal 70-543, 04510 M\'{e}xico, D.F., M\'{e}xico}
\author{Jos\'e F. Nieves}
\address{Laboratory of Theoretical Physics,
Department of Physics, P.O. Box 23343 \\
University of Puerto Rico,
R\'{\i}o Piedras, Puerto Rico 00931-3343}
\date{hep-ph/9710305}

\maketitle

\begin{abstract}

We calculate the nucleon contribution to the photon self-energy
in a plasma,
including the effect of the anomalous magnetic moment of the nucleons.
General formulas
for the transverse and longitudinal components of the self-energy 
are obtained and we give explicit results
in various limits of physical interest. 
The formulas are relevant
for the study of the photon dispersion relations and 
the dynamical susceptibility
in a nuclear medium such as the core of a supernova, 
and has implications with regard to the recent suggestion
that the Cerenkov process $\nu\rightarrow\nu\gamma$
can take place in such a system.

\end{abstract}

\section{Introduction}

The methods of Finite Temperature
Field Theory (FTFT)\cite{ftftreview} have proven to be
very useful in the study of the matter effects on
the propagation of  the elementary particles.
In these studies, a central role
is played by the inverse propagator, or equivalently, the
self-energy of the particle. The latter quantity
enters directly in the linear part of the
equations of motion for the effective classical field, and therefore
determines the modes that can propagate through the medium.
In a modern context, this point of view has been stimulated by
the influential work of  
Weldon\cite{weldon:cov,weldon:fermions,weldon:imag},
who emphasized the convenience of covariant, real-time,
calculations.   His  approach
has been applied to investigate the effects of a medium  on the
properties  of photons, gauge fields, and neutrinos,
in a variety of physical conditions.

The case of photons occupies an important place  within these
type of problems.
Its applications  range from the calculation of cross sections for
particle emission by stars, to the determination of the optical
properties of solids.
Under certain conditions of the matter background,
the photon dispersion relation in a plasma can be obtained
by a semiclassical approach based on the Maxwell-Boltzmann-Vlasov
equation\cite{LL:physkin}.   It seems that the
first complete quatum-field-theoretic treatment of the same problem
is due to Tsytovich\cite{tsytovich}.  He generalized earlier work along
similar lines\cite{silin}, obtaining results that are valid under more
general conditions than those based on the semiclassical approach.
The work of Weldon\cite{weldon:cov} demonstrated that the
real-time formulation of FTFT is well suited to study  this  
physical system
in an efficient and transparent way, and can be
extended to, for example, non-abelian gauge fields.

For photons that propagate in a hot plasma,
the medium typically  consists of an electron
gas superimposed over a static positive background.
It can be shown that the results of
the semiclassical calculation are recovered from those of
the field-theoretic treatment,  by taking the limit
in which the photon energy and momentum are small relative to the
electron mass.   Ordinarily, only the electrons play an  
important role, and
the dynamical effect of the nucleons in the background
are ignored. This is justified within
the context of a semiclassical approach.  In  fact,
the contributions of the ions and electrons
to the dielectric constant have similar form.
Since the plasma frequency associated with a certain particle is  
inversely
proportional
to its mass,
the contribution of the ions turns out to be negligible,
at least in the most common situation where all species share a common
temperature\cite{LL:physkin2}.

In a quantum-theoretic approach, however, the electromagnetic
couplings of the nucleons must include their anomalous magnetic moments.
For the case in which the nucleons constitute a degenerate
non-relativistic gas, a standard treatment yields the
Pauli formula for the paramagnetic spin susceptibility,
in the limit of zero photon frequency (static limit)\cite{fetter}.
Nevertheless, a general treatment of the effect
in the photon dispersion relation due to
the anomalous electromagmetic couplings of the nucleons
seems to be lacking in the literature.
Such a study has applications in some astrophysical contexts, where
systems with dense nuclear matter need to be considered.
For example, it has been recently pointed out by
Mohanty and Samal\cite{mohanty} that, in a supernova core,
the photon dispersion relation receives a contribution
from the nucleon magnetic moment which is of the opposite
sign to that of the usual electron plasma effect.  Under such   
circumstances,
the emission of Cerenkov radiation by neutrinos becomes possible,
which can have important implications for the energetics of the system.
More recently, Raffelt\cite{raffelt:critique} has noted that the
conclusions of Ref.\ \cite{mohanty} are based on the adoption
of  the static paramagnetic susceptibility formula mentioned above,
in order to determine the effect of the nucleons on the photon dispersion
relation.  However, for a proper account of this effect
what is required is the full nucleon
contribution to photon self-energy (or, equivalently, to the dielectric
constant
or the paramagnetic susceptibility) for any value
of the photon momentum and not just in the zero-frequency limit.
The necessary calculations along these lines are not in the literature,
to the best of our knowledge.

Motivated by  the above considerations, in the present work we
calculate the photon self-energy in a
matter background including the contribution from
the nucleons, and in particular the effect of their
anomalous electromagnetic couplings.  The calculation is
based on the application of FTFT to this problem
and complements the existing calculations of the same
quantity for the electron gas alone.  The results
for the photon self-energy can be equivalently
interpreted in terms of the dielectric constant and
the susceptibility of the system, and in that way we show that
the well known textbooks results for these quantities
are reproduced when the appropriate limits are taken.
On the other hand, the results we obtain are valid
for general conditions of the nucleon gas, whether
it is degenerate or not, and also for general values of
the photon momentum and not just for some particular limiting
values.  Therefore, they are relevant and useful in
the study of electromagnetic processes ocurring
within dense nuclear matter, such as the neutrino
Cerenkov radiation in a supernova core mentioned above.

In Section \ref{sec:photonselfenergy} we give the general
1-loop formulas for the generic contribution of a 
fermion to the photon self-energy, including the effect of
a possible magnetic moment coupling of the fermion to the 
electromagnetic field.  There it is shown that the contribution
from any given fermion can be written in terms of just
three independent functions, which are expressed as integrals over the
momentum distribution functions of the fermion.
In Section \ref{sec:limitingcases} 
explicit formulas are given for various limiting cases of physical interest.
The formulas so obtained are used in Section \ref{sec:discussion}
to study the photon dispersion relations in a background composed
of electrons and nucleons under various conditions, 
including those corresponding to a degenerate or a classical
nucleon background.  Although the focus
of our calculations is the real part of the self-energy, 
there we also summarize the results of the
calculation of the imaginary part, which
determines the damping of the propagating modes.  
Some details of that calculation are given in
Appendix \ref{appendix:impart}.
Section \ref{sec:conclusions} contains our conclusions,
and in Appendix \ref{appendix:dielectric} the relation
between the self-energy and the dielectric function 
of macroscopic electrodynamics is briefly reviewed.

\section{Photon self-energy}
\label{sec:photonselfenergy}

As shown in Ref.\ \cite{canonical}, the equation of motion
for the effective field in the medium, from which
the dispersion relations of the propagating modes are determined,
is given by
\begin{equation}\label{claseqmotion}
\left[-q^2\tilde g_{\mu\nu} + \pi^{(e\!f\!\!f)}_{\mu\nu}\right]A^\nu
= j_\mu \,,
\end{equation}
where
\begin{equation}\label{gtilde}
\tilde g_{\mu\nu} = g_{\mu\nu} - \frac{q_\mu q_\nu}{q^2} \,.
\end{equation}
and
\begin{equation}\label{pieff}
\pi^{(e\!f\!\!f)}_{\mu\nu} = \pi_{\mu\nu}\theta(q\cdot u) + \pi_{\nu\mu}^\ast\theta(-q\cdot u) \,.
\end{equation}
In Eq.\ (\ref{pieff}) $\theta$ is the step function, $u_\mu$ stands for the velocity
four-vector of the medium and $\pi_{\mu\nu}$ is determined
in terms of the elements of the $2\times 2$ self-energy matrix
in the form
\begin{equation}\label{repi}
\mbox{Re}\,\pi_{\mu\nu} = \mbox{Re}\,\pi_{11\mu\nu} \,,
\end{equation}
\begin{eqnarray}\label{impi}
\mbox{Im}\,\pi_{\mu\nu} & = & \tanh|\frac{1}{2}\beta q\cdot u|\mbox{Im}\,\pi_{11\mu\nu}\nonumber\\
& = & \frac{i\pi_{12\mu\nu}\varepsilon(q\cdot u)}{2n_\gamma(x)} \,,
\end{eqnarray}
where we have defined
\begin{eqnarray}\label{defreim}
\mbox{Re}\,\pi_{\mu\nu} & = & \frac{1}{2}(\pi_{\mu\nu} + \pi_{\nu\mu}^\ast)\nonumber\\
\mbox{Im}\,\pi_{\mu\nu} & = & \frac{1}{2i}(\pi_{\mu\nu} - \pi_{\nu\mu}^\ast) \,,
\end{eqnarray}
with similar definitions for $\mbox{Re}\,\pi_{11\mu\nu}$ and $\mbox{Im}\,\pi_{11\mu\nu}$.
In Eq.\ (\ref{impi}) $\beta$ is the inverse temperature, 
$\varepsilon(z) = \theta(z) - \theta(-z)$ and 
\begin{equation}\label{ngamma}
n_\gamma(x) = \frac{1}{e^x - 1} \,,
\end{equation}
where, for the photon,
\begin{equation}\label{x}
x = \beta q\cdot u \,.
\end{equation}
In our calculations we have adopted $u^\mu = (1,\vec 0)$,
so that all the three-dimensional quantities refer to the frame
in which the medium is at rest.
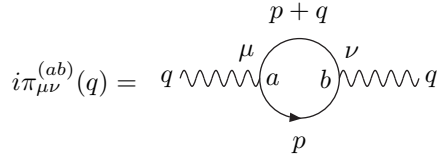
\begin{figure}
\begin{center}
\begin{picture}(160,120)(-50,0)
\Photon(30,60)(0,60){3}{5}
\Photon(90,60)(60,60){3}{5}
\LongArrowArc(45,60)(15,-93,273)
\Text(45,85)[]{$p + q$}
\Text(45,35)[]{$p$}
\Text(-40,60)[]{$i\pi^{(ab)}_{\mu\nu}(q) =$}
\Text(65,70)[]{$\nu$}
\Text(25,70)[]{$\mu$}
\Text(35,60)[]{$a$}
\Text(55,60)[]{$b$}
\Text(-5,60)[]{$q$}
\Text(95,60)[]{$q$}
\end{picture}
\end{center}
\caption{Diagram for the contribution to the photon self-energy
matrix from a generic fermion.\label{fig:pimunu}}
\end{figure}
In practical calculations it is much simpler to
determine the imaginary part of the self-energy
using the second form in Eq.\ (\ref{impi}), and therefore
we will determine $\pi^{(e\!f\!\!f)}_{\mu\nu}$ from the formulas
\begin{eqnarray}\label{pieff2}
\mbox{Re}\,\pi^{(e\!f\!\!f)}_{\mu\nu} & = & \mbox{Re}\,\pi_{11\mu\nu}\nonumber\\
\mbox{Im}\,\pi^{(e\!f\!\!f)}_{\mu\nu} & = & \frac{i\pi_{12\mu\nu}}{2n_\gamma(x)} \,.
\end{eqnarray}

The various components of the photon self-energy matrix are calculated from 
the diagram of Fig.\ \ref{fig:pimunu}.  For any fermion $f$ in the loop we
use the propagator
\begin{eqnarray}\label{SF}
S^{(f)}_{F11}(p) & = & (\not p + m_f)\left[\frac{1}{p^2 - m_f^2 + i\epsilon}
+ 2\pi i \delta(p^2 - m_f^2)\eta_f(p)\right]\nonumber\\
S^{(f)}_{F22}(p) & = & (\not p + m_f)\left[\frac{-1}{p^2 - m_f^2 - i\epsilon}
+ 2\pi i \delta(p^2 - m_f^2)\eta_f(p)\right]\nonumber\\
S^{(f)}_{F12}(p) & = & (\not p + m_f)2\pi i\delta(p^2 - m_f^2)\left[
\eta_f(p) - \theta(-p\cdot u)\right]\nonumber\\
S^{(f)}_{F21}(p) & = & (\not p + m_f)2\pi i\delta(p^2 - m_f^2)\left[
\eta_f(p) - \theta(p\cdot u)\right] \,.
\end{eqnarray}
In these formulas
\begin{equation}\label{etaf}
\eta_f(p) = \theta(p\cdot u)n_F(y_f) + \theta(-p\cdot u)n_F(-y_f) \,,
\end{equation}
with
\begin{equation}\label{nf}
n_F(y) = \frac{1}{e^y +1} 
\end{equation}
and
\begin{equation}\label{y}
y_f = \beta p\cdot u - \alpha_f \,,
\end{equation}
where $\alpha_f$ is the fermion chemical potential.  

We consider the calculation of $\pi_{11\mu\nu}$,
from which the real part of the physical self-energy
is determined by means of Eq.\ (\ref{pieff2}).  Referring
to Fig. \ref{fig:pimunu}, the contribution 
from one fermion in the loop is
\begin{equation}\label{pi11}
i\pi_{11\mu\nu}^{(f)} = (-1)(-i)^2 \mbox{Tr}\int\frac{d^4p}{(2\pi)^4}
j^{(em)}_{f\mu}(q) iS^{(f)}_{F11}(p + q)j^{(em)}_{f\nu}(-q) iS^{(f)}_{F11}(p) \,.
\end{equation}
The electromagnetic couplings
are defined by writing the matrix element
\begin{equation}\label{Fem}
\langle f(p^\prime)|J^{(em)}_\mu(0)|f(p)\rangle
= \overline u(p^\prime)j_{f\mu}^{(em)}(q)u(p) \,,
\end{equation}
where $q = p - p^\prime$ is the momentum of the outgoing photon, 
$u(p)$ is a Dirac spinor and $j^{(em)}_{f\mu}(q)$
is the total electromagnetic current of the fermion.
For the electron $j^{(em)}_{e\mu}(q) = e\gamma_\mu$, while
for the nucleons
\begin{eqnarray}\label{jnucl}
j^{(em)}_{p\mu}(q) & = & |e|\gamma_\mu
-i\kappa_p\sigma_{\mu\alpha}q^\alpha\nonumber\\
j^{(em)}_{n\mu}(q) & = & -i\kappa_n\sigma_{\mu\alpha}q^\alpha\,.
\end{eqnarray}
In Eq.\ (\ref{jnucl}) $\kappa_{n,p}$ are the anomalous part of the nucleon
magnetic moments, given by
\begin{eqnarray}\label{kappanp}
\kappa_p & = & 1.79\left(\frac{|e|}{2m_p}\right)\,,\nonumber\\
\kappa_n & = & -1.91\left(\frac{|e|}{2m_n}\right)\,,\nonumber\\
\end{eqnarray}
$e$ stands for the electron charge
and, as usual, $\sigma_{\mu\nu} = \frac{i}{2}[\gamma_\mu,\gamma_\nu]$.
For arbitrary values of $q$, Eq.\ (\ref{jnucl}) does not hold
and, in general, we have to parametrize the nucleon
electromagnetic current in terms of the momentum-dependent form factors.
However, in the applications of interest to us the photon momentum
is sufficiently small that the form factors are well approximated
by their values at $q^2 = 0$, and therefore Eq.\ (\ref{jnucl}) is
appropriate.

Substituting the formula for $S^{(f)}_{F11}$ given in
Eq.\ (\ref{SF}), $\pi_{11\mu\nu}^{(f)}$ can be written in the form
\begin{equation}\label{pi11decomp}
\pi_{11\mu\nu}^{(f)} = \pi_{11\mu\nu}^{(0)} + \pi_{11\mu\nu}^{\prime\,(f)}\,,
\end{equation}
where $\pi_{11\mu\nu}^{(0)}$ is the standard vacuum polarization term, while
the background dependent contribution is given by
\begin{eqnarray}\label{pi11T}
i\pi_{11\mu\nu}^{\prime\,(f)} & = & -(2\pi i)\int\frac{d^4p}{(2\pi)^4}
\mbox{Tr}
\left[j^{(em)}_{f\mu}(q)(\not p + \not q + m_f)
j^{(em)}_{f\nu}(-q)
(\not p + m_f)\right]\nonumber\\
& & \times\left\{\frac{\eta_f(p)\delta(p^2 - m_f^2)}
{(p + q)^2 - m_f^2 + i\epsilon} + 
\frac{\eta_f(p + q)\delta((p + q) ^2 - m_f^2)}
{p^2 - m_f^2 + i\epsilon}\right.\nonumber\\
& & \mbox{} \left.\vphantom{\frac{\eta_f}{p^2}}
+ 2\pi i\eta_f(p)\eta_f(p + q)\delta(p^2 - m_f^2)
\delta((p + q)^2 - m_f^2)\right\} \,.
\end{eqnarray}
The term in Eq.\ (\ref{pi11T}) that contains two factors of $\eta_f$
contributes only to the imaginary part of the self-energy.
We restrict our attention for the moment on the real part,
which receives contributions only from the terms linear
in $\eta_f$, and which we denote by $\mbox{Re}\,\pi^{(e\!f\!\!f)}_{f\mu\nu}$.
Then, making the change of variables $p + q \rightarrow p$
in the second of these terms in Eq.\ (\ref{pi11T}) and carrying out
the integral over $p_0$, we obtain from Eq.\ (\ref{pieff2})
\begin{equation}\label{repieffproton}
\mbox{Re}\,\pi^{(e\!f\!\!f)}_{p\mu\nu}  = \pi^{(\gamma)}_{\mu\nu}
+ \pi^{(\sigma)}_{\mu\nu} + \pi^{(\gamma\sigma)}_{\mu\nu}\,,
\end{equation}
where
\begin{eqnarray}\label{piproton}
\pi^{(\gamma)}_{\mu\nu} & = & -4e_f^2\int\frac{d^3{\cal P}}{(2\pi)^3 2{\cal E}}
(f_f + f_{\overline f})
\left[\frac{L_{\mu\nu}}{q^2 + 2p\cdot q} + (q \rightarrow
-q)\right]\,,\nonumber\\
\pi^{(\sigma)}_{\mu\nu} & = & -4\kappa_f^2\int\frac{d^3{\cal P}}{(2\pi)^3 2{\cal
E}}
(f_f + f_{\overline f})
\left[\frac{M_{\mu\nu}}{q^2 + 2p\cdot q} + (q \rightarrow -q)\right]
\,,\nonumber\\
\pi^{(\gamma\sigma)}_{\mu\nu} & = & 
-8(e_f\kappa_f m_f)q^2\tilde g_{\mu\nu}\int\frac{d^3{\cal P}}{(2\pi)^3 2{\cal
E}}
(f_f + f_{\overline f})
\left[\frac{1}{q^2 + 2p\cdot q} + (q \rightarrow -q)\right] \,.
\end{eqnarray}
In these formulas, 
\begin{eqnarray}\label{Lmunu}
L_{\mu\nu} & = & 2p_\mu p_\nu + p_\mu q_\nu + q_\mu p_\nu
- g_{\mu\nu}p\cdot q \,,\nonumber\\
\label{Mmunu}
M_{\mu\nu} & = & q^2(2m_f^2 - p\cdot q)\tilde g_{\mu\nu} - 2(p\cdot q)^2 g_{\mu\nu}
+ 2(p\cdot q)[p_\mu q_\nu + q_\mu p_\nu] - 2q^2 p_\mu p_\nu \,,
\end{eqnarray}
\begin{equation}\label{pE}
p^\mu = ({\cal E},\vec {\cal P})\,, \quad {\cal E} = 
\sqrt{\vec {\cal P}^{\,2} + m_f^2} \,,
\end{equation}
and $f_{f,\overline f}$ denote the particle and antiparticle number density
distributions given by
\begin{equation}\label{fe}
f_{f,\overline f} = \frac{1}{e^{\beta{\cal E} \mp \alpha_f} + 1}
\end{equation}
with the minus(plus) sign holding for the particles(antiparticles), 
respectively.
The integrals in Eq.\ (\ref{piproton}) is to be interpreted in the sense of its
principal value part. 

As we have already mentioned, for the imaginary part of $\pi^{(e\!f\!\!f)}_{\mu\nu}$
it is simpler to calculate $\pi_{12\mu\nu}$ first and then use the second form 
of the formula for $\mbox{Im}\,\pi^{(e\!f\!\!f)}_{\mu\nu}$ given in Eq.\ (\ref{pieff2}), 
instead of calculating $\mbox{Im}\,\pi_{11\mu\nu}$.  
Nevertheless, we have carried out the calculation in both ways and
explicitly verified that they yield the same result for $\pi^{(e\!f\!\!f)}_{\mu\nu}$.  
The results of that calculation are summarized in
Section \ref{subsec:impart}, and the main steps leading to it
are oulined in Appendix \ref{appendix:impart}.
In the remainder of this paper we focus on
the real part of $\pi^{(e\!f\!\!f)}_{\mu\nu}$.

In an isotropic medium, the most general form of the photon self-energy
is\cite{weldon:cov,pisubpi}
\begin{equation}\label{pieffgen}
\pi^{(e\!f\!\!f)}_{\mu\nu} = \pi_T R_{\mu\nu} + \pi_L Q_{\mu\nu} + \pi_P P_{\mu\nu} \,,
\end{equation}
where
\begin{eqnarray}\label{tensors}
R_{\mu\nu} & = & \tilde g_{\mu\nu} - Q_{\mu\nu} \nonumber\\
Q_{\mu\nu} & = & \frac{\tilde u_\mu\tilde u_\nu}{\tilde u^2}\nonumber\\
P_{\mu\nu} & = & \frac{i}{{\cal Q}}\epsilon_{\mu\nu\alpha\beta}q^\alpha u^\beta \,,
\end{eqnarray}
with
\begin{equation}\label{utilde}
\tilde u_\mu \equiv \tilde g_{\mu\nu}u^\nu \,.
\end{equation}
In general, $\pi_{T,L,P}$ are functions of the scalar variables
\begin{eqnarray}\label{QOmega}
\Omega & = & q\cdot u \nonumber\\
{\cal Q} & = & \sqrt{\Omega^2 - q^2} \,,
\end{eqnarray}
which have the interpretation of being the photon energy
and momentum (in the rest frame of the medium).  
The conditions under which the function $\pi_P$ can have a nonzero value, 
as well as some of its implications, have
been discussed in Ref.\ \cite{pisubpi}.  In the present case, it is easily seen 
that it is zero since all the terms in
Eq.\ (\ref{piproton}) are symmetric in $\mu,\nu$.  
The functions $\pi_{T,L}$ are determined by projecting Eq.\ (\ref{repieffproton})
with the tensors $R_{\mu\nu}$ and $Q_{\mu\nu}$.  
This procedure then leads to
\begin{eqnarray}\label{piprotonproj}
\pi^{(\gamma)}_{\mu\nu} & = & -4e_f^2\left[\frac{1}{2}\left(A_f - \frac{B_f}
{\tilde u^2}\right)R_{\mu\nu} + \frac{B_f}{\tilde u^2}Q_{\mu\nu}\right]
\,,\nonumber\\
\pi^{(\sigma)}_{\mu\nu} & = & -4\kappa_f^2\left[\frac{1}{2}
\left(A^\prime_f - \frac{B^\prime_f}
{\tilde u^2}\right)R_{\mu\nu} + \frac{B^\prime_f}{\tilde u^2}Q_{\mu\nu}\right]
\,,\nonumber\\
\pi^{(\gamma\sigma)}_{\mu\nu} & = & 
-8(e_f\kappa_f m_f)q^2 D_f\left[R_{\mu\nu} + Q_{\mu\nu}\right] \,,
\end{eqnarray}
where,
\begin{eqnarray}\label{ABCD}
A_f & \equiv &
\int\frac{d^3{\cal P}}{(2\pi)^3 2{\cal E}}
(f_f + f_{\overline f})
\left[\frac{L^\mu_\mu}{q^2 + 2p\cdot q} + (q \rightarrow
-q)\right]\,,\nonumber\\
B_f & \equiv &
\int\frac{d^3{\cal P}}{(2\pi)^3 2{\cal E}}
(f_f + f_{\overline f})
\left[\frac{u^\mu u^\nu L_{\mu\nu}}{q^2 + 2p\cdot q} + (q \rightarrow
-q)\right]\,,\nonumber\\
A^\prime_f & \equiv &
\int\frac{d^3{\cal P}}{(2\pi)^3 2{\cal E}}
(f_f + f_{\overline f})
\left[\frac{M^\mu_\mu}{q^2 + 2p\cdot q} + (q \rightarrow
-q)\right]\,,\nonumber\\
B^\prime_f & \equiv &
\int\frac{d^3{\cal P}}{(2\pi)^3 2{\cal E}}
(f_f + f_{\overline f})
\left[\frac{u^\mu u^\nu M_{\mu\nu}}{q^2 + 2p\cdot q} + (q \rightarrow
-q)\right]\,, \nonumber\\
D_{f} & \equiv & \int\frac{d^3{\cal P}}{(2\pi)^3 2{\cal E}}
(f_{f} + f_{\overline f})\left[
\frac{1}{q^2 + 2p\cdot q} + (q\rightarrow -q)\right] \,.
\end{eqnarray}
Substituting here the explicit
formulas for $L_{\mu\nu}$ and $M_{\mu\nu}$ given in Eqs.\ (\ref{Lmunu}) and (\ref{Mmunu}),
after some straightforward algebra we obtain the relations
\begin{eqnarray}\label{ABprime}
A^\prime_f & = & q^2\left(\frac{1}{2}A_f + 3m_f^2 D_f\right) \,,\nonumber\\
B^\prime_f & = & -{\cal Q}^2\left(\frac{1}{2}A_f + m_f^2 D_f\right)
- q^2 B_f \,,
\end{eqnarray} 
with $A_f, B_f$ being given by
\begin{eqnarray}\label{ApBp}
A_f & = & \int\frac{d^3{\cal P}}{(2\pi)^3 2{\cal E}}
(f_{f} + f_{\overline f})\left[
\frac{2m_f^2 - 2p\cdot q}{q^2 + 2p\cdot q} + (q\rightarrow -q)\right]\nonumber\\
B_f & = & \int\frac{d^3{\cal P}}{(2\pi)^3 2{\cal E}}
(f_{f} + f_{\overline f})
\left[
\frac{2(p\cdot v)^2 + 2(p\cdot v)(q\cdot v) - p\cdot q}
{q^2 + 2p\cdot q} + (q\rightarrow -q)\right]\,.
\end{eqnarray}
For later purposes, it is also useful to note here that,
by inspection of the formulas for 
$D_f$ and $A_f$ given in Eqs.\ (\ref{ABCD}) and (\ref{ApBp}) respectively, 
it follows immediately that
\begin{equation}\label{ADrelation}
A_f = -2{\cal J}_f + (2m_f^2 + q^2)D_f \,,
\end{equation}
where
\begin{equation}\label{funnyJ}
{\cal J}_f = \int\frac{d^3{\cal P}}{(2\pi)^3 2{\cal E}}
(f_f + f_{\overline f})\,.
\end{equation}
From Eq.\ (\ref{piprotonproj}) we can immediately identify
the contribution of any fermion to 
the transverse and longitudinal part of the
self-energy, which we denote by $\pi^{(f)}_{T,L}$. 
Using Eq.\ (\ref{ABprime}) they can be expressed in the form
\begin{eqnarray}\label{RepiTLfermion}
\mbox{Re}\,\pi^{(f)}_T & = & -2e_f^2\left(A_f
+ \frac{q^2}{{\cal Q}^2}B_f\right) 
- 8e_f\kappa_f m_f q^2 D_f
- 2\kappa^2_f q^2\left[2m_f^2 D_f - \frac{q^2}{{\cal Q}^2}B_f\right]
\,,\nonumber\\
\mbox{Re}\,\pi^{(f)}_L & = & 4e_f^2\frac{q^2}{{\cal Q}^2}B_f
- 8e_f\kappa_f m_f q^2 D_f
- 4\kappa^2_f q^2\left[\frac{1}{2}A_f
+ \frac{q^2}{{\cal Q}^2}B_f + m_f^2 D_f\right]\,.
\end{eqnarray}
%
%
%

In this way, the contribution of any fermion
to the photon self-energy is expressed in terms of
the three functions $A_f, B_f, D_f$.
While the evaluation of these functions is not possible
in the general case, some useful results can be obtained
by considering special cases.
Of particular interest to us are the approximate expressions  
obtained for small values of $q$, 
which we analyze in some detail
in Section \ref{sec:limitingcases}.

\section{Low photon-momentum limit}\label{sec:limitingcases}

The functions $A_f,B_f,D_f$ are the same ones that appear
in the calculation of the induced electromagnetic vertex 
of a neutrino in a matter background,
which were analyzed in Refs.\ \cite{dnp1,dnnuclmag}
with considerable detail for various limiting
values of the photon momentum, and for several conditions
of a background gas made of nucleons and 
electrons.\footnote{The quantities denoted by $A_e,B_e$ in
the present paper were denoted by $A,B$ in 
Refs.\ \cite{dnp1,dnnuclmag}.} 
The results
obtained there are directly applicable here.  In this section
we expand upon those results further.  

In most of the situations of practical interest, the photon momentum
is such that  $\Omega, {\cal Q} \ll m_{p,n}$.
In this case, from Eq.\ (\ref{ADrelation}), it follows that
\begin{equation}\label{Dqsmall}
D_f \simeq \frac{1}{2m_f^2} ( A_f + 2{\cal J}_f) \,.
\end{equation}
Later on we use  this equation to evaluate $D_{n,p}$.
On the other hand,
borrowing the results given in Eqs.\ (A6) and (A11)
of Ref.\ \cite{dnnuclmag},  for small values of q we have
\begin{eqnarray}\label{ABqsmall}
B_f & = & -\frac{1}{2}\int\frac{d^3{\cal P}}{(2\pi)^3}
\left(\frac{\vec v_{\cal P}\cdot\vec{\cal Q}}{\Omega - \vec v_{\cal P}
\cdot\vec{\cal Q}}
\right)\frac{d}{d{\cal E}}(f_f + f_{\overline f})\,,\nonumber\\
A_f & = & B_f + \frac{\Omega}{2}\int\frac{d^3{\cal P}}{(2\pi)^3}
\left(
\frac{v_{\cal P}^2}
{\Omega - \vec v_{\cal P}\cdot\vec{\cal Q}}
\right)
\frac{d}{d{\cal E}}
(f_f + f_{\overline f})\,,
\end{eqnarray}
with $\vec v_{\cal P} = \vec{\cal P} / \Omega$ denoting the velocity
of the particles in the background.
The above formula for $B_f$ can be rewritten by multiplying
the integrand by the factor
\begin{equation}\label{factor}
\frac{1}{\Omega}(\Omega - \vec v_{\cal P}\cdot\vec{\cal Q} + \vec v_{\cal
P}\cdot\vec{\cal Q}) \,.
\end{equation}
The first two terms integrate to zero, while the third one leads to
\begin{equation}\label{Bsmallqequiv}
B_f = -\frac{1}{2\Omega}\int\frac{d^3{\cal P}}{(2\pi)^3}
\left(\frac{(\vec v_{\cal P}\cdot\vec{\cal Q})^2}{\Omega - \vec v_{\cal P}
\cdot\vec{\cal Q}}
\right)\frac{d}{d{\cal E}}(f_f + f_{\overline f}) \,.
\end{equation}
We should note that expressions in Eq.\ (\ref{ABqsmall}) are 
derived from Eq.\ (\ref{ApBp})
by expanding the integrands in terms of $q/{\cal E}$
and retaining only the terms that are dominant when
$q/{\cal E} \rightarrow 0$.  Therefore, Eq.\ (\ref{ABqsmall})
is valid for values of $q$ such that
\begin{equation}\label{smallqcond}
q/\langle{\cal E}\rangle \ll 1 \,,
\end{equation}
where $\langle{\cal E}\rangle$ denotes a typical average energy
of the fermions in the gas. Thus, for a non-relativistic gas,
Eq.\ (\ref{ABqsmall}) holds for $\Omega,{\cal Q} \ll m_f$.  If the
gas is extremely relativistic, Eq.\ (\ref{ABqsmall}) also holds
for $\Omega,{\cal Q} > m_f$, subject to Eq.\ (\ref{smallqcond}).

For an isotropic fermion distribution, an elementary integration on
the angular variables yields
\begin{eqnarray}\label{ABqsmall2}
B_f & = & \frac {1}{4 \pi^2}  \int_{0}^{\infty} {d{\cal P} {\cal P}^2
{d \over d{\cal E}} (f_f + f_{\overline f})
\left(1 - \frac{{\cal E}\Omega}{2 {\cal P} {\cal Q}}
\ln\left|\frac{{\cal E}\Omega + {\cal P}{\cal Q}}{{\cal E}\Omega  - {\cal
P}{\cal Q}}\right|\right)
}\,,\nonumber\\
A_f & = & B_f + \frac {1}{4 \pi^2}  \int_{0}^{\infty} {d{\cal P}  
\frac{ {\cal
P}^3}{\cal E}
{d \over d{\cal E}} (f_f + f_{\overline f})
 \frac{\Omega}{2{\cal Q}}
\ln\left|\frac{{\cal E}\Omega + {\cal P}{\cal Q}}{{\cal E}\Omega - {\cal
P}{\cal Q}}\right| } \,.
\end{eqnarray}
Up to this point no assumption has
been made regarding the nature of the fermion background.
Apart from the restriction on $q$, 
Eqs.\ (\ref{ABqsmall}) and (\ref{ABqsmall2}) hold for a relativistic
or non-relativistic gas, whether it is degenerate or not.
Accordingly,  Eq.\ (\ref{ABqsmall2}) serves as
a convenient starting point  to find the dispersion relations of
photons that propagate through an isotropic medium, provided
Eq.\ (\ref{smallqcond}) is verified. In what follows, we present
the explicit results for several important physical situations.

\subsection{Degenerate gas}
\label{subsec:degengas}

As functions of the energy,
the distribution for a degenerate gas and its derivative
are given by
\begin{eqnarray}\label{degendist}
f_f = \theta({\cal E}_{Ff} - {\cal E})\,,\nonumber\\
\frac{df_f}{d{\cal E}} = -\delta({\cal E}_{Ff} - {\cal E})\,.
\end{eqnarray}
As a consequence, the integrations needed to determine
$A_f, B_f, {\cal J}_f$ become very simple.
The results can be expressed in the form
\begin{eqnarray}\label{ABJdegen}
{\cal J}_f & = & \frac{m_f {\cal P}_{Ff}\gamma_f}{8\pi^2}
\left[1 - \frac{1}{2\gamma_f^2 v_{Ff}}
\log\left(\frac{1 + v_{Ff}}{1 - v_{Ff}}\right)
\right]\,,\nonumber\\
B_f & = & \frac{m_f{\cal P}_{Ff}\gamma_f}{4\pi^2}\left[-1 +
\frac{1}{2}z_f\log\left|\frac{1 + z_f}{1 - z_f}\right|\right]\,,\nonumber\\
A_f & = & \frac{m_f{\cal P}_{Ff}\gamma_f}{4\pi^2}\left[-1 + 
\frac{1}{2\gamma_f^2}z_f\log\left|\frac{1 + z_f}{1 - z_f}\right|\right]\,,
\end{eqnarray}
where
\begin{eqnarray}\label{ABJdegenaux}
\gamma_f & = & \frac{1}{\sqrt{1 - v_{Ff}^2}}\,,\nonumber\\
z_f& = & \frac{\Omega}{v_{Ff}{\cal Q}}\,,
\end{eqnarray}
with $v_{Ff} = {\cal P}_{Ff} / {\cal E}_{Ff}$ being the Fermi velocity of  
the gas.
Let us  note that
\begin{equation}
A_f (\Omega =  {\cal Q}) = - 2{\cal J}_f \,,
\end{equation}
which is valid in general,  for any distribution function  
and any kinematic regime of the gas, as can be easily checked from 
Eqs.\ (\ref{ApBp}) and (\ref{funnyJ}).

\subsection{Relativistic gas}\label{subsec:relgas}

If the fermions are ultrarelativistic, then ${\cal E} \simeq {\cal P}$ and
Eq.\ (\ref{ABqsmall2}) reduces to
\begin{eqnarray}\label{ABultra}
B_f & = & - 3 \omega^2_{0 f} \left(1 - \frac{\Omega}{2  {\cal Q}}
 \ln\left|\frac{\Omega + {\cal Q}}{\Omega - {\cal
Q}}\right|\right)\,,\nonumber\\
A_f & = & - 3  \omega^2_{0 f}  \,.
\end{eqnarray}
The quantity $\omega_{0f}$ is given by
\begin{equation}\label{omega0}
\omega^2_{0f} = \int\frac{d^3{\cal P}}{(2\pi)^3 2{\cal E}}
(f_f + f_{\overline f})\left[1 - \frac{{\cal P}^2}{3{\cal E}^2}
\right]\,,
\end{equation}
and in Eq.\ (\ref {ABultra}) we have used its limiting form
\begin{equation}\label{omega0ultra}
\omega^2_{0 f} =  \frac{1}{ 6\pi^2 }  \int_{0}^{\infty} {d{\cal P}
{\cal P} (f_f + f_{\overline f})}\,,
\end{equation}
for  ${\cal E} = { \cal P}$. In terms of  Eq.\ (\ref {omega0ultra})  
${\cal J}_f$ becomes
\begin{equation}\label{jota0ultra}
{\cal J}_f = \frac{3}{2} \omega^2_{0 f}\,.
\end{equation}

The remaining integral in Eq.\ (\ref{omega0ultra}) cannot be performed
without specifying the distribution function.  For a classical
(relativistic) gas $f_f = \mbox{exp} (- \beta_f  {\cal P} +  
\alpha_f)$  and

\begin{equation}\label{omega0ultracl}
\omega^2_{0 f} =  \frac{\beta_f}{12} n_f \,,
\end{equation}
while for a degenerate gas the integration over ${\cal P}$ yields
\begin{equation}\label{omega0ultradeg}
\omega^2_{0 f} =  \frac{1}{12} \left( \frac{3 n_f} {\pi} \right)^{2/3}
= \frac{ {\cal P}^2_{Ff}} {12 \pi^2}\,.
\end{equation}
The formulas for $A_f$ and $B_f$  given in Eq.\ (\ref{ABultra}),
with $\omega^2_{0 f}$ determined by Eq.\ (\ref{omega0ultradeg}), coincide
with the relativistic limit (${\cal E}_F \simeq {\cal P}_F$) of  
Eq.\ (\ref{ABJdegen}), as it should be.

\subsection{Non-relativistic, non-degenerate gas}
\label{subsec:nrnd}

The integrals cannot be carried out explicitly in this case.  However,
some useful approximate results can be obtained by considering
the two regions $\Omega \gg {\overline v}_f{\cal Q}$ and 
$\Omega \ll {\overline v}_f{\cal Q}$ separately, where 
\begin{equation}\label{vbar}
{\overline v}_f^2 \equiv \frac{1}{\beta_f m_f}\,,
\end{equation}
We have denoted the inverse temperature of the gas by $\beta_f$
in order to allow for the possibility that different components
of the background may be at different temperatures.

The results that we have obtained in this way can be written
in the form
\begin{eqnarray}\label{Bqsmallclass}
B_f & = & \frac{\beta_f n_f}{4}\left\{
\begin{array}{ll}
-1 + {\overline z}_f^2 &
\mbox{(for ${\overline z}_f\ll 1$)}\\
\frac{1}{{\overline z}_f^2} + \frac{3}{{\overline z}_f^4} &
\mbox{(for ${\overline z}_f \gg 1$)}\,,
\end{array}
\right. \\
\label{Aqsmallclass}
A_f & = & \frac{\beta_f n_f}{4}\left\{
\begin{array}{ll}
-1 + {\overline z}_f^2 - {\overline v}_f^2{\overline z}_f^2 &
\mbox{(for ${\overline z}_f\ll 1$)}\\
\frac{1}{{\overline z}_f^2} + \frac{3}{{\overline z}_f^4}
- {\overline v}_f^2\left(3 + \frac{5}{{\overline z}_f^2}\right) &
\mbox{(for ${\overline z}_f \gg 1$)}\,,
\end{array}
\right. \\
\label{J0class}
{\cal J}_f & = & \frac{n_f}{4m_f}\,,
\end{eqnarray}
with ${\overline v}_f$ as defined in Eq.\ (\ref{vbar}), and
\begin{equation}\label{zbar}
{\overline z}_f \equiv \frac{\Omega}{{\overline v}_f{\cal Q}}\,.
\end{equation}

As an illustration of how we have proceeded, let us consider 
in some detail the calculation of $B_f$.
For a classical non-relativistic gas, 
\begin{equation}\label{dfdE}
\frac{df_f}{d{\cal E}} = -\beta_f f_f \,,
\end{equation}
and from Eq.\ (\ref{ABqsmall}) 
\begin{equation}\label{Bclassaux}
B_f = -\frac{\beta_f n_f}{4} + \frac{\Omega\beta_f m_f}{4\pi^2{\cal Q}}
\int_0^\infty d{\cal P}{\cal P}f_f\log\left|\frac{1 + \xi}{1 - \xi}\right| \,,
\end{equation}
where
\begin{equation}\label{xi}
\xi\equiv \frac{\Omega}{v_{\cal P}{\cal Q}}\,.
\end{equation}
For the low frequency region, we proceed by putting
\begin{equation}\label{logsmallxi}
\log\left|\frac{1 + \xi}{1 - \xi}\right| \simeq 2\xi
\end{equation}
and then using the result
\begin{equation}\label{Isubp}
\int_0^\infty d{\cal P} f_f = \frac{\pi^2\beta_f n_f}{m_f}
\end{equation}
in Eq.\ (\ref{Bclassaux}).  In this way the result 
for $\Omega \ll {\overline v}_f{\cal Q}$ given in Eq.\ (\ref{Bqsmallclass}) follows.  
For the high frequency region, we put
\begin{equation}\label{loglargexi}
\log\left|\frac{1 + \xi}{1 - \xi}\right| \simeq
\frac{2}{\xi}\left(1 + \frac{1}{3\xi^2} + \frac{1}{5\xi^4}\right) \,,
\end{equation}
and then use, in addition to Eq.\ (\ref{Isubp}), the results
\begin{eqnarray}\label{Isubpderivs}
\int_0^\infty d{\cal P} {\cal P}^2 f_f & = & \pi^2 n_f \,,\nonumber\\
\int_0^\infty d{\cal P} {\cal P}^4 f_f & = & \frac{3\pi^2 m_f n_f}{\beta_f} \,,\nonumber\\
\int_0^\infty d{\cal P} {\cal P}^6 f_f & = & \frac{15\pi^2 m_f^2 n_f}{\beta_f^2}\,.
\end{eqnarray}
in Eq.\ (\ref{Bclassaux}).  

Similarly for $A_f$, following a similar procedure it is easy to show 
from Eq.\ (\ref{ABqsmall}) that
\begin{equation}\label{AminusB}
A_f - B_f = -\frac{n_f}{4m_f}\left\{
\begin{array}{ll}
\left(\frac{\Omega}{{\overline v}_f{\cal Q}}\right)^2\,, & 
\mbox{(for $\Omega\ll{\overline v}_f{\cal Q}$)}\\
3 + 5\left(\frac{{\overline v}_f{\cal Q}}{\Omega}\right)^2
& \mbox{(for $\Omega\gg{\overline v}_f{\cal Q}$)}
\end{array}
\right.
\end{equation}
Therefore, comparing with the formula for $B_f$ in Eq.\ (\ref{Bqsmallclass}), we arrive
at Eq.\ (\ref{Aqsmallclass}).
\section{Discussion and Applications}\label{sec:discussion}

In general, the dispersion relations are obtained by solving
Eq.\ (\ref{claseqmotion}) with $j_\mu = 0$ which, using
the decomposition in Eq.\ (\ref{pieffgen}) and the fact
that $\pi_P = 0$ in the present context, is equivalent
to solve
\begin{equation}\label{claseqmotionfree}
\left[R_{\mu\nu}(-q^2 + \pi_T) + Q_{\mu\nu}(-q^2 + \pi_L)\right]A^\nu = 0 \,.
\end{equation}
Since $R_{\mu\nu}$ and $Q_{\mu\nu}$ are orthogonal, the solutions
to this equation exist only for 
$q = (\omega_{T,L}({\cal Q}),\vec {\cal Q})$, where
$\omega_{T,L}({\cal Q})$ satisfy
\begin{eqnarray}\label{disprel}
\omega_T^2 - {\cal Q}^2 & = & \pi_T(\omega_T,{\cal Q})\,,\nonumber\\
\omega_L^2 - {\cal Q}^2 & = & \pi_L(\omega_L,{\cal Q}) \,.
\end{eqnarray}
In general, $\omega_{T,L}$ are complex functions of $\cal Q$
and can be written in the form
$\omega_{T,L} = \Omega_{T,L} - i\gamma_{T,L}/2$.   The quantities
$\Omega_{T,L}$ and $\gamma_{T,L}$ are real and have the interpretation
of being the dispersion relation and damping rate of the  
propagating mode, respectively.  
Retaining terms that are at most linear in  
$\gamma_{T, L}$ it follows that
\begin{eqnarray}
\Omega_{T,L}^2 - {\cal Q}^2 & = &\mbox{Re}\, \pi_{T,L}(\Omega_{T,L},{\cal
Q})\,,\nonumber\\
\frac {\gamma_{T,L} }{2}& = & -  \left[ \frac{\mbox{Im}\,\pi_{T,L}}
{{\partial \over \partial \omega}{(\omega^2 - \mbox{Re}\,\pi_{T,L})}}
\right]_{\omega = \Omega_{T,L}}
\end{eqnarray}

For a background  containing several fermions species,
\begin{equation}\label{realpitl}
\mbox {Re}\,\pi^{}_{T,L} = \sum_{f}^{}{\mbox{Re}\,\pi^{(f)}_{T,L} }\,.
\end{equation}
Each individual contribution in Eq.\ (\ref{realpitl}) can be immediately
identified from Eq.\ (\ref{piprotonproj}).
Thus, for the electron component we set $\kappa_e = 0$ while
for the neutron $e_n = 0$.
The corresponding expressions for the proton are more complicated due to
the presence of two terms in $j^{(em)}_{p\mu}$.

The explicit formulas given in Section \ref{sec:limitingcases}
allow us to study in some detail the photon self-energy
in a nuclear medium, provided of course that
the physical conditions are such that the approximations
and idealizations that led to them are valid in the particular
context in which they are being applied.  As an illustration, we consider
some possible situations.

\subsection{Electron background}\label{subsec:electronbackground}

Although the results for this case are well known,
we briefly review them here in a form that will be useful
for our later purposes.  Setting $\kappa_e = 0$
in Eq.\ (\ref{RepiTLfermion}),
\begin{eqnarray}\label{RepiTLe}
\mbox{Re}\,\pi^{(e)}_T & = & -2e^2\left(A_e + \frac{q^2}{{\cal Q}^2}B_e\right)
\,,\nonumber\\
\mbox{Re}\,\pi^{(e)}_L & = & 4e^2\frac{q^2}{{\cal Q}^2}B_e \,.
\end{eqnarray}

The functions $A_e$ and $B_e$ are in given in general
by Eq.\ (\ref{ApBp}) and, in the low $q$ regime, by Eq.\ (\ref{ABqsmall}).
By specializing these formulas
further to the zero frequency (static) limit or zero momentum (long wavelength)
limit,
we obtain explicit formulas for them, given explicitly in  
Refs.\ \cite{dnp1,dnnuclmag}, which are
useful in practical applications.  For example, from Eq.\ (3.9) of
Ref.\ \cite{dnnuclmag} we have
\begin{eqnarray}\label{ABQ0}
A_e(\Omega,{\cal Q}\rightarrow 0) & = & -3\omega^2_{0e} \,,\nonumber\\
B_e(\Omega,{\cal Q}\rightarrow 0) & = & 
\frac{{\cal Q}^2\omega^2_{0e}}{\Omega^2}\,.
\end{eqnarray}
Substituting Eq.\ (\ref{ABQ0}) into Eq.\ (\ref{RepiTLe}) we then obtain
\begin{equation}\label{RepiTLeQ0}
\mbox{Re}\,\pi^{(e)}_{T,L}(\Omega,0) = 4e^2\omega_{0e}^2 \,,
\end{equation}
which in turns implies the
well known result that the longitudinal and transverse permitivity
have the same long wavelength limit
\begin{equation}\label{permitQ0}
\epsilon_{t,l}(\Omega, 0) = 1 -
\frac{4e^2\omega_{0e}^2}{\Omega^2}\,.
\end{equation}
Eq.\ (\ref{RepiTLeQ0}), or equivalently \ref{permitQ0},
implies the well known expression for the electron
contribution to the plasma frequency, namely
\begin{equation}\label{plasmafreq}
\Omega_e = \sqrt{4e^2\omega_{0e}^2}\,,
\end{equation}
defined as the value $\Omega_{T,L}(0)$ of the dispersion relation
in the limit ${\cal Q}\rightarrow 0$. In particular,  
if the electrons are nonrelativistic, then  
$\omega^2_{0e} = n_e / 4 m_e $, as is immediately
seen from Eq.\ (\ref{omega0}), and $\Omega_e = \sqrt{e^2 n_e  / m_e}$.
In the relativistic regime, $\omega^2_{0e}$ is given
by Eqs.\ (\ref{omega0ultracl}) and (\ref{omega0ultradeg}).

For values of ${\cal Q} \not = 0$, using Eqs.\ (\ref{ABqsmall}) and (\ref{Bsmallqequiv}) 
in Eq.\ (\ref{RepiTLe}) we obtain
\begin{eqnarray}\label{Repiqsmall}
\mbox{Re}\,\pi^{(e)}_T & = &
-e^2\Omega\int\frac{d^3{\cal P}}{(2\pi)^3}
\left(\frac{\vec v_\perp\cdot\vec v_{\cal P}}{\Omega - \vec v_{\cal P}
\cdot\vec{\cal Q}}
\right)\frac{d}{d{\cal E}}(f_e + f_{\overline e}) \,, \nonumber\\
\mbox{Re}\,\pi^{(e)}_L & = & -\frac{2e^2q^2}{\Omega{\cal Q}^2}\int\frac{d^3{\cal
P}}{(2\pi)^3}
\left(\frac{(\vec v_{\cal P}\cdot\vec{\cal Q})^2}{\Omega - \vec v_{\cal P}
\cdot\vec{\cal Q}}
\right)\frac{d}{d{\cal E}}(f_e + f_{\overline e}) \,,
\end{eqnarray}
where
\begin{equation}\label{vperp}
\vec v_\perp = \vec v_{{\cal P}} - (\vec v_{{\cal P}}\cdot\hat{\cal Q})\hat{\cal
Q} \,.
\end{equation}
Substituting Eq.\ (\ref{RepiTLeqsmallur}) into Eq.\ (\ref{epsilonpirel})
we obtain the explicit formulas for the transverse
and longitudinal components $\epsilon_{t,l}$
of the dielectric constant,
that coincide precisely with those obtained 
by the semiclassical approach based on the Boltzman equation\cite{LL:physkin3}.
For the particular case of an ultrarelativistic electron gas,
using the result in Eq.\ (\ref{ABultra}) we obtain
\begin{eqnarray}\label{RepiTLeqsmallur}
\mbox{Re}\,\pi^{(e)}_T & = & 3 e^2\omega_{0e}^2\frac{\Omega}{{\cal Q}}
\left(\frac{2\Omega}{{\cal Q}} - \frac{q^2}{{\cal Q}^2}
\ln\left|\frac{\Omega + {\cal Q}}{\Omega - {\cal Q}}\right|\right)
\,,\nonumber\\
\mbox{Re}\,\pi^{(e)}_L & = & -12 e^2\omega_{0e}^2\frac{q^2}{{\cal Q}^2}
\left(1 - \frac{\Omega}{2{\cal Q}}
\ln\left|\frac{\Omega + {\cal Q}}{\Omega - {\cal Q}}\right|\right) \,.
\end{eqnarray}

Another useful limit corresponds to set $m_e\rightarrow 0$
in Eq.\ (\ref{ApBp}).  The formulas obtained
for $\mbox{Re}\,\pi^{(e)}_{T,L}$ by substituting the expressions so obtained
for $A_e$ and $B_e$ into Eq.\ (\ref{RepiTLe}),
coincide with those given by Weldon\cite{weldon:cov}.
Those formulas are valid in the regime $T,q\gg m_e$,
and neglecting the terms of order $q/T$ they coincide
precisely with the result given in Eq.\ (\ref{RepiTLeqsmallur}).  
In contrast, the formulas
given in Eq.\ (\ref{ABqsmall}) and the corresponding expressions
for $\mbox{Re}\,\pi^{(e)}_{T,L}$ obtained from Eq.\ (\ref{RepiTLe}) are valid
for $q \ll \langle{\cal E}\rangle$, but they hold whether $T$ is larger
or smaller than $m_e$.

\subsection{Degenerate neutron gas}
\label{subsec:degenneutgas}

The neutron contribution to the photon self-energy is
given by setting $e_n = 0$ in Eq.\ (\ref{RepiTLfermion}).  
Remembering Eq.\ (\ref{Dqsmall})
and using Eq.\ (\ref{ABJdegen}), we obtain for this case
\begin{eqnarray}\label{piTLdegenneut}
\mbox{Re}\,\pi^{(n)}_T & = & q^2\chi^{(n)}_0\gamma_n\left\{
\frac{1}{2\gamma_n^2}\left[\frac{1}{2v_{Fn}}
\log\left(\frac{1 + v_{Fn}}{1 - v_{Fn}}\right) -1\right]
+ \left[1 - \frac{1}{2}v_{Fn}^2(1 + z_n^2)\right]
\left[1 - \frac{1}{2}z_n\log\left|\frac{1 + z_n}{1 - z_n}\right|\right]\right\}
\,,\nonumber\\
\mbox{Re}\,\pi^{(n)}_L & = & q^2\chi^{(n)}_0\gamma_n v_{Fn}^2\left\{
\frac{1}{2v_{Fn}^2}\left[\frac{1}{2v_{Fn}\gamma_n^2}
\log\left(\frac{1 + v_{Fn}}{1 - v_{Fn}}\right) -1\right]
+ z_n^2 + \frac{1}{2}z_n(1 - z_n^2)\log\left|\frac{1 + z_n}{1 - z_n}\right|\right\}\,,
\end{eqnarray}
where we have defined
\begin{equation}\label{chi0}
\chi^{(n)}_0 \equiv \frac{\kappa_n^2 m_n{\cal P}_{Fn}}{\pi^2} \,.
\end{equation}
We emphasize that the above formulas are valid for all values of
$q$ subject only to $\Omega,{\cal Q}\ll m_n$, and also they hold
in the non-relativistic as well as the relativistic case. 
However, in the non-relativistic limit, they reduce to
\begin{eqnarray}\label{piTLdegenneutnr}
\mbox{Re}\,\pi^{(n)}_T & = & q^2\chi^{(n)}_0 f_T(z_n) \,,\nonumber\\
\mbox{Re}\,\pi^{(n)}_L & = & q^2\chi_0^{(n)}v_{Fn}^2 f_L(z_n)\,,
\end{eqnarray}
where
\begin{eqnarray}\label{fTL}
f_T(z) & = & \frac{1}{6}v_{F}^2 +
\left[1 - \frac{1}{2}v_{F}^2 z^2\right]
\left[1 - \frac{1}{2}z\log\left|\frac{1 + z}{1 - z}\right|\right]\,,\nonumber\\
f_L(z) & = & -\frac{1}{3} + z^2 + \frac{1}{2}z(1 - z^2)
\log\left|\frac{1 + z}{1 - z}\right| \,.
\end{eqnarray}
In the limits $\Omega\rightarrow 0$ or ${\cal Q}\rightarrow 0$
we obtain from Eq.\ (\ref{piTLdegenneutnr}) the limiting values
\begin{eqnarray}\label{piTLdegenneutnrstatic}
\mbox{Re}\,\pi^{(n)}_T(0,{\cal Q})& = & -{\cal Q}^2\chi^{(n)}_0\,,\nonumber\\
\mbox{Re}\,\pi^{(n)}_L(0,{\cal Q}) & = & \frac{1}{3}{\cal Q}^2\chi^{(n)}_0
v_{Fn}^2\,,\nonumber\\
\mbox{Re}\,\pi^{(n)}_T(\Omega,0) = 
\mbox{Re}\,\pi^{(n)}_L(\Omega,0) & = & \frac{1}{3}{\Omega}^2\chi^{(n)}_0v_{Fn}^2\,.
\end{eqnarray}
In particular, it is instructive to observe that
taking the static limit ($\Omega = 0$) in Eq.\ (\ref{mu}) and
using the results of Eq.\ (\ref{piTLdegenneutnrstatic}), we obtain
\begin{equation}\label{pauli}
\frac{1}{\mu(0,{\cal Q})} = 1 - \chi^{(n)}_0\,,
\end{equation}
which, glanzing at Eq.\ (\ref{chi0}), is recognized as
the classic Pauli formula for the static magnetic permeability
of a degenerate fermion gas.  For later reference, it is useful
to observe that in the nonrelativistic limit
\begin{equation}\label{chi0nr}
\chi^{(n)}_0 \simeq (3.6)\alpha v_{Fn}\,,
\end{equation}
which follows from Eq.\ (\ref{chi0}) by using Eq.\ (\ref{kappanp})
and setting ${\cal P}_{Fn}\approx m_n v_{Fn}$.

\begin{figure}
\begin{center}
\setlength{\unitlength}{0.240900pt}
\ifx\plotpoint\undefined\newsavebox{\plotpoint}\fi
\sbox{\plotpoint}{\rule[-0.200pt]{0.400pt}{0.400pt}}%
\begin{picture}(1125,675)(0,0)
\font\gnuplot=cmr10 at 10pt
\gnuplot
\sbox{\plotpoint}{\rule[-0.200pt]{0.400pt}{0.400pt}}%
\put(176.0,329.0){\rule[-0.200pt]{213.196pt}{0.400pt}}
\put(176.0,113.0){\rule[-0.200pt]{0.400pt}{129.845pt}}
\put(176.0,113.0){\rule[-0.200pt]{4.818pt}{0.400pt}}
\put(154,113){\makebox(0,0)[r]{-1}}
\put(1041.0,113.0){\rule[-0.200pt]{4.818pt}{0.400pt}}
\put(176.0,221.0){\rule[-0.200pt]{4.818pt}{0.400pt}}
\put(154,221){\makebox(0,0)[r]{-0.5}}
\put(1041.0,221.0){\rule[-0.200pt]{4.818pt}{0.400pt}}
\put(176.0,329.0){\rule[-0.200pt]{4.818pt}{0.400pt}}
\put(154,329){\makebox(0,0)[r]{0}}
\put(1041.0,329.0){\rule[-0.200pt]{4.818pt}{0.400pt}}
\put(176.0,436.0){\rule[-0.200pt]{4.818pt}{0.400pt}}
\put(154,436){\makebox(0,0)[r]{0.5}}
\put(1041.0,436.0){\rule[-0.200pt]{4.818pt}{0.400pt}}
\put(176.0,544.0){\rule[-0.200pt]{4.818pt}{0.400pt}}
\put(154,544){\makebox(0,0)[r]{1}}
\put(1041.0,544.0){\rule[-0.200pt]{4.818pt}{0.400pt}}
\put(176.0,652.0){\rule[-0.200pt]{4.818pt}{0.400pt}}
\put(154,652){\makebox(0,0)[r]{1.5}}
\put(1041.0,652.0){\rule[-0.200pt]{4.818pt}{0.400pt}}
\put(176.0,113.0){\rule[-0.200pt]{0.400pt}{4.818pt}}
\put(176,68){\makebox(0,0){0}}
\put(176.0,632.0){\rule[-0.200pt]{0.400pt}{4.818pt}}
\put(353.0,113.0){\rule[-0.200pt]{0.400pt}{4.818pt}}
\put(353,68){\makebox(0,0){2}}
\put(353.0,632.0){\rule[-0.200pt]{0.400pt}{4.818pt}}
\put(530.0,113.0){\rule[-0.200pt]{0.400pt}{4.818pt}}
\put(530,68){\makebox(0,0){4}}
\put(530.0,632.0){\rule[-0.200pt]{0.400pt}{4.818pt}}
\put(707.0,113.0){\rule[-0.200pt]{0.400pt}{4.818pt}}
\put(707,68){\makebox(0,0){6}}
\put(707.0,632.0){\rule[-0.200pt]{0.400pt}{4.818pt}}
\put(884.0,113.0){\rule[-0.200pt]{0.400pt}{4.818pt}}
\put(884,68){\makebox(0,0){8}}
\put(884.0,632.0){\rule[-0.200pt]{0.400pt}{4.818pt}}
\put(1061.0,113.0){\rule[-0.200pt]{0.400pt}{4.818pt}}
\put(1061,68){\makebox(0,0){10}}
\put(1061.0,632.0){\rule[-0.200pt]{0.400pt}{4.818pt}}
\put(176.0,113.0){\rule[-0.200pt]{213.196pt}{0.400pt}}
\put(1061.0,113.0){\rule[-0.200pt]{0.400pt}{129.845pt}}
\put(176.0,652.0){\rule[-0.200pt]{213.196pt}{0.400pt}}
\put(618,23){\makebox(0,0){$z$}}
\put(176.0,113.0){\rule[-0.200pt]{0.400pt}{129.845pt}}
\put(931,587){\makebox(0,0)[r]{$f_T, v_F = 0.1$}}
\put(953.0,587.0){\rule[-0.200pt]{15.899pt}{0.400pt}}
\put(176,545){\usebox{\plotpoint}}
\multiput(176.00,543.95)(1.802,-0.447){3}{\rule{1.300pt}{0.108pt}}
\multiput(176.00,544.17)(6.302,-3.000){2}{\rule{0.650pt}{0.400pt}}
\multiput(185.00,540.93)(0.762,-0.482){9}{\rule{0.700pt}{0.116pt}}
\multiput(185.00,541.17)(7.547,-6.000){2}{\rule{0.350pt}{0.400pt}}
\multiput(194.59,533.37)(0.489,-0.669){15}{\rule{0.118pt}{0.633pt}}
\multiput(193.17,534.69)(9.000,-10.685){2}{\rule{0.400pt}{0.317pt}}
\multiput(203.59,520.45)(0.489,-0.961){15}{\rule{0.118pt}{0.856pt}}
\multiput(202.17,522.22)(9.000,-15.224){2}{\rule{0.400pt}{0.428pt}}
\multiput(212.59,502.34)(0.489,-1.310){15}{\rule{0.118pt}{1.122pt}}
\multiput(211.17,504.67)(9.000,-20.671){2}{\rule{0.400pt}{0.561pt}}
\multiput(221.59,477.87)(0.489,-1.776){15}{\rule{0.118pt}{1.478pt}}
\multiput(220.17,480.93)(9.000,-27.933){2}{\rule{0.400pt}{0.739pt}}
\multiput(230.59,444.65)(0.489,-2.475){15}{\rule{0.118pt}{2.011pt}}
\multiput(229.17,448.83)(9.000,-38.826){2}{\rule{0.400pt}{1.006pt}}
\multiput(239.59,398.33)(0.489,-3.524){15}{\rule{0.118pt}{2.811pt}}
\multiput(238.17,404.17)(9.000,-55.165){2}{\rule{0.400pt}{1.406pt}}
\multiput(248.59,327.41)(0.488,-6.701){13}{\rule{0.117pt}{5.200pt}}
\multiput(247.17,338.21)(8.000,-91.207){2}{\rule{0.400pt}{2.600pt}}
\multiput(256.59,202.09)(0.477,-14.848){7}{\rule{0.115pt}{10.820pt}}
\multiput(255.17,224.54)(5.000,-111.543){2}{\rule{0.400pt}{5.410pt}}
\multiput(271.61,113.00)(0.447,17.653){3}{\rule{0.108pt}{10.767pt}}
\multiput(270.17,113.00)(3.000,57.653){2}{\rule{0.400pt}{5.383pt}}
\multiput(274.59,193.00)(0.489,2.650){15}{\rule{0.118pt}{2.144pt}}
\multiput(273.17,193.00)(9.000,41.549){2}{\rule{0.400pt}{1.072pt}}
\multiput(283.59,239.00)(0.489,1.310){15}{\rule{0.118pt}{1.122pt}}
\multiput(282.17,239.00)(9.000,20.671){2}{\rule{0.400pt}{0.561pt}}
\multiput(292.59,262.00)(0.489,0.786){15}{\rule{0.118pt}{0.722pt}}
\multiput(291.17,262.00)(9.000,12.501){2}{\rule{0.400pt}{0.361pt}}
\multiput(301.59,276.00)(0.489,0.553){15}{\rule{0.118pt}{0.544pt}}
\multiput(300.17,276.00)(9.000,8.870){2}{\rule{0.400pt}{0.272pt}}
\multiput(310.00,286.59)(0.645,0.485){11}{\rule{0.614pt}{0.117pt}}
\multiput(310.00,285.17)(7.725,7.000){2}{\rule{0.307pt}{0.400pt}}
\multiput(319.00,293.59)(0.933,0.477){7}{\rule{0.820pt}{0.115pt}}
\multiput(319.00,292.17)(7.298,5.000){2}{\rule{0.410pt}{0.400pt}}
\multiput(328.00,298.59)(0.933,0.477){7}{\rule{0.820pt}{0.115pt}}
\multiput(328.00,297.17)(7.298,5.000){2}{\rule{0.410pt}{0.400pt}}
\multiput(337.00,303.61)(1.802,0.447){3}{\rule{1.300pt}{0.108pt}}
\multiput(337.00,302.17)(6.302,3.000){2}{\rule{0.650pt}{0.400pt}}
\multiput(346.00,306.61)(1.802,0.447){3}{\rule{1.300pt}{0.108pt}}
\multiput(346.00,305.17)(6.302,3.000){2}{\rule{0.650pt}{0.400pt}}
\put(355,309.17){\rule{1.900pt}{0.400pt}}
\multiput(355.00,308.17)(5.056,2.000){2}{\rule{0.950pt}{0.400pt}}
\put(364,311.17){\rule{1.900pt}{0.400pt}}
\multiput(364.00,310.17)(5.056,2.000){2}{\rule{0.950pt}{0.400pt}}
\put(373,312.67){\rule{2.168pt}{0.400pt}}
\multiput(373.00,312.17)(4.500,1.000){2}{\rule{1.084pt}{0.400pt}}
\put(382,314.17){\rule{1.900pt}{0.400pt}}
\multiput(382.00,313.17)(5.056,2.000){2}{\rule{0.950pt}{0.400pt}}
\put(391,315.67){\rule{1.927pt}{0.400pt}}
\multiput(391.00,315.17)(4.000,1.000){2}{\rule{0.964pt}{0.400pt}}
\put(399,316.67){\rule{2.168pt}{0.400pt}}
\multiput(399.00,316.17)(4.500,1.000){2}{\rule{1.084pt}{0.400pt}}
\put(408,317.67){\rule{2.168pt}{0.400pt}}
\multiput(408.00,317.17)(4.500,1.000){2}{\rule{1.084pt}{0.400pt}}
\put(417,318.67){\rule{2.168pt}{0.400pt}}
\multiput(417.00,318.17)(4.500,1.000){2}{\rule{1.084pt}{0.400pt}}
\put(435,319.67){\rule{2.168pt}{0.400pt}}
\multiput(435.00,319.17)(4.500,1.000){2}{\rule{1.084pt}{0.400pt}}
\put(444,320.67){\rule{2.168pt}{0.400pt}}
\multiput(444.00,320.17)(4.500,1.000){2}{\rule{1.084pt}{0.400pt}}
\put(426.0,320.0){\rule[-0.200pt]{2.168pt}{0.400pt}}
\put(471,321.67){\rule{2.168pt}{0.400pt}}
\multiput(471.00,321.17)(4.500,1.000){2}{\rule{1.084pt}{0.400pt}}
\put(453.0,322.0){\rule[-0.200pt]{4.336pt}{0.400pt}}
\put(489,322.67){\rule{2.168pt}{0.400pt}}
\multiput(489.00,322.17)(4.500,1.000){2}{\rule{1.084pt}{0.400pt}}
\put(480.0,323.0){\rule[-0.200pt]{2.168pt}{0.400pt}}
\put(516,323.67){\rule{2.168pt}{0.400pt}}
\multiput(516.00,323.17)(4.500,1.000){2}{\rule{1.084pt}{0.400pt}}
\put(498.0,324.0){\rule[-0.200pt]{4.336pt}{0.400pt}}
\put(560,324.67){\rule{2.168pt}{0.400pt}}
\multiput(560.00,324.17)(4.500,1.000){2}{\rule{1.084pt}{0.400pt}}
\put(525.0,325.0){\rule[-0.200pt]{8.431pt}{0.400pt}}
\put(623,325.67){\rule{2.168pt}{0.400pt}}
\multiput(623.00,325.17)(4.500,1.000){2}{\rule{1.084pt}{0.400pt}}
\put(569.0,326.0){\rule[-0.200pt]{13.009pt}{0.400pt}}
\put(730,326.67){\rule{2.168pt}{0.400pt}}
\multiput(730.00,326.17)(4.500,1.000){2}{\rule{1.084pt}{0.400pt}}
\put(632.0,327.0){\rule[-0.200pt]{23.608pt}{0.400pt}}
\put(998,327.67){\rule{2.168pt}{0.400pt}}
\multiput(998.00,327.17)(4.500,1.000){2}{\rule{1.084pt}{0.400pt}}
\put(739.0,328.0){\rule[-0.200pt]{62.393pt}{0.400pt}}
\put(1007.0,329.0){\rule[-0.200pt]{13.009pt}{0.400pt}}
\put(931,542){\makebox(0,0)[r]{$f_T, v_F = 0.3$}}
\multiput(953,542)(20.756,0.000){4}{\usebox{\plotpoint}}
\put(1019,542){\usebox{\plotpoint}}
\put(176,547){\usebox{\plotpoint}}
\put(176.00,547.00){\usebox{\plotpoint}}
\multiput(185,545)(16.383,-12.743){0}{\usebox{\plotpoint}}
\put(194.08,537.89){\usebox{\plotpoint}}
\put(205.76,520.79){\usebox{\plotpoint}}
\put(214.60,502.06){\usebox{\plotpoint}}
\multiput(221,485)(5.787,-19.932){2}{\usebox{\plotpoint}}
\multiput(230,454)(4.252,-20.315){2}{\usebox{\plotpoint}}
\multiput(239,411)(3.079,-20.526){3}{\usebox{\plotpoint}}
\multiput(248,351)(1.672,-20.688){5}{\usebox{\plotpoint}}
\multiput(256,252)(0.746,-20.742){6}{\usebox{\plotpoint}}
\put(261,113){\usebox{\plotpoint}}
\multiput(271,113)(0.699,20.744){5}{\usebox{\plotpoint}}
\multiput(274,202)(4.070,20.352){2}{\usebox{\plotpoint}}
\put(286.77,256.63){\usebox{\plotpoint}}
\put(295.64,275.26){\usebox{\plotpoint}}
\put(308.59,291.43){\usebox{\plotpoint}}
\multiput(310,293)(16.383,12.743){0}{\usebox{\plotpoint}}
\put(325.34,303.52){\usebox{\plotpoint}}
\multiput(328,305)(18.967,8.430){0}{\usebox{\plotpoint}}
\put(344.46,311.49){\usebox{\plotpoint}}
\multiput(346,312)(19.690,6.563){0}{\usebox{\plotpoint}}
\multiput(355,315)(20.261,4.503){0}{\usebox{\plotpoint}}
\put(364.41,317.09){\usebox{\plotpoint}}
\multiput(373,319)(20.629,2.292){0}{\usebox{\plotpoint}}
\put(384.83,320.63){\usebox{\plotpoint}}
\multiput(391,322)(20.595,2.574){0}{\usebox{\plotpoint}}
\put(405.34,323.70){\usebox{\plotpoint}}
\multiput(408,324)(20.629,2.292){0}{\usebox{\plotpoint}}
\put(425.96,326.00){\usebox{\plotpoint}}
\multiput(426,326)(20.756,0.000){0}{\usebox{\plotpoint}}
\multiput(435,326)(20.629,2.292){0}{\usebox{\plotpoint}}
\put(446.66,327.00){\usebox{\plotpoint}}
\multiput(453,327)(20.629,2.292){0}{\usebox{\plotpoint}}
\put(467.36,328.00){\usebox{\plotpoint}}
\multiput(471,328)(20.629,2.292){0}{\usebox{\plotpoint}}
\put(488.06,329.00){\usebox{\plotpoint}}
\multiput(489,329)(20.629,2.292){0}{\usebox{\plotpoint}}
\multiput(498,330)(20.756,0.000){0}{\usebox{\plotpoint}}
\put(508.76,330.00){\usebox{\plotpoint}}
\multiput(516,330)(20.756,0.000){0}{\usebox{\plotpoint}}
\put(529.49,330.50){\usebox{\plotpoint}}
\multiput(534,331)(20.756,0.000){0}{\usebox{\plotpoint}}
\put(550.22,331.00){\usebox{\plotpoint}}
\multiput(551,331)(20.756,0.000){0}{\usebox{\plotpoint}}
\multiput(560,331)(20.756,0.000){0}{\usebox{\plotpoint}}
\put(570.96,331.22){\usebox{\plotpoint}}
\multiput(578,332)(20.756,0.000){0}{\usebox{\plotpoint}}
\put(591.68,332.00){\usebox{\plotpoint}}
\multiput(596,332)(20.756,0.000){0}{\usebox{\plotpoint}}
\put(612.43,332.00){\usebox{\plotpoint}}
\multiput(614,332)(20.756,0.000){0}{\usebox{\plotpoint}}
\multiput(623,332)(20.756,0.000){0}{\usebox{\plotpoint}}
\put(633.19,332.00){\usebox{\plotpoint}}
\multiput(641,332)(20.629,2.292){0}{\usebox{\plotpoint}}
\put(653.89,333.00){\usebox{\plotpoint}}
\multiput(659,333)(20.756,0.000){0}{\usebox{\plotpoint}}
\put(674.64,333.00){\usebox{\plotpoint}}
\multiput(677,333)(20.756,0.000){0}{\usebox{\plotpoint}}
\multiput(686,333)(20.756,0.000){0}{\usebox{\plotpoint}}
\put(695.40,333.00){\usebox{\plotpoint}}
\multiput(703,333)(20.756,0.000){0}{\usebox{\plotpoint}}
\put(716.15,333.00){\usebox{\plotpoint}}
\multiput(721,333)(20.756,0.000){0}{\usebox{\plotpoint}}
\put(736.91,333.00){\usebox{\plotpoint}}
\multiput(739,333)(20.756,0.000){0}{\usebox{\plotpoint}}
\multiput(748,333)(20.756,0.000){0}{\usebox{\plotpoint}}
\put(757.66,333.00){\usebox{\plotpoint}}
\multiput(766,333)(20.629,2.292){0}{\usebox{\plotpoint}}
\put(778.36,334.00){\usebox{\plotpoint}}
\multiput(784,334)(20.756,0.000){0}{\usebox{\plotpoint}}
\put(799.12,334.00){\usebox{\plotpoint}}
\multiput(802,334)(20.756,0.000){0}{\usebox{\plotpoint}}
\put(819.88,334.00){\usebox{\plotpoint}}
\multiput(820,334)(20.756,0.000){0}{\usebox{\plotpoint}}
\multiput(829,334)(20.756,0.000){0}{\usebox{\plotpoint}}
\put(840.63,334.00){\usebox{\plotpoint}}
\multiput(846,334)(20.756,0.000){0}{\usebox{\plotpoint}}
\put(861.39,334.00){\usebox{\plotpoint}}
\multiput(864,334)(20.756,0.000){0}{\usebox{\plotpoint}}
\multiput(873,334)(20.756,0.000){0}{\usebox{\plotpoint}}
\put(882.14,334.00){\usebox{\plotpoint}}
\multiput(891,334)(20.756,0.000){0}{\usebox{\plotpoint}}
\put(902.90,334.00){\usebox{\plotpoint}}
\multiput(909,334)(20.756,0.000){0}{\usebox{\plotpoint}}
\put(923.65,334.00){\usebox{\plotpoint}}
\multiput(927,334)(20.756,0.000){0}{\usebox{\plotpoint}}
\put(944.41,334.00){\usebox{\plotpoint}}
\multiput(945,334)(20.756,0.000){0}{\usebox{\plotpoint}}
\multiput(954,334)(20.756,0.000){0}{\usebox{\plotpoint}}
\put(965.16,334.00){\usebox{\plotpoint}}
\multiput(972,334)(20.756,0.000){0}{\usebox{\plotpoint}}
\put(985.92,334.00){\usebox{\plotpoint}}
\multiput(989,334)(20.756,0.000){0}{\usebox{\plotpoint}}
\put(1006.68,334.00){\usebox{\plotpoint}}
\multiput(1007,334)(20.756,0.000){0}{\usebox{\plotpoint}}
\multiput(1016,334)(20.756,0.000){0}{\usebox{\plotpoint}}
\put(1027.43,334.00){\usebox{\plotpoint}}
\multiput(1034,334)(20.756,0.000){0}{\usebox{\plotpoint}}
\put(1048.19,334.00){\usebox{\plotpoint}}
\multiput(1052,334)(20.756,0.000){0}{\usebox{\plotpoint}}
\put(1061,334){\usebox{\plotpoint}}
\sbox{\plotpoint}{\rule[-0.500pt]{1.000pt}{1.000pt}}%
\put(931,497){\makebox(0,0)[r]{$f_L$}}
\multiput(953,497)(20.756,0.000){4}{\usebox{\plotpoint}}
\put(1019,497){\usebox{\plotpoint}}
\put(176,257){\usebox{\plotpoint}}
\put(176.00,257.00){\usebox{\plotpoint}}
\put(191.21,269.97){\usebox{\plotpoint}}
\put(200.24,288.57){\usebox{\plotpoint}}
\put(207.21,308.10){\usebox{\plotpoint}}
\multiput(212,323)(5.311,20.064){2}{\usebox{\plotpoint}}
\multiput(221,357)(4.906,20.167){2}{\usebox{\plotpoint}}
\multiput(230,394)(4.783,20.197){2}{\usebox{\plotpoint}}
\put(243.56,448.72){\usebox{\plotpoint}}
\put(249.33,468.65){\usebox{\plotpoint}}
\multiput(256,487)(7.859,-19.210){2}{\usebox{\plotpoint}}
\put(271.43,447.15){\usebox{\plotpoint}}
\put(282.80,430.22){\usebox{\plotpoint}}
\multiput(283,430)(16.383,-12.743){0}{\usebox{\plotpoint}}
\put(300.28,419.32){\usebox{\plotpoint}}
\multiput(301,419)(19.690,-6.563){0}{\usebox{\plotpoint}}
\multiput(310,416)(20.261,-4.503){0}{\usebox{\plotpoint}}
\put(320.23,413.73){\usebox{\plotpoint}}
\multiput(328,412)(20.629,-2.292){0}{\usebox{\plotpoint}}
\put(340.65,410.19){\usebox{\plotpoint}}
\multiput(346,409)(20.629,-2.292){0}{\usebox{\plotpoint}}
\put(361.22,408.00){\usebox{\plotpoint}}
\multiput(364,408)(20.629,-2.292){0}{\usebox{\plotpoint}}
\put(381.87,406.01){\usebox{\plotpoint}}
\multiput(382,406)(20.756,0.000){0}{\usebox{\plotpoint}}
\multiput(391,406)(20.595,-2.574){0}{\usebox{\plotpoint}}
\put(402.56,405.00){\usebox{\plotpoint}}
\multiput(408,405)(20.756,0.000){0}{\usebox{\plotpoint}}
\put(423.27,404.30){\usebox{\plotpoint}}
\multiput(426,404)(20.756,0.000){0}{\usebox{\plotpoint}}
\multiput(435,404)(20.756,0.000){0}{\usebox{\plotpoint}}
\put(444.01,404.00){\usebox{\plotpoint}}
\multiput(453,404)(20.629,-2.292){0}{\usebox{\plotpoint}}
\put(464.71,403.00){\usebox{\plotpoint}}
\multiput(471,403)(20.756,0.000){0}{\usebox{\plotpoint}}
\put(485.47,403.00){\usebox{\plotpoint}}
\multiput(489,403)(20.756,0.000){0}{\usebox{\plotpoint}}
\put(506.22,403.00){\usebox{\plotpoint}}
\multiput(507,403)(20.629,-2.292){0}{\usebox{\plotpoint}}
\multiput(516,402)(20.756,0.000){0}{\usebox{\plotpoint}}
\put(526.92,402.00){\usebox{\plotpoint}}
\multiput(534,402)(20.756,0.000){0}{\usebox{\plotpoint}}
\put(547.68,402.00){\usebox{\plotpoint}}
\multiput(551,402)(20.756,0.000){0}{\usebox{\plotpoint}}
\put(568.44,402.00){\usebox{\plotpoint}}
\multiput(569,402)(20.756,0.000){0}{\usebox{\plotpoint}}
\multiput(578,402)(20.756,0.000){0}{\usebox{\plotpoint}}
\put(589.19,402.00){\usebox{\plotpoint}}
\multiput(596,402)(20.756,0.000){0}{\usebox{\plotpoint}}
\put(609.95,402.00){\usebox{\plotpoint}}
\multiput(614,402)(20.756,0.000){0}{\usebox{\plotpoint}}
\put(630.70,402.00){\usebox{\plotpoint}}
\multiput(632,402)(20.756,0.000){0}{\usebox{\plotpoint}}
\multiput(641,402)(20.629,-2.292){0}{\usebox{\plotpoint}}
\put(651.40,401.00){\usebox{\plotpoint}}
\multiput(659,401)(20.756,0.000){0}{\usebox{\plotpoint}}
\put(672.16,401.00){\usebox{\plotpoint}}
\multiput(677,401)(20.756,0.000){0}{\usebox{\plotpoint}}
\put(692.91,401.00){\usebox{\plotpoint}}
\multiput(694,401)(20.756,0.000){0}{\usebox{\plotpoint}}
\multiput(703,401)(20.756,0.000){0}{\usebox{\plotpoint}}
\put(713.67,401.00){\usebox{\plotpoint}}
\multiput(721,401)(20.756,0.000){0}{\usebox{\plotpoint}}
\put(734.42,401.00){\usebox{\plotpoint}}
\multiput(739,401)(20.756,0.000){0}{\usebox{\plotpoint}}
\put(755.18,401.00){\usebox{\plotpoint}}
\multiput(757,401)(20.756,0.000){0}{\usebox{\plotpoint}}
\multiput(766,401)(20.756,0.000){0}{\usebox{\plotpoint}}
\put(775.93,401.00){\usebox{\plotpoint}}
\multiput(784,401)(20.756,0.000){0}{\usebox{\plotpoint}}
\put(796.69,401.00){\usebox{\plotpoint}}
\multiput(802,401)(20.756,0.000){0}{\usebox{\plotpoint}}
\put(817.45,401.00){\usebox{\plotpoint}}
\multiput(820,401)(20.756,0.000){0}{\usebox{\plotpoint}}
\multiput(829,401)(20.756,0.000){0}{\usebox{\plotpoint}}
\put(838.20,401.00){\usebox{\plotpoint}}
\multiput(846,401)(20.756,0.000){0}{\usebox{\plotpoint}}
\put(858.96,401.00){\usebox{\plotpoint}}
\multiput(864,401)(20.756,0.000){0}{\usebox{\plotpoint}}
\put(879.71,401.00){\usebox{\plotpoint}}
\multiput(882,401)(20.756,0.000){0}{\usebox{\plotpoint}}
\multiput(891,401)(20.756,0.000){0}{\usebox{\plotpoint}}
\put(900.47,401.00){\usebox{\plotpoint}}
\multiput(909,401)(20.756,0.000){0}{\usebox{\plotpoint}}
\put(921.22,401.00){\usebox{\plotpoint}}
\multiput(927,401)(20.756,0.000){0}{\usebox{\plotpoint}}
\put(941.98,401.00){\usebox{\plotpoint}}
\multiput(945,401)(20.756,0.000){0}{\usebox{\plotpoint}}
\put(962.73,401.00){\usebox{\plotpoint}}
\multiput(963,401)(20.756,0.000){0}{\usebox{\plotpoint}}
\multiput(972,401)(20.756,0.000){0}{\usebox{\plotpoint}}
\put(983.49,401.00){\usebox{\plotpoint}}
\multiput(989,401)(20.756,0.000){0}{\usebox{\plotpoint}}
\put(1004.25,401.00){\usebox{\plotpoint}}
\multiput(1007,401)(20.756,0.000){0}{\usebox{\plotpoint}}
\multiput(1016,401)(20.756,0.000){0}{\usebox{\plotpoint}}
\put(1025.00,401.00){\usebox{\plotpoint}}
\multiput(1034,401)(20.756,0.000){0}{\usebox{\plotpoint}}
\put(1045.76,401.00){\usebox{\plotpoint}}
\multiput(1052,401)(20.756,0.000){0}{\usebox{\plotpoint}}
\put(1061,401){\usebox{\plotpoint}}
\end{picture}
\caption[Functions $f_{T,L}$]{
Plot of the functions $f_L$ and $f_T$ defined in Eq.\ (\ref{fTL}).  
At $z = 1$, $f_T$ becomes infinitely negative.
Even for relatively large values of $v_{F}$ the function $f_T$
has approximately the same form, and for small values of $v_{F}$
the curves fall almost on top of each other.
\label{fig:plotfTL}
}
\end{center}
\end{figure}
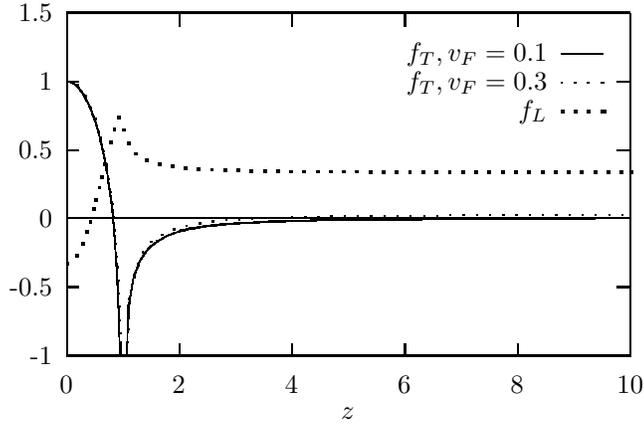

For our purposes, the primary quantities of interest
are $\pi_{T,L}$, which determine the dispersion relations
of the photon modes through Eq.\ (\ref{disprel}).
If the electron and proton contribution to the self-energy
can be neglected and we use Eq.\ (\ref{piTLdegenneutnr}) for
the neutron contribution, the dispersion relations
are obtained by solving the equations
\begin{eqnarray}\label{dispreldegneutgas}
\left(\Omega_{T}^2 - {\cal Q}^2\right)
\left[1 - \chi^{(n)}_0 f_{T}
\left(\frac{\Omega_{T}}{v_{Fn}{\cal Q}}\right)\right] & = & 0 \,,\nonumber\\
\left(\Omega_{L}^2 - {\cal Q}^2\right)
\left[1 - \chi^{(n)}_0 v_{Fn}^2 f_{L}
\left(\frac{\Omega_{L}}{v_{Fn}{\cal Q}}\right)\right] & = & 0 \,.
\end{eqnarray}
The functions $f_{T,L}$ are ploted in Fig.\ \ref{fig:plotfTL}.
As can be seen from the figure, both functions satisfy
$f_{T,L} \leq 1$.  On the other hand,
from Eq.\ (\ref{chi0nr}) we have $\chi^{(n)}_0 < 1$.
Therefore Eq.\ (\ref{dispreldegneutgas}) has no other solutions
than $\Omega_{T,L} = {\cal Q}$, so that the photon dispersion
relation is not modified by the presence of the
background in this case.

\subsection{Relativistic electron and degenerate neutron gas}
\label{subsec:neutelectron}

In general,  
the total background contribution to the
transverse and longitudinal components
of the photon self-energy in a matter background is
\begin{equation}\label{pitotal}
\pi_{T,L} = \pi^{(e)}_{T,L} + \pi^{(p)}_{T,L} + \pi^{(n)}_{T,L}\,.
\end{equation}
Let us neglect for the moment the contribution
from the protons, and assume that the electrons
can be represented by a relativistic gas and the neutrons
by a degenerate non-relativistic gas as considered in
Section \ref{subsec:degenneutgas}.  In this case
$\pi_{T,L}^{(e,n)}$ are given by Eqs.\ (\ref{RepiTLeqsmallur}) and (\ref{piTLdegenneutnr}), 
respectively.

The dispersion relations for the transverse modes
in this case are obtained by solving the equations
\begin{equation}\label{disprelneeq}
q^2 = \mbox{Re}\,\pi^{(n)}_T + \mbox{Re}\,\pi^{(e)}_T \,,
\end{equation}
which can be written in the form
\begin{equation}\label{disprelneeq2}
q^2\left(1 - \chi^{(n)}_0 f_T\right) = 3e^2\omega_{0e}^2\frac{\Omega}{{\cal Q}}
\left[\frac{2\Omega}{{\cal Q}} - \frac{q^2}{{\cal Q}^2}
\log\left|\frac{\Omega + {\cal Q}}{\Omega - {\cal Q}}\right|\right]\,.
\end{equation}
While it is not possible to find the general solution
to this equation, some useful conclusions can be drawn
from it.  For example, for values of $\Omega < {\cal Q}$,
the right hand side of Eq.\ (\ref{disprelneeq2}) is a positive
quantity, while the left hand side is negative since,
as we have seen above, $\chi^{(n)}_0 f_T < 1$.  Therefore,
the solution to Eq.\ (\ref{disprelneeq2}) is such that
$\Omega_T > {\cal Q}$.  This implies, in particular,
that a particle propagating such a medium cannot
emit transverse photons in the form of Cerenkov radiation.
Furthermore, since $z_n$ is considerably greater
than unity for $\Omega > {\cal Q}$, 
the value of $f_T(z_n)$ is negligible for that range 
and therefore the solution
to Eq.\ (\ref{disprelneeq2}) is well approximated by
the corresponding solution for the electron background only.

Notice that a possibly different, and erroneous,
conclusion would have been obtained if we were
to approximate the neutron contribution
in Eq.\ (\ref{disprelneeq}) by the value in the static
limit $\mbox{Re}\,\pi^{(n)}_T(0,{\cal Q})$ given in Eq.\ (\ref{piTLdegenneutnrstatic}).
Then, instead of Eq.\ (\ref{disprelneeq2}), such an approach
leads to the equation
\begin{equation}\label{disprelfake}
q^2 = -{\cal Q}^2\chi^{(n)}_0 + \mbox{Re}\,\pi_T^{(e)}
\end{equation}
for the dispersion relation,
which can have a solution such that 
$\Omega_T < {\cal Q}$ depending
on the relative size of the two competing
terms in the right hand side.

\begin{figure}
\begin{center}
\setlength{\unitlength}{0.240900pt}
\ifx\plotpoint\undefined\newsavebox{\plotpoint}\fi
\sbox{\plotpoint}{\rule[-0.200pt]{0.400pt}{0.400pt}}%
\begin{picture}(750,450)(0,0)
\font\gnuplot=cmr10 at 10pt
\gnuplot
\sbox{\plotpoint}{\rule[-0.200pt]{0.400pt}{0.400pt}}%
\put(176.0,113.0){\rule[-0.200pt]{122.859pt}{0.400pt}}
\put(176.0,113.0){\rule[-0.200pt]{0.400pt}{75.643pt}}
\put(176.0,113.0){\rule[-0.200pt]{4.818pt}{0.400pt}}
\put(154,113){\makebox(0,0)[r]{0}}
\put(666.0,113.0){\rule[-0.200pt]{4.818pt}{0.400pt}}
\put(176.0,176.0){\rule[-0.200pt]{4.818pt}{0.400pt}}
\put(154,176){\makebox(0,0)[r]{0.5}}
\put(666.0,176.0){\rule[-0.200pt]{4.818pt}{0.400pt}}
\put(176.0,239.0){\rule[-0.200pt]{4.818pt}{0.400pt}}
\put(154,239){\makebox(0,0)[r]{1}}
\put(666.0,239.0){\rule[-0.200pt]{4.818pt}{0.400pt}}
\put(176.0,301.0){\rule[-0.200pt]{4.818pt}{0.400pt}}
\put(154,301){\makebox(0,0)[r]{1.5}}
\put(666.0,301.0){\rule[-0.200pt]{4.818pt}{0.400pt}}
\put(176.0,364.0){\rule[-0.200pt]{4.818pt}{0.400pt}}
\put(154,364){\makebox(0,0)[r]{2}}
\put(666.0,364.0){\rule[-0.200pt]{4.818pt}{0.400pt}}
\put(176.0,427.0){\rule[-0.200pt]{4.818pt}{0.400pt}}
\put(154,427){\makebox(0,0)[r]{2.5}}
\put(666.0,427.0){\rule[-0.200pt]{4.818pt}{0.400pt}}
\put(176.0,113.0){\rule[-0.200pt]{0.400pt}{4.818pt}}
\put(176,68){\makebox(0,0){0}}
\put(176.0,407.0){\rule[-0.200pt]{0.400pt}{4.818pt}}
\put(278.0,113.0){\rule[-0.200pt]{0.400pt}{4.818pt}}
\put(278,68){\makebox(0,0){0.2}}
\put(278.0,407.0){\rule[-0.200pt]{0.400pt}{4.818pt}}
\put(380.0,113.0){\rule[-0.200pt]{0.400pt}{4.818pt}}
\put(380,68){\makebox(0,0){0.4}}
\put(380.0,407.0){\rule[-0.200pt]{0.400pt}{4.818pt}}
\put(482.0,113.0){\rule[-0.200pt]{0.400pt}{4.818pt}}
\put(482,68){\makebox(0,0){0.6}}
\put(482.0,407.0){\rule[-0.200pt]{0.400pt}{4.818pt}}
\put(584.0,113.0){\rule[-0.200pt]{0.400pt}{4.818pt}}
\put(584,68){\makebox(0,0){0.8}}
\put(584.0,407.0){\rule[-0.200pt]{0.400pt}{4.818pt}}
\put(686.0,113.0){\rule[-0.200pt]{0.400pt}{4.818pt}}
\put(686,68){\makebox(0,0){1}}
\put(686.0,407.0){\rule[-0.200pt]{0.400pt}{4.818pt}}
\put(176.0,113.0){\rule[-0.200pt]{122.859pt}{0.400pt}}
\put(686.0,113.0){\rule[-0.200pt]{0.400pt}{75.643pt}}
\put(176.0,427.0){\rule[-0.200pt]{122.859pt}{0.400pt}}
\put(431,23){\makebox(0,0){$\frac{\Omega}{{\cal Q}}$}}
\put(176.0,113.0){\rule[-0.200pt]{0.400pt}{75.643pt}}
\put(176,113){\usebox{\plotpoint}}
\put(207,112.67){\rule{1.204pt}{0.400pt}}
\multiput(207.00,112.17)(2.500,1.000){2}{\rule{0.602pt}{0.400pt}}
\put(176.0,113.0){\rule[-0.200pt]{7.468pt}{0.400pt}}
\put(228,113.67){\rule{1.204pt}{0.400pt}}
\multiput(228.00,113.17)(2.500,1.000){2}{\rule{0.602pt}{0.400pt}}
\put(212.0,114.0){\rule[-0.200pt]{3.854pt}{0.400pt}}
\put(243,114.67){\rule{1.204pt}{0.400pt}}
\multiput(243.00,114.17)(2.500,1.000){2}{\rule{0.602pt}{0.400pt}}
\put(233.0,115.0){\rule[-0.200pt]{2.409pt}{0.400pt}}
\put(258,115.67){\rule{1.445pt}{0.400pt}}
\multiput(258.00,115.17)(3.000,1.000){2}{\rule{0.723pt}{0.400pt}}
\put(248.0,116.0){\rule[-0.200pt]{2.409pt}{0.400pt}}
\put(269,116.67){\rule{1.204pt}{0.400pt}}
\multiput(269.00,116.17)(2.500,1.000){2}{\rule{0.602pt}{0.400pt}}
\put(264.0,117.0){\rule[-0.200pt]{1.204pt}{0.400pt}}
\put(279,117.67){\rule{1.204pt}{0.400pt}}
\multiput(279.00,117.17)(2.500,1.000){2}{\rule{0.602pt}{0.400pt}}
\put(274.0,118.0){\rule[-0.200pt]{1.204pt}{0.400pt}}
\put(289,118.67){\rule{1.204pt}{0.400pt}}
\multiput(289.00,118.17)(2.500,1.000){2}{\rule{0.602pt}{0.400pt}}
\put(294,119.67){\rule{1.445pt}{0.400pt}}
\multiput(294.00,119.17)(3.000,1.000){2}{\rule{0.723pt}{0.400pt}}
\put(284.0,119.0){\rule[-0.200pt]{1.204pt}{0.400pt}}
\put(305,120.67){\rule{1.204pt}{0.400pt}}
\multiput(305.00,120.17)(2.500,1.000){2}{\rule{0.602pt}{0.400pt}}
\put(310,121.67){\rule{1.204pt}{0.400pt}}
\multiput(310.00,121.17)(2.500,1.000){2}{\rule{0.602pt}{0.400pt}}
\put(300.0,121.0){\rule[-0.200pt]{1.204pt}{0.400pt}}
\put(320,122.67){\rule{1.204pt}{0.400pt}}
\multiput(320.00,122.17)(2.500,1.000){2}{\rule{0.602pt}{0.400pt}}
\put(325,123.67){\rule{1.445pt}{0.400pt}}
\multiput(325.00,123.17)(3.000,1.000){2}{\rule{0.723pt}{0.400pt}}
\put(331,124.67){\rule{1.204pt}{0.400pt}}
\multiput(331.00,124.17)(2.500,1.000){2}{\rule{0.602pt}{0.400pt}}
\put(336,125.67){\rule{1.204pt}{0.400pt}}
\multiput(336.00,125.17)(2.500,1.000){2}{\rule{0.602pt}{0.400pt}}
\put(341,126.67){\rule{1.204pt}{0.400pt}}
\multiput(341.00,126.17)(2.500,1.000){2}{\rule{0.602pt}{0.400pt}}
\put(315.0,123.0){\rule[-0.200pt]{1.204pt}{0.400pt}}
\put(351,127.67){\rule{1.204pt}{0.400pt}}
\multiput(351.00,127.17)(2.500,1.000){2}{\rule{0.602pt}{0.400pt}}
\put(356,128.67){\rule{1.204pt}{0.400pt}}
\multiput(356.00,128.17)(2.500,1.000){2}{\rule{0.602pt}{0.400pt}}
\put(361,129.67){\rule{1.445pt}{0.400pt}}
\multiput(361.00,129.17)(3.000,1.000){2}{\rule{0.723pt}{0.400pt}}
\put(367,131.17){\rule{1.100pt}{0.400pt}}
\multiput(367.00,130.17)(2.717,2.000){2}{\rule{0.550pt}{0.400pt}}
\put(372,132.67){\rule{1.204pt}{0.400pt}}
\multiput(372.00,132.17)(2.500,1.000){2}{\rule{0.602pt}{0.400pt}}
\put(377,133.67){\rule{1.204pt}{0.400pt}}
\multiput(377.00,133.17)(2.500,1.000){2}{\rule{0.602pt}{0.400pt}}
\put(382,134.67){\rule{1.204pt}{0.400pt}}
\multiput(382.00,134.17)(2.500,1.000){2}{\rule{0.602pt}{0.400pt}}
\put(387,135.67){\rule{1.204pt}{0.400pt}}
\multiput(387.00,135.17)(2.500,1.000){2}{\rule{0.602pt}{0.400pt}}
\put(392,136.67){\rule{1.445pt}{0.400pt}}
\multiput(392.00,136.17)(3.000,1.000){2}{\rule{0.723pt}{0.400pt}}
\put(398,138.17){\rule{1.100pt}{0.400pt}}
\multiput(398.00,137.17)(2.717,2.000){2}{\rule{0.550pt}{0.400pt}}
\put(403,139.67){\rule{1.204pt}{0.400pt}}
\multiput(403.00,139.17)(2.500,1.000){2}{\rule{0.602pt}{0.400pt}}
\put(408,140.67){\rule{1.204pt}{0.400pt}}
\multiput(408.00,140.17)(2.500,1.000){2}{\rule{0.602pt}{0.400pt}}
\put(413,142.17){\rule{1.100pt}{0.400pt}}
\multiput(413.00,141.17)(2.717,2.000){2}{\rule{0.550pt}{0.400pt}}
\put(418,143.67){\rule{1.204pt}{0.400pt}}
\multiput(418.00,143.17)(2.500,1.000){2}{\rule{0.602pt}{0.400pt}}
\put(423,145.17){\rule{1.100pt}{0.400pt}}
\multiput(423.00,144.17)(2.717,2.000){2}{\rule{0.550pt}{0.400pt}}
\put(428,146.67){\rule{1.445pt}{0.400pt}}
\multiput(428.00,146.17)(3.000,1.000){2}{\rule{0.723pt}{0.400pt}}
\put(434,148.17){\rule{1.100pt}{0.400pt}}
\multiput(434.00,147.17)(2.717,2.000){2}{\rule{0.550pt}{0.400pt}}
\put(439,149.67){\rule{1.204pt}{0.400pt}}
\multiput(439.00,149.17)(2.500,1.000){2}{\rule{0.602pt}{0.400pt}}
\put(444,151.17){\rule{1.100pt}{0.400pt}}
\multiput(444.00,150.17)(2.717,2.000){2}{\rule{0.550pt}{0.400pt}}
\put(449,153.17){\rule{1.100pt}{0.400pt}}
\multiput(449.00,152.17)(2.717,2.000){2}{\rule{0.550pt}{0.400pt}}
\put(454,155.17){\rule{1.100pt}{0.400pt}}
\multiput(454.00,154.17)(2.717,2.000){2}{\rule{0.550pt}{0.400pt}}
\put(459,157.17){\rule{1.100pt}{0.400pt}}
\multiput(459.00,156.17)(2.717,2.000){2}{\rule{0.550pt}{0.400pt}}
\put(464,158.67){\rule{1.445pt}{0.400pt}}
\multiput(464.00,158.17)(3.000,1.000){2}{\rule{0.723pt}{0.400pt}}
\put(470,160.17){\rule{1.100pt}{0.400pt}}
\multiput(470.00,159.17)(2.717,2.000){2}{\rule{0.550pt}{0.400pt}}
\put(475,162.17){\rule{1.100pt}{0.400pt}}
\multiput(475.00,161.17)(2.717,2.000){2}{\rule{0.550pt}{0.400pt}}
\put(480,164.17){\rule{1.100pt}{0.400pt}}
\multiput(480.00,163.17)(2.717,2.000){2}{\rule{0.550pt}{0.400pt}}
\multiput(485.00,166.61)(0.909,0.447){3}{\rule{0.767pt}{0.108pt}}
\multiput(485.00,165.17)(3.409,3.000){2}{\rule{0.383pt}{0.400pt}}
\put(490,169.17){\rule{1.100pt}{0.400pt}}
\multiput(490.00,168.17)(2.717,2.000){2}{\rule{0.550pt}{0.400pt}}
\put(495,171.17){\rule{1.300pt}{0.400pt}}
\multiput(495.00,170.17)(3.302,2.000){2}{\rule{0.650pt}{0.400pt}}
\put(501,173.17){\rule{1.100pt}{0.400pt}}
\multiput(501.00,172.17)(2.717,2.000){2}{\rule{0.550pt}{0.400pt}}
\multiput(506.00,175.61)(0.909,0.447){3}{\rule{0.767pt}{0.108pt}}
\multiput(506.00,174.17)(3.409,3.000){2}{\rule{0.383pt}{0.400pt}}
\put(511,178.17){\rule{1.100pt}{0.400pt}}
\multiput(511.00,177.17)(2.717,2.000){2}{\rule{0.550pt}{0.400pt}}
\multiput(516.00,180.61)(0.909,0.447){3}{\rule{0.767pt}{0.108pt}}
\multiput(516.00,179.17)(3.409,3.000){2}{\rule{0.383pt}{0.400pt}}
\multiput(521.00,183.61)(0.909,0.447){3}{\rule{0.767pt}{0.108pt}}
\multiput(521.00,182.17)(3.409,3.000){2}{\rule{0.383pt}{0.400pt}}
\put(526,186.17){\rule{1.100pt}{0.400pt}}
\multiput(526.00,185.17)(2.717,2.000){2}{\rule{0.550pt}{0.400pt}}
\multiput(531.00,188.61)(1.132,0.447){3}{\rule{0.900pt}{0.108pt}}
\multiput(531.00,187.17)(4.132,3.000){2}{\rule{0.450pt}{0.400pt}}
\multiput(537.00,191.61)(0.909,0.447){3}{\rule{0.767pt}{0.108pt}}
\multiput(537.00,190.17)(3.409,3.000){2}{\rule{0.383pt}{0.400pt}}
\multiput(542.00,194.61)(0.909,0.447){3}{\rule{0.767pt}{0.108pt}}
\multiput(542.00,193.17)(3.409,3.000){2}{\rule{0.383pt}{0.400pt}}
\multiput(547.00,197.61)(0.909,0.447){3}{\rule{0.767pt}{0.108pt}}
\multiput(547.00,196.17)(3.409,3.000){2}{\rule{0.383pt}{0.400pt}}
\multiput(552.00,200.60)(0.627,0.468){5}{\rule{0.600pt}{0.113pt}}
\multiput(552.00,199.17)(3.755,4.000){2}{\rule{0.300pt}{0.400pt}}
\multiput(557.00,204.61)(0.909,0.447){3}{\rule{0.767pt}{0.108pt}}
\multiput(557.00,203.17)(3.409,3.000){2}{\rule{0.383pt}{0.400pt}}
\multiput(562.00,207.60)(0.774,0.468){5}{\rule{0.700pt}{0.113pt}}
\multiput(562.00,206.17)(4.547,4.000){2}{\rule{0.350pt}{0.400pt}}
\multiput(568.00,211.60)(0.627,0.468){5}{\rule{0.600pt}{0.113pt}}
\multiput(568.00,210.17)(3.755,4.000){2}{\rule{0.300pt}{0.400pt}}
\multiput(573.00,215.61)(0.909,0.447){3}{\rule{0.767pt}{0.108pt}}
\multiput(573.00,214.17)(3.409,3.000){2}{\rule{0.383pt}{0.400pt}}
\multiput(578.00,218.59)(0.487,0.477){7}{\rule{0.500pt}{0.115pt}}
\multiput(578.00,217.17)(3.962,5.000){2}{\rule{0.250pt}{0.400pt}}
\multiput(583.00,223.60)(0.627,0.468){5}{\rule{0.600pt}{0.113pt}}
\multiput(583.00,222.17)(3.755,4.000){2}{\rule{0.300pt}{0.400pt}}
\multiput(588.00,227.60)(0.627,0.468){5}{\rule{0.600pt}{0.113pt}}
\multiput(588.00,226.17)(3.755,4.000){2}{\rule{0.300pt}{0.400pt}}
\multiput(593.00,231.59)(0.487,0.477){7}{\rule{0.500pt}{0.115pt}}
\multiput(593.00,230.17)(3.962,5.000){2}{\rule{0.250pt}{0.400pt}}
\multiput(598.00,236.59)(0.599,0.477){7}{\rule{0.580pt}{0.115pt}}
\multiput(598.00,235.17)(4.796,5.000){2}{\rule{0.290pt}{0.400pt}}
\multiput(604.00,241.59)(0.487,0.477){7}{\rule{0.500pt}{0.115pt}}
\multiput(604.00,240.17)(3.962,5.000){2}{\rule{0.250pt}{0.400pt}}
\multiput(609.59,246.00)(0.477,0.599){7}{\rule{0.115pt}{0.580pt}}
\multiput(608.17,246.00)(5.000,4.796){2}{\rule{0.400pt}{0.290pt}}
\multiput(614.59,252.00)(0.477,0.599){7}{\rule{0.115pt}{0.580pt}}
\multiput(613.17,252.00)(5.000,4.796){2}{\rule{0.400pt}{0.290pt}}
\multiput(619.59,258.00)(0.477,0.599){7}{\rule{0.115pt}{0.580pt}}
\multiput(618.17,258.00)(5.000,4.796){2}{\rule{0.400pt}{0.290pt}}
\multiput(624.59,264.00)(0.477,0.710){7}{\rule{0.115pt}{0.660pt}}
\multiput(623.17,264.00)(5.000,5.630){2}{\rule{0.400pt}{0.330pt}}
\multiput(629.59,271.00)(0.477,0.821){7}{\rule{0.115pt}{0.740pt}}
\multiput(628.17,271.00)(5.000,6.464){2}{\rule{0.400pt}{0.370pt}}
\multiput(634.59,279.00)(0.482,0.671){9}{\rule{0.116pt}{0.633pt}}
\multiput(633.17,279.00)(6.000,6.685){2}{\rule{0.400pt}{0.317pt}}
\multiput(640.59,287.00)(0.477,0.933){7}{\rule{0.115pt}{0.820pt}}
\multiput(639.17,287.00)(5.000,7.298){2}{\rule{0.400pt}{0.410pt}}
\multiput(645.59,296.00)(0.477,1.044){7}{\rule{0.115pt}{0.900pt}}
\multiput(644.17,296.00)(5.000,8.132){2}{\rule{0.400pt}{0.450pt}}
\multiput(650.59,306.00)(0.477,1.155){7}{\rule{0.115pt}{0.980pt}}
\multiput(649.17,306.00)(5.000,8.966){2}{\rule{0.400pt}{0.490pt}}
\multiput(655.59,317.00)(0.477,1.489){7}{\rule{0.115pt}{1.220pt}}
\multiput(654.17,317.00)(5.000,11.468){2}{\rule{0.400pt}{0.610pt}}
\multiput(660.59,331.00)(0.477,1.712){7}{\rule{0.115pt}{1.380pt}}
\multiput(659.17,331.00)(5.000,13.136){2}{\rule{0.400pt}{0.690pt}}
\multiput(665.59,347.00)(0.482,1.756){9}{\rule{0.116pt}{1.433pt}}
\multiput(664.17,347.00)(6.000,17.025){2}{\rule{0.400pt}{0.717pt}}
\multiput(671.59,367.00)(0.477,3.048){7}{\rule{0.115pt}{2.340pt}}
\multiput(670.17,367.00)(5.000,23.143){2}{\rule{0.400pt}{1.170pt}}
\multiput(676.61,395.00)(0.447,6.937){3}{\rule{0.108pt}{4.367pt}}
\multiput(675.17,395.00)(3.000,22.937){2}{\rule{0.400pt}{2.183pt}}
\put(346.0,128.0){\rule[-0.200pt]{1.204pt}{0.400pt}}
\put(176,239){\usebox{\plotpoint}}
\put(176.00,239.00){\usebox{\plotpoint}}
\multiput(181,239)(20.756,0.000){0}{\usebox{\plotpoint}}
\multiput(186,239)(20.756,0.000){0}{\usebox{\plotpoint}}
\put(196.76,239.00){\usebox{\plotpoint}}
\multiput(197,239)(20.756,0.000){0}{\usebox{\plotpoint}}
\multiput(202,239)(20.756,0.000){0}{\usebox{\plotpoint}}
\multiput(207,239)(20.756,0.000){0}{\usebox{\plotpoint}}
\multiput(212,239)(20.756,0.000){0}{\usebox{\plotpoint}}
\put(217.51,239.00){\usebox{\plotpoint}}
\multiput(222,239)(20.756,0.000){0}{\usebox{\plotpoint}}
\multiput(228,239)(20.756,0.000){0}{\usebox{\plotpoint}}
\multiput(233,239)(20.756,0.000){0}{\usebox{\plotpoint}}
\put(238.27,239.00){\usebox{\plotpoint}}
\multiput(243,239)(20.756,0.000){0}{\usebox{\plotpoint}}
\multiput(248,239)(20.756,0.000){0}{\usebox{\plotpoint}}
\multiput(253,239)(20.756,0.000){0}{\usebox{\plotpoint}}
\put(259.02,239.00){\usebox{\plotpoint}}
\multiput(264,239)(20.756,0.000){0}{\usebox{\plotpoint}}
\multiput(269,239)(20.756,0.000){0}{\usebox{\plotpoint}}
\multiput(274,239)(20.756,0.000){0}{\usebox{\plotpoint}}
\put(279.78,239.00){\usebox{\plotpoint}}
\multiput(284,239)(20.756,0.000){0}{\usebox{\plotpoint}}
\multiput(289,239)(20.756,0.000){0}{\usebox{\plotpoint}}
\multiput(294,239)(20.756,0.000){0}{\usebox{\plotpoint}}
\put(300.53,239.00){\usebox{\plotpoint}}
\multiput(305,239)(20.756,0.000){0}{\usebox{\plotpoint}}
\multiput(310,239)(20.756,0.000){0}{\usebox{\plotpoint}}
\multiput(315,239)(20.756,0.000){0}{\usebox{\plotpoint}}
\put(321.29,239.00){\usebox{\plotpoint}}
\multiput(325,239)(20.756,0.000){0}{\usebox{\plotpoint}}
\multiput(331,239)(20.756,0.000){0}{\usebox{\plotpoint}}
\multiput(336,239)(20.756,0.000){0}{\usebox{\plotpoint}}
\put(342.04,239.00){\usebox{\plotpoint}}
\multiput(346,239)(20.756,0.000){0}{\usebox{\plotpoint}}
\multiput(351,239)(20.756,0.000){0}{\usebox{\plotpoint}}
\multiput(356,239)(20.756,0.000){0}{\usebox{\plotpoint}}
\put(362.80,239.00){\usebox{\plotpoint}}
\multiput(367,239)(20.756,0.000){0}{\usebox{\plotpoint}}
\multiput(372,239)(20.756,0.000){0}{\usebox{\plotpoint}}
\multiput(377,239)(20.756,0.000){0}{\usebox{\plotpoint}}
\put(383.55,239.00){\usebox{\plotpoint}}
\multiput(387,239)(20.756,0.000){0}{\usebox{\plotpoint}}
\multiput(392,239)(20.756,0.000){0}{\usebox{\plotpoint}}
\multiput(398,239)(20.756,0.000){0}{\usebox{\plotpoint}}
\put(404.31,239.00){\usebox{\plotpoint}}
\multiput(408,239)(20.756,0.000){0}{\usebox{\plotpoint}}
\multiput(413,239)(20.756,0.000){0}{\usebox{\plotpoint}}
\multiput(418,239)(20.756,0.000){0}{\usebox{\plotpoint}}
\put(425.07,239.00){\usebox{\plotpoint}}
\multiput(428,239)(20.756,0.000){0}{\usebox{\plotpoint}}
\multiput(434,239)(20.756,0.000){0}{\usebox{\plotpoint}}
\multiput(439,239)(20.756,0.000){0}{\usebox{\plotpoint}}
\put(445.82,239.00){\usebox{\plotpoint}}
\multiput(449,239)(20.756,0.000){0}{\usebox{\plotpoint}}
\multiput(454,239)(20.756,0.000){0}{\usebox{\plotpoint}}
\multiput(459,239)(20.756,0.000){0}{\usebox{\plotpoint}}
\put(466.58,239.00){\usebox{\plotpoint}}
\multiput(470,239)(20.756,0.000){0}{\usebox{\plotpoint}}
\multiput(475,239)(20.756,0.000){0}{\usebox{\plotpoint}}
\multiput(480,239)(20.756,0.000){0}{\usebox{\plotpoint}}
\put(487.33,239.00){\usebox{\plotpoint}}
\multiput(490,239)(20.756,0.000){0}{\usebox{\plotpoint}}
\multiput(495,239)(20.756,0.000){0}{\usebox{\plotpoint}}
\multiput(501,239)(20.756,0.000){0}{\usebox{\plotpoint}}
\put(508.09,239.00){\usebox{\plotpoint}}
\multiput(511,239)(20.756,0.000){0}{\usebox{\plotpoint}}
\multiput(516,239)(20.756,0.000){0}{\usebox{\plotpoint}}
\multiput(521,239)(20.756,0.000){0}{\usebox{\plotpoint}}
\put(528.84,239.00){\usebox{\plotpoint}}
\multiput(531,239)(20.756,0.000){0}{\usebox{\plotpoint}}
\multiput(537,239)(20.756,0.000){0}{\usebox{\plotpoint}}
\multiput(542,239)(20.756,0.000){0}{\usebox{\plotpoint}}
\put(549.60,239.00){\usebox{\plotpoint}}
\multiput(552,239)(20.756,0.000){0}{\usebox{\plotpoint}}
\multiput(557,239)(20.756,0.000){0}{\usebox{\plotpoint}}
\multiput(562,239)(20.756,0.000){0}{\usebox{\plotpoint}}
\put(570.35,239.00){\usebox{\plotpoint}}
\multiput(573,239)(20.756,0.000){0}{\usebox{\plotpoint}}
\multiput(578,239)(20.756,0.000){0}{\usebox{\plotpoint}}
\multiput(583,239)(20.756,0.000){0}{\usebox{\plotpoint}}
\put(591.11,239.00){\usebox{\plotpoint}}
\multiput(593,239)(20.756,0.000){0}{\usebox{\plotpoint}}
\multiput(598,239)(20.756,0.000){0}{\usebox{\plotpoint}}
\multiput(604,239)(20.756,0.000){0}{\usebox{\plotpoint}}
\put(611.87,239.00){\usebox{\plotpoint}}
\multiput(614,239)(20.756,0.000){0}{\usebox{\plotpoint}}
\multiput(619,239)(20.756,0.000){0}{\usebox{\plotpoint}}
\multiput(624,239)(20.756,0.000){0}{\usebox{\plotpoint}}
\put(632.62,239.00){\usebox{\plotpoint}}
\multiput(634,239)(20.756,0.000){0}{\usebox{\plotpoint}}
\multiput(640,239)(20.756,0.000){0}{\usebox{\plotpoint}}
\multiput(645,239)(20.756,0.000){0}{\usebox{\plotpoint}}
\put(653.38,239.00){\usebox{\plotpoint}}
\multiput(655,239)(20.756,0.000){0}{\usebox{\plotpoint}}
\multiput(660,239)(20.756,0.000){0}{\usebox{\plotpoint}}
\multiput(665,239)(20.756,0.000){0}{\usebox{\plotpoint}}
\put(674.13,239.00){\usebox{\plotpoint}}
\multiput(676,239)(20.756,0.000){0}{\usebox{\plotpoint}}
\multiput(681,239)(20.756,0.000){0}{\usebox{\plotpoint}}
\put(686,239){\usebox{\plotpoint}}
\end{picture}
\caption[Factor in $\pi^{(e)}_L$]{
The factor $\frac{2\Omega}{{\cal Q}}\log\left|\frac{\Omega + {\cal Q}}
{\Omega - {\cal Q}}\right|$ is plotted as a function of $\frac{\Omega}{{\cal Q}}$.
\label{fig:fefactor}}
\end{center}
\end{figure}
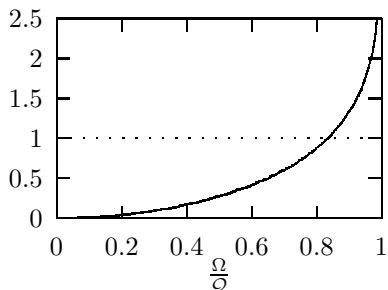

For the longitudinal modes on the other hand,
the dispersion relation is determined by solving the equation
\begin{equation}\label{disprelLne}
q^2\left(1 - \chi^{(n)}_0 v_{Fn}^2 f_L\right) = \mbox{Re}\,\pi_L^{(e)}\,,
\end{equation}
with $\mbox{Re}\,\pi_L^{(e)}$ given by Eq.\ (\ref{RepiTLeqsmallur}).    
The factor
\begin{equation}\label{logpart}
\frac{\Omega}{2{\cal Q}}
\ln\left|\frac{\Omega + {\cal Q}}{\Omega - {\cal Q}}\right| \,,
\end{equation}
which is plotted in Fig.\ \ref{fig:fefactor},
is larger than unity for values of $\Omega/{\cal Q}$ larger
than about 0.84.
Therefore, 
there is a small range
\begin{equation}\label{cerenkrange}
0.84\stackrel{<}{\sim} \frac{\Omega}{{\cal Q}} < 1 \,,
\end{equation}
for which the solution to Eq.\ (\ref{disprelLne}) is such that $\Omega_L < {\cal Q}$.
Hence, the Cerenkov radiation of longitudinal photons
is allowed.  For those values of $\Omega/{\cal Q}$, the value of
$z_n$ corresponds to the region in Fig.\ \ref{fig:plotfTL} 
where the function $f_L$ is negligible. Therefore, 
the neutron component of the background
does not play a significant role in the photon dispersion
relation in this case either.

\subsection{Relativistic electron and degenerate neutron and proton gases}
\label{subsec:neutprotelectron}

\begin{figure}
\begin{center}
\setlength{\unitlength}{0.240900pt}
\ifx\plotpoint\undefined\newsavebox{\plotpoint}\fi
\sbox{\plotpoint}{\rule[-0.200pt]{0.400pt}{0.400pt}}%
\begin{picture}(900,540)(0,0)
\font\gnuplot=cmr10 at 10pt
\gnuplot
\sbox{\plotpoint}{\rule[-0.200pt]{0.400pt}{0.400pt}}%
\put(176.0,248.0){\rule[-0.200pt]{158.994pt}{0.400pt}}
\put(176.0,113.0){\rule[-0.200pt]{0.400pt}{97.324pt}}
\put(176.0,113.0){\rule[-0.200pt]{4.818pt}{0.400pt}}
\put(154,113){\makebox(0,0)[r]{-1}}
\put(816.0,113.0){\rule[-0.200pt]{4.818pt}{0.400pt}}
\put(176.0,180.0){\rule[-0.200pt]{4.818pt}{0.400pt}}
\put(154,180){\makebox(0,0)[r]{-0.5}}
\put(816.0,180.0){\rule[-0.200pt]{4.818pt}{0.400pt}}
\put(176.0,248.0){\rule[-0.200pt]{4.818pt}{0.400pt}}
\put(154,248){\makebox(0,0)[r]{0}}
\put(816.0,248.0){\rule[-0.200pt]{4.818pt}{0.400pt}}
\put(176.0,315.0){\rule[-0.200pt]{4.818pt}{0.400pt}}
\put(154,315){\makebox(0,0)[r]{0.5}}
\put(816.0,315.0){\rule[-0.200pt]{4.818pt}{0.400pt}}
\put(176.0,382.0){\rule[-0.200pt]{4.818pt}{0.400pt}}
\put(154,382){\makebox(0,0)[r]{1}}
\put(816.0,382.0){\rule[-0.200pt]{4.818pt}{0.400pt}}
\put(176.0,450.0){\rule[-0.200pt]{4.818pt}{0.400pt}}
\put(154,450){\makebox(0,0)[r]{1.5}}
\put(816.0,450.0){\rule[-0.200pt]{4.818pt}{0.400pt}}
\put(176.0,517.0){\rule[-0.200pt]{4.818pt}{0.400pt}}
\put(154,517){\makebox(0,0)[r]{2}}
\put(816.0,517.0){\rule[-0.200pt]{4.818pt}{0.400pt}}
\put(176.0,113.0){\rule[-0.200pt]{0.400pt}{4.818pt}}
\put(176,68){\makebox(0,0){0}}
\put(176.0,497.0){\rule[-0.200pt]{0.400pt}{4.818pt}}
\put(259.0,113.0){\rule[-0.200pt]{0.400pt}{4.818pt}}
\put(259,68){\makebox(0,0){0.5}}
\put(259.0,497.0){\rule[-0.200pt]{0.400pt}{4.818pt}}
\put(341.0,113.0){\rule[-0.200pt]{0.400pt}{4.818pt}}
\put(341,68){\makebox(0,0){1}}
\put(341.0,497.0){\rule[-0.200pt]{0.400pt}{4.818pt}}
\put(424.0,113.0){\rule[-0.200pt]{0.400pt}{4.818pt}}
\put(424,68){\makebox(0,0){1.5}}
\put(424.0,497.0){\rule[-0.200pt]{0.400pt}{4.818pt}}
\put(506.0,113.0){\rule[-0.200pt]{0.400pt}{4.818pt}}
\put(506,68){\makebox(0,0){2}}
\put(506.0,497.0){\rule[-0.200pt]{0.400pt}{4.818pt}}
\put(589.0,113.0){\rule[-0.200pt]{0.400pt}{4.818pt}}
\put(589,68){\makebox(0,0){2.5}}
\put(589.0,497.0){\rule[-0.200pt]{0.400pt}{4.818pt}}
\put(671.0,113.0){\rule[-0.200pt]{0.400pt}{4.818pt}}
\put(671,68){\makebox(0,0){3}}
\put(671.0,497.0){\rule[-0.200pt]{0.400pt}{4.818pt}}
\put(754.0,113.0){\rule[-0.200pt]{0.400pt}{4.818pt}}
\put(754,68){\makebox(0,0){3.5}}
\put(754.0,497.0){\rule[-0.200pt]{0.400pt}{4.818pt}}
\put(836.0,113.0){\rule[-0.200pt]{0.400pt}{4.818pt}}
\put(836,68){\makebox(0,0){4}}
\put(836.0,497.0){\rule[-0.200pt]{0.400pt}{4.818pt}}
\put(176.0,113.0){\rule[-0.200pt]{158.994pt}{0.400pt}}
\put(836.0,113.0){\rule[-0.200pt]{0.400pt}{97.324pt}}
\put(176.0,517.0){\rule[-0.200pt]{158.994pt}{0.400pt}}
\put(506,23){\makebox(0,0){$z$}}
\put(176.0,113.0){\rule[-0.200pt]{0.400pt}{97.324pt}}
\put(706,452){\makebox(0,0)[r]{$g_T$}}
\put(728.0,452.0){\rule[-0.200pt]{15.899pt}{0.400pt}}
\put(176,248){\usebox{\plotpoint}}
\put(183,247.67){\rule{1.445pt}{0.400pt}}
\multiput(183.00,247.17)(3.000,1.000){2}{\rule{0.723pt}{0.400pt}}
\multiput(189.00,249.61)(1.355,0.447){3}{\rule{1.033pt}{0.108pt}}
\multiput(189.00,248.17)(4.855,3.000){2}{\rule{0.517pt}{0.400pt}}
\multiput(196.00,252.61)(1.355,0.447){3}{\rule{1.033pt}{0.108pt}}
\multiput(196.00,251.17)(4.855,3.000){2}{\rule{0.517pt}{0.400pt}}
\multiput(203.00,255.60)(0.774,0.468){5}{\rule{0.700pt}{0.113pt}}
\multiput(203.00,254.17)(4.547,4.000){2}{\rule{0.350pt}{0.400pt}}
\multiput(209.00,259.60)(0.920,0.468){5}{\rule{0.800pt}{0.113pt}}
\multiput(209.00,258.17)(5.340,4.000){2}{\rule{0.400pt}{0.400pt}}
\multiput(216.00,263.59)(0.581,0.482){9}{\rule{0.567pt}{0.116pt}}
\multiput(216.00,262.17)(5.824,6.000){2}{\rule{0.283pt}{0.400pt}}
\multiput(223.00,269.59)(0.491,0.482){9}{\rule{0.500pt}{0.116pt}}
\multiput(223.00,268.17)(4.962,6.000){2}{\rule{0.250pt}{0.400pt}}
\multiput(229.00,275.59)(0.492,0.485){11}{\rule{0.500pt}{0.117pt}}
\multiput(229.00,274.17)(5.962,7.000){2}{\rule{0.250pt}{0.400pt}}
\multiput(236.00,282.59)(0.492,0.485){11}{\rule{0.500pt}{0.117pt}}
\multiput(236.00,281.17)(5.962,7.000){2}{\rule{0.250pt}{0.400pt}}
\multiput(243.59,289.00)(0.482,0.671){9}{\rule{0.116pt}{0.633pt}}
\multiput(242.17,289.00)(6.000,6.685){2}{\rule{0.400pt}{0.317pt}}
\multiput(249.59,297.00)(0.485,0.645){11}{\rule{0.117pt}{0.614pt}}
\multiput(248.17,297.00)(7.000,7.725){2}{\rule{0.400pt}{0.307pt}}
\multiput(256.59,306.00)(0.485,0.645){11}{\rule{0.117pt}{0.614pt}}
\multiput(255.17,306.00)(7.000,7.725){2}{\rule{0.400pt}{0.307pt}}
\multiput(263.59,315.00)(0.482,0.762){9}{\rule{0.116pt}{0.700pt}}
\multiput(262.17,315.00)(6.000,7.547){2}{\rule{0.400pt}{0.350pt}}
\multiput(269.59,324.00)(0.485,0.645){11}{\rule{0.117pt}{0.614pt}}
\multiput(268.17,324.00)(7.000,7.725){2}{\rule{0.400pt}{0.307pt}}
\multiput(276.59,333.00)(0.485,0.721){11}{\rule{0.117pt}{0.671pt}}
\multiput(275.17,333.00)(7.000,8.606){2}{\rule{0.400pt}{0.336pt}}
\multiput(283.59,343.00)(0.482,0.762){9}{\rule{0.116pt}{0.700pt}}
\multiput(282.17,343.00)(6.000,7.547){2}{\rule{0.400pt}{0.350pt}}
\multiput(289.59,352.00)(0.485,0.645){11}{\rule{0.117pt}{0.614pt}}
\multiput(288.17,352.00)(7.000,7.725){2}{\rule{0.400pt}{0.307pt}}
\multiput(296.59,361.00)(0.485,0.645){11}{\rule{0.117pt}{0.614pt}}
\multiput(295.17,361.00)(7.000,7.725){2}{\rule{0.400pt}{0.307pt}}
\multiput(303.59,370.00)(0.482,0.671){9}{\rule{0.116pt}{0.633pt}}
\multiput(302.17,370.00)(6.000,6.685){2}{\rule{0.400pt}{0.317pt}}
\multiput(309.00,378.59)(0.492,0.485){11}{\rule{0.500pt}{0.117pt}}
\multiput(309.00,377.17)(5.962,7.000){2}{\rule{0.250pt}{0.400pt}}
\multiput(316.00,385.59)(0.710,0.477){7}{\rule{0.660pt}{0.115pt}}
\multiput(316.00,384.17)(5.630,5.000){2}{\rule{0.330pt}{0.400pt}}
\put(323,390.17){\rule{1.300pt}{0.400pt}}
\multiput(323.00,389.17)(3.302,2.000){2}{\rule{0.650pt}{0.400pt}}
\put(329,390.67){\rule{1.686pt}{0.400pt}}
\multiput(329.00,391.17)(3.500,-1.000){2}{\rule{0.843pt}{0.400pt}}
\multiput(336.59,387.50)(0.485,-0.950){11}{\rule{0.117pt}{0.843pt}}
\multiput(335.17,389.25)(7.000,-11.251){2}{\rule{0.400pt}{0.421pt}}
\multiput(343.59,375.09)(0.482,-0.762){9}{\rule{0.116pt}{0.700pt}}
\multiput(342.17,376.55)(6.000,-7.547){2}{\rule{0.400pt}{0.350pt}}
\multiput(349.00,367.93)(0.710,-0.477){7}{\rule{0.660pt}{0.115pt}}
\multiput(349.00,368.17)(5.630,-5.000){2}{\rule{0.330pt}{0.400pt}}
\multiput(356.00,362.95)(1.355,-0.447){3}{\rule{1.033pt}{0.108pt}}
\multiput(356.00,363.17)(4.855,-3.000){2}{\rule{0.517pt}{0.400pt}}
\multiput(363.00,359.95)(1.132,-0.447){3}{\rule{0.900pt}{0.108pt}}
\multiput(363.00,360.17)(4.132,-3.000){2}{\rule{0.450pt}{0.400pt}}
\put(369,356.17){\rule{1.500pt}{0.400pt}}
\multiput(369.00,357.17)(3.887,-2.000){2}{\rule{0.750pt}{0.400pt}}
\put(376,354.17){\rule{1.500pt}{0.400pt}}
\multiput(376.00,355.17)(3.887,-2.000){2}{\rule{0.750pt}{0.400pt}}
\put(383,352.17){\rule{1.300pt}{0.400pt}}
\multiput(383.00,353.17)(3.302,-2.000){2}{\rule{0.650pt}{0.400pt}}
\put(389,350.67){\rule{1.686pt}{0.400pt}}
\multiput(389.00,351.17)(3.500,-1.000){2}{\rule{0.843pt}{0.400pt}}
\put(396,349.67){\rule{1.686pt}{0.400pt}}
\multiput(396.00,350.17)(3.500,-1.000){2}{\rule{0.843pt}{0.400pt}}
\put(403,348.67){\rule{1.445pt}{0.400pt}}
\multiput(403.00,349.17)(3.000,-1.000){2}{\rule{0.723pt}{0.400pt}}
\put(409,347.67){\rule{1.686pt}{0.400pt}}
\multiput(409.00,348.17)(3.500,-1.000){2}{\rule{0.843pt}{0.400pt}}
\put(176.0,248.0){\rule[-0.200pt]{1.686pt}{0.400pt}}
\put(423,346.67){\rule{1.445pt}{0.400pt}}
\multiput(423.00,347.17)(3.000,-1.000){2}{\rule{0.723pt}{0.400pt}}
\put(429,345.67){\rule{1.686pt}{0.400pt}}
\multiput(429.00,346.17)(3.500,-1.000){2}{\rule{0.843pt}{0.400pt}}
\put(416.0,348.0){\rule[-0.200pt]{1.686pt}{0.400pt}}
\put(443,344.67){\rule{1.445pt}{0.400pt}}
\multiput(443.00,345.17)(3.000,-1.000){2}{\rule{0.723pt}{0.400pt}}
\put(436.0,346.0){\rule[-0.200pt]{1.686pt}{0.400pt}}
\put(456,343.67){\rule{1.686pt}{0.400pt}}
\multiput(456.00,344.17)(3.500,-1.000){2}{\rule{0.843pt}{0.400pt}}
\put(449.0,345.0){\rule[-0.200pt]{1.686pt}{0.400pt}}
\put(476,342.67){\rule{1.686pt}{0.400pt}}
\multiput(476.00,343.17)(3.500,-1.000){2}{\rule{0.843pt}{0.400pt}}
\put(463.0,344.0){\rule[-0.200pt]{3.132pt}{0.400pt}}
\put(503,341.67){\rule{1.445pt}{0.400pt}}
\multiput(503.00,342.17)(3.000,-1.000){2}{\rule{0.723pt}{0.400pt}}
\put(483.0,343.0){\rule[-0.200pt]{4.818pt}{0.400pt}}
\put(536,340.67){\rule{1.686pt}{0.400pt}}
\multiput(536.00,341.17)(3.500,-1.000){2}{\rule{0.843pt}{0.400pt}}
\put(509.0,342.0){\rule[-0.200pt]{6.504pt}{0.400pt}}
\put(589,339.67){\rule{1.686pt}{0.400pt}}
\multiput(589.00,340.17)(3.500,-1.000){2}{\rule{0.843pt}{0.400pt}}
\put(543.0,341.0){\rule[-0.200pt]{11.081pt}{0.400pt}}
\put(669,338.67){\rule{1.686pt}{0.400pt}}
\multiput(669.00,339.17)(3.500,-1.000){2}{\rule{0.843pt}{0.400pt}}
\put(596.0,340.0){\rule[-0.200pt]{17.586pt}{0.400pt}}
\put(676.0,339.0){\rule[-0.200pt]{38.544pt}{0.400pt}}
\put(706,407){\makebox(0,0)[r]{$h,v_F = 0.2$}}
\multiput(728,407)(20.756,0.000){4}{\usebox{\plotpoint}}
\put(794,407){\usebox{\plotpoint}}
\put(176,381){\usebox{\plotpoint}}
\put(176.00,381.00){\usebox{\plotpoint}}
\multiput(183,381)(20.756,0.000){0}{\usebox{\plotpoint}}
\multiput(189,381)(19.957,-5.702){0}{\usebox{\plotpoint}}
\put(196.47,378.93){\usebox{\plotpoint}}
\multiput(203,378)(19.690,-6.563){0}{\usebox{\plotpoint}}
\multiput(209,376)(19.077,-8.176){0}{\usebox{\plotpoint}}
\put(216.20,372.91){\usebox{\plotpoint}}
\multiput(223,370)(18.564,-9.282){0}{\usebox{\plotpoint}}
\put(234.77,363.70){\usebox{\plotpoint}}
\multiput(236,363)(16.889,-12.064){0}{\usebox{\plotpoint}}
\multiput(243,358)(15.945,-13.287){0}{\usebox{\plotpoint}}
\put(251.23,351.09){\usebox{\plotpoint}}
\multiput(256,347)(14.676,-14.676){0}{\usebox{\plotpoint}}
\put(265.97,336.53){\usebox{\plotpoint}}
\multiput(269,333)(12.743,-16.383){0}{\usebox{\plotpoint}}
\put(278.70,320.15){\usebox{\plotpoint}}
\multiput(283,314)(10.679,-17.798){0}{\usebox{\plotpoint}}
\put(289.75,302.60){\usebox{\plotpoint}}
\put(299.39,284.22){\usebox{\plotpoint}}
\put(307.01,264.97){\usebox{\plotpoint}}
\put(313.78,245.35){\usebox{\plotpoint}}
\put(319.52,225.42){\usebox{\plotpoint}}
\multiput(323,212)(3.322,-20.488){2}{\usebox{\plotpoint}}
\multiput(329,175)(2.329,-20.624){3}{\usebox{\plotpoint}}
\put(336,113){\usebox{\plotpoint}}
\put(349.00,113.00){\usebox{\plotpoint}}
\put(352.15,133.41){\usebox{\plotpoint}}
\put(357.00,153.57){\usebox{\plotpoint}}
\put(364.88,172.75){\usebox{\plotpoint}}
\multiput(369,181)(12.743,16.383){0}{\usebox{\plotpoint}}
\put(376.09,190.09){\usebox{\plotpoint}}
\multiput(383,197)(15.945,13.287){0}{\usebox{\plotpoint}}
\put(391.59,203.85){\usebox{\plotpoint}}
\multiput(396,207)(19.077,8.176){0}{\usebox{\plotpoint}}
\multiput(403,210)(17.270,11.513){0}{\usebox{\plotpoint}}
\put(409.49,214.14){\usebox{\plotpoint}}
\multiput(416,216)(19.077,8.176){0}{\usebox{\plotpoint}}
\multiput(423,219)(19.690,6.563){0}{\usebox{\plotpoint}}
\put(429.04,221.01){\usebox{\plotpoint}}
\multiput(436,222)(19.957,5.702){0}{\usebox{\plotpoint}}
\multiput(443,224)(19.690,6.563){0}{\usebox{\plotpoint}}
\put(449.12,226.02){\usebox{\plotpoint}}
\multiput(456,227)(20.547,2.935){0}{\usebox{\plotpoint}}
\multiput(463,228)(20.473,3.412){0}{\usebox{\plotpoint}}
\put(469.64,229.09){\usebox{\plotpoint}}
\multiput(476,230)(20.547,2.935){0}{\usebox{\plotpoint}}
\multiput(483,231)(20.473,3.412){0}{\usebox{\plotpoint}}
\put(490.18,232.00){\usebox{\plotpoint}}
\multiput(496,232)(20.547,2.935){0}{\usebox{\plotpoint}}
\multiput(503,233)(20.473,3.412){0}{\usebox{\plotpoint}}
\put(510.78,234.00){\usebox{\plotpoint}}
\multiput(516,234)(20.547,2.935){0}{\usebox{\plotpoint}}
\multiput(523,235)(20.756,0.000){0}{\usebox{\plotpoint}}
\put(531.44,235.35){\usebox{\plotpoint}}
\multiput(536,236)(20.756,0.000){0}{\usebox{\plotpoint}}
\multiput(543,236)(20.473,3.412){0}{\usebox{\plotpoint}}
\put(552.07,237.00){\usebox{\plotpoint}}
\multiput(556,237)(20.547,2.935){0}{\usebox{\plotpoint}}
\multiput(563,238)(20.756,0.000){0}{\usebox{\plotpoint}}
\put(572.75,238.00){\usebox{\plotpoint}}
\multiput(576,238)(20.547,2.935){0}{\usebox{\plotpoint}}
\multiput(583,239)(20.756,0.000){0}{\usebox{\plotpoint}}
\put(593.44,239.00){\usebox{\plotpoint}}
\multiput(596,239)(20.756,0.000){0}{\usebox{\plotpoint}}
\multiput(603,239)(20.473,3.412){0}{\usebox{\plotpoint}}
\put(614.11,240.00){\usebox{\plotpoint}}
\multiput(616,240)(20.756,0.000){0}{\usebox{\plotpoint}}
\multiput(623,240)(20.756,0.000){0}{\usebox{\plotpoint}}
\put(634.81,240.83){\usebox{\plotpoint}}
\multiput(636,241)(20.756,0.000){0}{\usebox{\plotpoint}}
\multiput(643,241)(20.756,0.000){0}{\usebox{\plotpoint}}
\put(655.55,241.00){\usebox{\plotpoint}}
\multiput(656,241)(20.756,0.000){0}{\usebox{\plotpoint}}
\multiput(663,241)(20.756,0.000){0}{\usebox{\plotpoint}}
\multiput(669,241)(20.547,2.935){0}{\usebox{\plotpoint}}
\put(676.24,242.00){\usebox{\plotpoint}}
\multiput(683,242)(20.756,0.000){0}{\usebox{\plotpoint}}
\multiput(689,242)(20.756,0.000){0}{\usebox{\plotpoint}}
\put(696.99,242.00){\usebox{\plotpoint}}
\multiput(703,242)(20.756,0.000){0}{\usebox{\plotpoint}}
\multiput(709,242)(20.756,0.000){0}{\usebox{\plotpoint}}
\put(717.75,242.00){\usebox{\plotpoint}}
\multiput(723,242)(20.473,3.412){0}{\usebox{\plotpoint}}
\multiput(729,243)(20.756,0.000){0}{\usebox{\plotpoint}}
\put(738.42,243.00){\usebox{\plotpoint}}
\multiput(743,243)(20.756,0.000){0}{\usebox{\plotpoint}}
\multiput(749,243)(20.756,0.000){0}{\usebox{\plotpoint}}
\put(759.17,243.00){\usebox{\plotpoint}}
\multiput(763,243)(20.756,0.000){0}{\usebox{\plotpoint}}
\multiput(769,243)(20.756,0.000){0}{\usebox{\plotpoint}}
\put(779.93,243.00){\usebox{\plotpoint}}
\multiput(783,243)(20.756,0.000){0}{\usebox{\plotpoint}}
\multiput(789,243)(20.756,0.000){0}{\usebox{\plotpoint}}
\put(800.64,243.66){\usebox{\plotpoint}}
\multiput(803,244)(20.756,0.000){0}{\usebox{\plotpoint}}
\multiput(809,244)(20.756,0.000){0}{\usebox{\plotpoint}}
\put(821.37,244.00){\usebox{\plotpoint}}
\multiput(823,244)(20.756,0.000){0}{\usebox{\plotpoint}}
\multiput(829,244)(20.756,0.000){0}{\usebox{\plotpoint}}
\put(836,244){\usebox{\plotpoint}}
\sbox{\plotpoint}{\rule[-0.500pt]{1.000pt}{1.000pt}}%
\put(706,362){\makebox(0,0)[r]{$h,v_F = 0.5$}}
\multiput(728,362)(20.756,0.000){4}{\usebox{\plotpoint}}
\put(794,362){\usebox{\plotpoint}}
\put(176,377){\usebox{\plotpoint}}
\put(176.00,377.00){\usebox{\plotpoint}}
\multiput(183,377)(20.473,-3.412){0}{\usebox{\plotpoint}}
\multiput(189,376)(20.547,-2.935){0}{\usebox{\plotpoint}}
\put(196.58,374.83){\usebox{\plotpoint}}
\multiput(203,373)(19.690,-6.563){0}{\usebox{\plotpoint}}
\multiput(209,371)(19.957,-5.702){0}{\usebox{\plotpoint}}
\put(216.43,368.81){\usebox{\plotpoint}}
\multiput(223,366)(17.270,-11.513){0}{\usebox{\plotpoint}}
\put(234.56,358.82){\usebox{\plotpoint}}
\multiput(236,358)(16.889,-12.064){0}{\usebox{\plotpoint}}
\multiput(243,353)(15.945,-13.287){0}{\usebox{\plotpoint}}
\put(251.04,346.25){\usebox{\plotpoint}}
\multiput(256,342)(14.676,-14.676){0}{\usebox{\plotpoint}}
\put(265.81,331.72){\usebox{\plotpoint}}
\multiput(269,328)(12.743,-16.383){0}{\usebox{\plotpoint}}
\put(278.73,315.49){\usebox{\plotpoint}}
\multiput(283,310)(9.939,-18.221){0}{\usebox{\plotpoint}}
\put(289.61,297.88){\usebox{\plotpoint}}
\put(299.25,279.50){\usebox{\plotpoint}}
\put(307.12,260.34){\usebox{\plotpoint}}
\put(313.77,240.68){\usebox{\plotpoint}}
\put(319.44,220.72){\usebox{\plotpoint}}
\multiput(323,207)(3.322,-20.488){2}{\usebox{\plotpoint}}
\multiput(329,170)(2.173,-20.641){3}{\usebox{\plotpoint}}
\put(335,113){\usebox{\plotpoint}}
\put(349.00,113.00){\usebox{\plotpoint}}
\put(353.13,133.29){\usebox{\plotpoint}}
\put(358.80,153.20){\usebox{\plotpoint}}
\put(366.84,172.33){\usebox{\plotpoint}}
\multiput(369,177)(13.668,15.620){0}{\usebox{\plotpoint}}
\put(379.52,188.52){\usebox{\plotpoint}}
\multiput(383,192)(14.676,14.676){0}{\usebox{\plotpoint}}
\put(395.38,201.65){\usebox{\plotpoint}}
\multiput(396,202)(18.021,10.298){0}{\usebox{\plotpoint}}
\multiput(403,206)(18.564,9.282){0}{\usebox{\plotpoint}}
\put(413.85,211.08){\usebox{\plotpoint}}
\multiput(416,212)(19.957,5.702){0}{\usebox{\plotpoint}}
\multiput(423,214)(19.690,6.563){0}{\usebox{\plotpoint}}
\put(433.62,217.32){\usebox{\plotpoint}}
\multiput(436,218)(20.547,2.935){0}{\usebox{\plotpoint}}
\multiput(443,219)(19.690,6.563){0}{\usebox{\plotpoint}}
\put(453.84,221.69){\usebox{\plotpoint}}
\multiput(456,222)(20.547,2.935){0}{\usebox{\plotpoint}}
\multiput(463,223)(20.473,3.412){0}{\usebox{\plotpoint}}
\put(474.36,224.77){\usebox{\plotpoint}}
\multiput(476,225)(20.547,2.935){0}{\usebox{\plotpoint}}
\multiput(483,226)(20.473,3.412){0}{\usebox{\plotpoint}}
\put(494.89,227.84){\usebox{\plotpoint}}
\multiput(496,228)(20.756,0.000){0}{\usebox{\plotpoint}}
\multiput(503,228)(20.473,3.412){0}{\usebox{\plotpoint}}
\put(515.49,229.93){\usebox{\plotpoint}}
\multiput(516,230)(20.756,0.000){0}{\usebox{\plotpoint}}
\multiput(523,230)(20.473,3.412){0}{\usebox{\plotpoint}}
\multiput(529,231)(20.756,0.000){0}{\usebox{\plotpoint}}
\put(536.15,231.02){\usebox{\plotpoint}}
\multiput(543,232)(20.756,0.000){0}{\usebox{\plotpoint}}
\multiput(549,232)(20.756,0.000){0}{\usebox{\plotpoint}}
\put(556.83,232.12){\usebox{\plotpoint}}
\multiput(563,233)(20.756,0.000){0}{\usebox{\plotpoint}}
\multiput(569,233)(20.547,2.935){0}{\usebox{\plotpoint}}
\put(577.45,234.00){\usebox{\plotpoint}}
\multiput(583,234)(20.756,0.000){0}{\usebox{\plotpoint}}
\multiput(589,234)(20.756,0.000){0}{\usebox{\plotpoint}}
\put(598.18,234.31){\usebox{\plotpoint}}
\multiput(603,235)(20.756,0.000){0}{\usebox{\plotpoint}}
\multiput(609,235)(20.756,0.000){0}{\usebox{\plotpoint}}
\put(618.89,235.00){\usebox{\plotpoint}}
\multiput(623,235)(20.473,3.412){0}{\usebox{\plotpoint}}
\multiput(629,236)(20.756,0.000){0}{\usebox{\plotpoint}}
\put(639.56,236.00){\usebox{\plotpoint}}
\multiput(643,236)(20.756,0.000){0}{\usebox{\plotpoint}}
\multiput(649,236)(20.756,0.000){0}{\usebox{\plotpoint}}
\put(660.28,236.61){\usebox{\plotpoint}}
\multiput(663,237)(20.756,0.000){0}{\usebox{\plotpoint}}
\multiput(669,237)(20.756,0.000){0}{\usebox{\plotpoint}}
\put(681.00,237.00){\usebox{\plotpoint}}
\multiput(683,237)(20.756,0.000){0}{\usebox{\plotpoint}}
\multiput(689,237)(20.756,0.000){0}{\usebox{\plotpoint}}
\put(701.76,237.00){\usebox{\plotpoint}}
\multiput(703,237)(20.756,0.000){0}{\usebox{\plotpoint}}
\multiput(709,237)(20.547,2.935){0}{\usebox{\plotpoint}}
\put(722.44,238.00){\usebox{\plotpoint}}
\multiput(723,238)(20.756,0.000){0}{\usebox{\plotpoint}}
\multiput(729,238)(20.756,0.000){0}{\usebox{\plotpoint}}
\multiput(736,238)(20.756,0.000){0}{\usebox{\plotpoint}}
\put(743.20,238.00){\usebox{\plotpoint}}
\multiput(749,238)(20.756,0.000){0}{\usebox{\plotpoint}}
\multiput(756,238)(20.756,0.000){0}{\usebox{\plotpoint}}
\put(763.95,238.00){\usebox{\plotpoint}}
\multiput(769,238)(20.756,0.000){0}{\usebox{\plotpoint}}
\multiput(776,238)(20.547,2.935){0}{\usebox{\plotpoint}}
\put(784.64,239.00){\usebox{\plotpoint}}
\multiput(789,239)(20.756,0.000){0}{\usebox{\plotpoint}}
\multiput(796,239)(20.756,0.000){0}{\usebox{\plotpoint}}
\put(805.39,239.00){\usebox{\plotpoint}}
\multiput(809,239)(20.756,0.000){0}{\usebox{\plotpoint}}
\multiput(816,239)(20.756,0.000){0}{\usebox{\plotpoint}}
\put(826.15,239.00){\usebox{\plotpoint}}
\multiput(829,239)(20.756,0.000){0}{\usebox{\plotpoint}}
\put(836,239){\usebox{\plotpoint}}
\end{picture}
\setlength{\unitlength}{0.240900pt}
\ifx\plotpoint\undefined\newsavebox{\plotpoint}\fi
\sbox{\plotpoint}{\rule[-0.200pt]{0.400pt}{0.400pt}}%
\begin{picture}(900,540)(0,0)
\font\gnuplot=cmr10 at 10pt
\gnuplot
\sbox{\plotpoint}{\rule[-0.200pt]{0.400pt}{0.400pt}}%
\put(176.0,315.0){\rule[-0.200pt]{158.994pt}{0.400pt}}
\put(176.0,113.0){\rule[-0.200pt]{0.400pt}{97.324pt}}
\put(176.0,113.0){\rule[-0.200pt]{4.818pt}{0.400pt}}
\put(154,113){\makebox(0,0)[r]{-100}}
\put(816.0,113.0){\rule[-0.200pt]{4.818pt}{0.400pt}}
\put(176.0,214.0){\rule[-0.200pt]{4.818pt}{0.400pt}}
\put(154,214){\makebox(0,0)[r]{-50}}
\put(816.0,214.0){\rule[-0.200pt]{4.818pt}{0.400pt}}
\put(176.0,315.0){\rule[-0.200pt]{4.818pt}{0.400pt}}
\put(154,315){\makebox(0,0)[r]{0}}
\put(816.0,315.0){\rule[-0.200pt]{4.818pt}{0.400pt}}
\put(176.0,416.0){\rule[-0.200pt]{4.818pt}{0.400pt}}
\put(154,416){\makebox(0,0)[r]{50}}
\put(816.0,416.0){\rule[-0.200pt]{4.818pt}{0.400pt}}
\put(176.0,517.0){\rule[-0.200pt]{4.818pt}{0.400pt}}
\put(154,517){\makebox(0,0)[r]{100}}
\put(816.0,517.0){\rule[-0.200pt]{4.818pt}{0.400pt}}
\put(176.0,113.0){\rule[-0.200pt]{0.400pt}{4.818pt}}
\put(176,68){\makebox(0,0){0}}
\put(176.0,497.0){\rule[-0.200pt]{0.400pt}{4.818pt}}
\put(259.0,113.0){\rule[-0.200pt]{0.400pt}{4.818pt}}
\put(259,68){\makebox(0,0){0.5}}
\put(259.0,497.0){\rule[-0.200pt]{0.400pt}{4.818pt}}
\put(341.0,113.0){\rule[-0.200pt]{0.400pt}{4.818pt}}
\put(341,68){\makebox(0,0){1}}
\put(341.0,497.0){\rule[-0.200pt]{0.400pt}{4.818pt}}
\put(424.0,113.0){\rule[-0.200pt]{0.400pt}{4.818pt}}
\put(424,68){\makebox(0,0){1.5}}
\put(424.0,497.0){\rule[-0.200pt]{0.400pt}{4.818pt}}
\put(506.0,113.0){\rule[-0.200pt]{0.400pt}{4.818pt}}
\put(506,68){\makebox(0,0){2}}
\put(506.0,497.0){\rule[-0.200pt]{0.400pt}{4.818pt}}
\put(589.0,113.0){\rule[-0.200pt]{0.400pt}{4.818pt}}
\put(589,68){\makebox(0,0){2.5}}
\put(589.0,497.0){\rule[-0.200pt]{0.400pt}{4.818pt}}
\put(671.0,113.0){\rule[-0.200pt]{0.400pt}{4.818pt}}
\put(671,68){\makebox(0,0){3}}
\put(671.0,497.0){\rule[-0.200pt]{0.400pt}{4.818pt}}
\put(754.0,113.0){\rule[-0.200pt]{0.400pt}{4.818pt}}
\put(754,68){\makebox(0,0){3.5}}
\put(754.0,497.0){\rule[-0.200pt]{0.400pt}{4.818pt}}
\put(836.0,113.0){\rule[-0.200pt]{0.400pt}{4.818pt}}
\put(836,68){\makebox(0,0){4}}
\put(836.0,497.0){\rule[-0.200pt]{0.400pt}{4.818pt}}
\put(176.0,113.0){\rule[-0.200pt]{158.994pt}{0.400pt}}
\put(836.0,113.0){\rule[-0.200pt]{0.400pt}{97.324pt}}
\put(176.0,517.0){\rule[-0.200pt]{158.994pt}{0.400pt}}
\put(506,23){\makebox(0,0){$z$}}
\put(176.0,113.0){\rule[-0.200pt]{0.400pt}{97.324pt}}
\put(706,452){\makebox(0,0)[r]{$g_L,v_F = 0.1$}}
\put(728.0,452.0){\rule[-0.200pt]{15.899pt}{0.400pt}}
\put(176,517){\usebox{\plotpoint}}
\put(183,515.67){\rule{0.723pt}{0.400pt}}
\multiput(183.00,516.17)(1.500,-1.000){2}{\rule{0.361pt}{0.400pt}}
\put(176.0,517.0){\rule[-0.200pt]{1.686pt}{0.400pt}}
\put(189,514.67){\rule{0.964pt}{0.400pt}}
\multiput(189.00,515.17)(2.000,-1.000){2}{\rule{0.482pt}{0.400pt}}
\put(193,513.67){\rule{0.723pt}{0.400pt}}
\multiput(193.00,514.17)(1.500,-1.000){2}{\rule{0.361pt}{0.400pt}}
\put(196,512.67){\rule{0.723pt}{0.400pt}}
\multiput(196.00,513.17)(1.500,-1.000){2}{\rule{0.361pt}{0.400pt}}
\put(199,511.67){\rule{0.723pt}{0.400pt}}
\multiput(199.00,512.17)(1.500,-1.000){2}{\rule{0.361pt}{0.400pt}}
\put(202,510.17){\rule{0.900pt}{0.400pt}}
\multiput(202.00,511.17)(2.132,-2.000){2}{\rule{0.450pt}{0.400pt}}
\put(206,508.67){\rule{0.723pt}{0.400pt}}
\multiput(206.00,509.17)(1.500,-1.000){2}{\rule{0.361pt}{0.400pt}}
\put(209,507.17){\rule{0.700pt}{0.400pt}}
\multiput(209.00,508.17)(1.547,-2.000){2}{\rule{0.350pt}{0.400pt}}
\put(212,505.17){\rule{0.900pt}{0.400pt}}
\multiput(212.00,506.17)(2.132,-2.000){2}{\rule{0.450pt}{0.400pt}}
\put(216,503.17){\rule{0.700pt}{0.400pt}}
\multiput(216.00,504.17)(1.547,-2.000){2}{\rule{0.350pt}{0.400pt}}
\put(219,501.17){\rule{0.700pt}{0.400pt}}
\multiput(219.00,502.17)(1.547,-2.000){2}{\rule{0.350pt}{0.400pt}}
\multiput(222.00,499.95)(0.685,-0.447){3}{\rule{0.633pt}{0.108pt}}
\multiput(222.00,500.17)(2.685,-3.000){2}{\rule{0.317pt}{0.400pt}}
\multiput(226.00,496.95)(0.462,-0.447){3}{\rule{0.500pt}{0.108pt}}
\multiput(226.00,497.17)(1.962,-3.000){2}{\rule{0.250pt}{0.400pt}}
\multiput(229.00,493.95)(0.462,-0.447){3}{\rule{0.500pt}{0.108pt}}
\multiput(229.00,494.17)(1.962,-3.000){2}{\rule{0.250pt}{0.400pt}}
\multiput(232.00,490.95)(0.462,-0.447){3}{\rule{0.500pt}{0.108pt}}
\multiput(232.00,491.17)(1.962,-3.000){2}{\rule{0.250pt}{0.400pt}}
\multiput(235.00,487.95)(0.685,-0.447){3}{\rule{0.633pt}{0.108pt}}
\multiput(235.00,488.17)(2.685,-3.000){2}{\rule{0.317pt}{0.400pt}}
\multiput(239.00,484.95)(0.462,-0.447){3}{\rule{0.500pt}{0.108pt}}
\multiput(239.00,485.17)(1.962,-3.000){2}{\rule{0.250pt}{0.400pt}}
\multiput(242.61,480.37)(0.447,-0.685){3}{\rule{0.108pt}{0.633pt}}
\multiput(241.17,481.69)(3.000,-2.685){2}{\rule{0.400pt}{0.317pt}}
\multiput(245.00,477.94)(0.481,-0.468){5}{\rule{0.500pt}{0.113pt}}
\multiput(245.00,478.17)(2.962,-4.000){2}{\rule{0.250pt}{0.400pt}}
\multiput(249.61,471.82)(0.447,-0.909){3}{\rule{0.108pt}{0.767pt}}
\multiput(248.17,473.41)(3.000,-3.409){2}{\rule{0.400pt}{0.383pt}}
\multiput(252.61,467.37)(0.447,-0.685){3}{\rule{0.108pt}{0.633pt}}
\multiput(251.17,468.69)(3.000,-2.685){2}{\rule{0.400pt}{0.317pt}}
\multiput(255.60,463.51)(0.468,-0.627){5}{\rule{0.113pt}{0.600pt}}
\multiput(254.17,464.75)(4.000,-3.755){2}{\rule{0.400pt}{0.300pt}}
\multiput(259.61,457.82)(0.447,-0.909){3}{\rule{0.108pt}{0.767pt}}
\multiput(258.17,459.41)(3.000,-3.409){2}{\rule{0.400pt}{0.383pt}}
\multiput(262.61,452.82)(0.447,-0.909){3}{\rule{0.108pt}{0.767pt}}
\multiput(261.17,454.41)(3.000,-3.409){2}{\rule{0.400pt}{0.383pt}}
\multiput(265.61,447.26)(0.447,-1.132){3}{\rule{0.108pt}{0.900pt}}
\multiput(264.17,449.13)(3.000,-4.132){2}{\rule{0.400pt}{0.450pt}}
\multiput(268.60,442.09)(0.468,-0.774){5}{\rule{0.113pt}{0.700pt}}
\multiput(267.17,443.55)(4.000,-4.547){2}{\rule{0.400pt}{0.350pt}}
\multiput(272.61,435.26)(0.447,-1.132){3}{\rule{0.108pt}{0.900pt}}
\multiput(271.17,437.13)(3.000,-4.132){2}{\rule{0.400pt}{0.450pt}}
\multiput(275.61,428.71)(0.447,-1.355){3}{\rule{0.108pt}{1.033pt}}
\multiput(274.17,430.86)(3.000,-4.855){2}{\rule{0.400pt}{0.517pt}}
\multiput(278.60,422.68)(0.468,-0.920){5}{\rule{0.113pt}{0.800pt}}
\multiput(277.17,424.34)(4.000,-5.340){2}{\rule{0.400pt}{0.400pt}}
\multiput(282.61,414.16)(0.447,-1.579){3}{\rule{0.108pt}{1.167pt}}
\multiput(281.17,416.58)(3.000,-5.579){2}{\rule{0.400pt}{0.583pt}}
\multiput(285.61,406.16)(0.447,-1.579){3}{\rule{0.108pt}{1.167pt}}
\multiput(284.17,408.58)(3.000,-5.579){2}{\rule{0.400pt}{0.583pt}}
\multiput(288.60,398.85)(0.468,-1.212){5}{\rule{0.113pt}{1.000pt}}
\multiput(287.17,400.92)(4.000,-6.924){2}{\rule{0.400pt}{0.500pt}}
\multiput(292.61,388.60)(0.447,-1.802){3}{\rule{0.108pt}{1.300pt}}
\multiput(291.17,391.30)(3.000,-6.302){2}{\rule{0.400pt}{0.650pt}}
\multiput(295.61,379.05)(0.447,-2.025){3}{\rule{0.108pt}{1.433pt}}
\multiput(294.17,382.03)(3.000,-7.025){2}{\rule{0.400pt}{0.717pt}}
\multiput(298.61,368.50)(0.447,-2.248){3}{\rule{0.108pt}{1.567pt}}
\multiput(297.17,371.75)(3.000,-7.748){2}{\rule{0.400pt}{0.783pt}}
\multiput(301.60,358.60)(0.468,-1.651){5}{\rule{0.113pt}{1.300pt}}
\multiput(300.17,361.30)(4.000,-9.302){2}{\rule{0.400pt}{0.650pt}}
\multiput(305.61,344.39)(0.447,-2.695){3}{\rule{0.108pt}{1.833pt}}
\multiput(304.17,348.19)(3.000,-9.195){2}{\rule{0.400pt}{0.917pt}}
\multiput(308.61,330.84)(0.447,-2.918){3}{\rule{0.108pt}{1.967pt}}
\multiput(307.17,334.92)(3.000,-9.918){2}{\rule{0.400pt}{0.983pt}}
\multiput(311.60,318.36)(0.468,-2.090){5}{\rule{0.113pt}{1.600pt}}
\multiput(310.17,321.68)(4.000,-11.679){2}{\rule{0.400pt}{0.800pt}}
\multiput(315.61,299.62)(0.447,-3.811){3}{\rule{0.108pt}{2.500pt}}
\multiput(314.17,304.81)(3.000,-12.811){2}{\rule{0.400pt}{1.250pt}}
\multiput(318.61,281.07)(0.447,-4.034){3}{\rule{0.108pt}{2.633pt}}
\multiput(317.17,286.53)(3.000,-13.534){2}{\rule{0.400pt}{1.317pt}}
\multiput(321.60,263.04)(0.468,-3.259){5}{\rule{0.113pt}{2.400pt}}
\multiput(320.17,268.02)(4.000,-18.019){2}{\rule{0.400pt}{1.200pt}}
\multiput(325.61,234.09)(0.447,-6.044){3}{\rule{0.108pt}{3.833pt}}
\multiput(324.17,242.04)(3.000,-20.044){2}{\rule{0.400pt}{1.917pt}}
\multiput(328.61,202.77)(0.447,-7.383){3}{\rule{0.108pt}{4.633pt}}
\multiput(327.17,212.38)(3.000,-24.383){2}{\rule{0.400pt}{2.317pt}}
\multiput(331.61,161.57)(0.447,-10.286){3}{\rule{0.108pt}{6.367pt}}
\multiput(330.17,174.79)(3.000,-33.786){2}{\rule{0.400pt}{3.183pt}}
\put(334.17,113){\rule{0.400pt}{5.700pt}}
\multiput(333.17,129.17)(2.000,-16.169){2}{\rule{0.400pt}{2.850pt}}
\multiput(348.61,113.00)(0.447,5.820){3}{\rule{0.108pt}{3.700pt}}
\multiput(347.17,113.00)(3.000,19.320){2}{\rule{0.400pt}{1.850pt}}
\multiput(351.61,140.00)(0.447,4.927){3}{\rule{0.108pt}{3.167pt}}
\multiput(350.17,140.00)(3.000,16.427){2}{\rule{0.400pt}{1.583pt}}
\multiput(354.60,163.00)(0.468,2.382){5}{\rule{0.113pt}{1.800pt}}
\multiput(353.17,163.00)(4.000,13.264){2}{\rule{0.400pt}{0.900pt}}
\multiput(358.61,180.00)(0.447,2.918){3}{\rule{0.108pt}{1.967pt}}
\multiput(357.17,180.00)(3.000,9.918){2}{\rule{0.400pt}{0.983pt}}
\multiput(361.61,194.00)(0.447,2.025){3}{\rule{0.108pt}{1.433pt}}
\multiput(360.17,194.00)(3.000,7.025){2}{\rule{0.400pt}{0.717pt}}
\multiput(364.61,204.00)(0.447,1.802){3}{\rule{0.108pt}{1.300pt}}
\multiput(363.17,204.00)(3.000,6.302){2}{\rule{0.400pt}{0.650pt}}
\multiput(367.60,213.00)(0.468,1.066){5}{\rule{0.113pt}{0.900pt}}
\multiput(366.17,213.00)(4.000,6.132){2}{\rule{0.400pt}{0.450pt}}
\multiput(371.61,221.00)(0.447,1.355){3}{\rule{0.108pt}{1.033pt}}
\multiput(370.17,221.00)(3.000,4.855){2}{\rule{0.400pt}{0.517pt}}
\multiput(374.61,228.00)(0.447,0.909){3}{\rule{0.108pt}{0.767pt}}
\multiput(373.17,228.00)(3.000,3.409){2}{\rule{0.400pt}{0.383pt}}
\multiput(377.60,233.00)(0.468,0.627){5}{\rule{0.113pt}{0.600pt}}
\multiput(376.17,233.00)(4.000,3.755){2}{\rule{0.400pt}{0.300pt}}
\multiput(381.61,238.00)(0.447,0.909){3}{\rule{0.108pt}{0.767pt}}
\multiput(380.17,238.00)(3.000,3.409){2}{\rule{0.400pt}{0.383pt}}
\multiput(384.61,243.00)(0.447,0.685){3}{\rule{0.108pt}{0.633pt}}
\multiput(383.17,243.00)(3.000,2.685){2}{\rule{0.400pt}{0.317pt}}
\multiput(387.00,247.60)(0.481,0.468){5}{\rule{0.500pt}{0.113pt}}
\multiput(387.00,246.17)(2.962,4.000){2}{\rule{0.250pt}{0.400pt}}
\multiput(391.00,251.61)(0.462,0.447){3}{\rule{0.500pt}{0.108pt}}
\multiput(391.00,250.17)(1.962,3.000){2}{\rule{0.250pt}{0.400pt}}
\multiput(394.00,254.61)(0.462,0.447){3}{\rule{0.500pt}{0.108pt}}
\multiput(394.00,253.17)(1.962,3.000){2}{\rule{0.250pt}{0.400pt}}
\multiput(397.00,257.61)(0.462,0.447){3}{\rule{0.500pt}{0.108pt}}
\multiput(397.00,256.17)(1.962,3.000){2}{\rule{0.250pt}{0.400pt}}
\put(400,260.17){\rule{0.900pt}{0.400pt}}
\multiput(400.00,259.17)(2.132,2.000){2}{\rule{0.450pt}{0.400pt}}
\multiput(404.00,262.61)(0.462,0.447){3}{\rule{0.500pt}{0.108pt}}
\multiput(404.00,261.17)(1.962,3.000){2}{\rule{0.250pt}{0.400pt}}
\put(407,265.17){\rule{0.700pt}{0.400pt}}
\multiput(407.00,264.17)(1.547,2.000){2}{\rule{0.350pt}{0.400pt}}
\put(410,267.17){\rule{0.900pt}{0.400pt}}
\multiput(410.00,266.17)(2.132,2.000){2}{\rule{0.450pt}{0.400pt}}
\put(414,269.17){\rule{0.700pt}{0.400pt}}
\multiput(414.00,268.17)(1.547,2.000){2}{\rule{0.350pt}{0.400pt}}
\put(417,270.67){\rule{0.723pt}{0.400pt}}
\multiput(417.00,270.17)(1.500,1.000){2}{\rule{0.361pt}{0.400pt}}
\put(420,272.17){\rule{0.900pt}{0.400pt}}
\multiput(420.00,271.17)(2.132,2.000){2}{\rule{0.450pt}{0.400pt}}
\put(424,274.17){\rule{0.700pt}{0.400pt}}
\multiput(424.00,273.17)(1.547,2.000){2}{\rule{0.350pt}{0.400pt}}
\put(427,275.67){\rule{0.723pt}{0.400pt}}
\multiput(427.00,275.17)(1.500,1.000){2}{\rule{0.361pt}{0.400pt}}
\put(430,276.67){\rule{0.723pt}{0.400pt}}
\multiput(430.00,276.17)(1.500,1.000){2}{\rule{0.361pt}{0.400pt}}
\put(433,278.17){\rule{0.900pt}{0.400pt}}
\multiput(433.00,277.17)(2.132,2.000){2}{\rule{0.450pt}{0.400pt}}
\put(437,279.67){\rule{0.723pt}{0.400pt}}
\multiput(437.00,279.17)(1.500,1.000){2}{\rule{0.361pt}{0.400pt}}
\put(440,280.67){\rule{0.723pt}{0.400pt}}
\multiput(440.00,280.17)(1.500,1.000){2}{\rule{0.361pt}{0.400pt}}
\put(443,281.67){\rule{0.964pt}{0.400pt}}
\multiput(443.00,281.17)(2.000,1.000){2}{\rule{0.482pt}{0.400pt}}
\put(447,282.67){\rule{0.723pt}{0.400pt}}
\multiput(447.00,282.17)(1.500,1.000){2}{\rule{0.361pt}{0.400pt}}
\put(450,283.67){\rule{0.723pt}{0.400pt}}
\multiput(450.00,283.17)(1.500,1.000){2}{\rule{0.361pt}{0.400pt}}
\put(453,284.67){\rule{0.964pt}{0.400pt}}
\multiput(453.00,284.17)(2.000,1.000){2}{\rule{0.482pt}{0.400pt}}
\put(457,285.67){\rule{0.723pt}{0.400pt}}
\multiput(457.00,285.17)(1.500,1.000){2}{\rule{0.361pt}{0.400pt}}
\put(460,286.67){\rule{0.723pt}{0.400pt}}
\multiput(460.00,286.17)(1.500,1.000){2}{\rule{0.361pt}{0.400pt}}
\put(463,287.67){\rule{0.723pt}{0.400pt}}
\multiput(463.00,287.17)(1.500,1.000){2}{\rule{0.361pt}{0.400pt}}
\put(186.0,516.0){\rule[-0.200pt]{0.723pt}{0.400pt}}
\put(470,288.67){\rule{0.723pt}{0.400pt}}
\multiput(470.00,288.17)(1.500,1.000){2}{\rule{0.361pt}{0.400pt}}
\put(473,289.67){\rule{0.723pt}{0.400pt}}
\multiput(473.00,289.17)(1.500,1.000){2}{\rule{0.361pt}{0.400pt}}
\put(466.0,289.0){\rule[-0.200pt]{0.964pt}{0.400pt}}
\put(480,290.67){\rule{0.723pt}{0.400pt}}
\multiput(480.00,290.17)(1.500,1.000){2}{\rule{0.361pt}{0.400pt}}
\put(483,291.67){\rule{0.723pt}{0.400pt}}
\multiput(483.00,291.17)(1.500,1.000){2}{\rule{0.361pt}{0.400pt}}
\put(476.0,291.0){\rule[-0.200pt]{0.964pt}{0.400pt}}
\put(490,292.67){\rule{0.723pt}{0.400pt}}
\multiput(490.00,292.17)(1.500,1.000){2}{\rule{0.361pt}{0.400pt}}
\put(486.0,293.0){\rule[-0.200pt]{0.964pt}{0.400pt}}
\put(496,293.67){\rule{0.723pt}{0.400pt}}
\multiput(496.00,293.17)(1.500,1.000){2}{\rule{0.361pt}{0.400pt}}
\put(493.0,294.0){\rule[-0.200pt]{0.723pt}{0.400pt}}
\put(503,294.67){\rule{0.723pt}{0.400pt}}
\multiput(503.00,294.17)(1.500,1.000){2}{\rule{0.361pt}{0.400pt}}
\put(499.0,295.0){\rule[-0.200pt]{0.964pt}{0.400pt}}
\put(509,295.67){\rule{0.964pt}{0.400pt}}
\multiput(509.00,295.17)(2.000,1.000){2}{\rule{0.482pt}{0.400pt}}
\put(506.0,296.0){\rule[-0.200pt]{0.723pt}{0.400pt}}
\put(516,296.67){\rule{0.723pt}{0.400pt}}
\multiput(516.00,296.17)(1.500,1.000){2}{\rule{0.361pt}{0.400pt}}
\put(513.0,297.0){\rule[-0.200pt]{0.723pt}{0.400pt}}
\put(526,297.67){\rule{0.723pt}{0.400pt}}
\multiput(526.00,297.17)(1.500,1.000){2}{\rule{0.361pt}{0.400pt}}
\put(519.0,298.0){\rule[-0.200pt]{1.686pt}{0.400pt}}
\put(536,298.67){\rule{0.723pt}{0.400pt}}
\multiput(536.00,298.17)(1.500,1.000){2}{\rule{0.361pt}{0.400pt}}
\put(529.0,299.0){\rule[-0.200pt]{1.686pt}{0.400pt}}
\put(546,299.67){\rule{0.723pt}{0.400pt}}
\multiput(546.00,299.17)(1.500,1.000){2}{\rule{0.361pt}{0.400pt}}
\put(539.0,300.0){\rule[-0.200pt]{1.686pt}{0.400pt}}
\put(556,300.67){\rule{0.723pt}{0.400pt}}
\multiput(556.00,300.17)(1.500,1.000){2}{\rule{0.361pt}{0.400pt}}
\put(549.0,301.0){\rule[-0.200pt]{1.686pt}{0.400pt}}
\put(569,301.67){\rule{0.723pt}{0.400pt}}
\multiput(569.00,301.17)(1.500,1.000){2}{\rule{0.361pt}{0.400pt}}
\put(559.0,302.0){\rule[-0.200pt]{2.409pt}{0.400pt}}
\put(582,302.67){\rule{0.723pt}{0.400pt}}
\multiput(582.00,302.17)(1.500,1.000){2}{\rule{0.361pt}{0.400pt}}
\put(572.0,303.0){\rule[-0.200pt]{2.409pt}{0.400pt}}
\put(598,303.67){\rule{0.964pt}{0.400pt}}
\multiput(598.00,303.17)(2.000,1.000){2}{\rule{0.482pt}{0.400pt}}
\put(585.0,304.0){\rule[-0.200pt]{3.132pt}{0.400pt}}
\put(618,304.67){\rule{0.964pt}{0.400pt}}
\multiput(618.00,304.17)(2.000,1.000){2}{\rule{0.482pt}{0.400pt}}
\put(602.0,305.0){\rule[-0.200pt]{3.854pt}{0.400pt}}
\put(638,305.67){\rule{0.723pt}{0.400pt}}
\multiput(638.00,305.17)(1.500,1.000){2}{\rule{0.361pt}{0.400pt}}
\put(622.0,306.0){\rule[-0.200pt]{3.854pt}{0.400pt}}
\put(664,306.67){\rule{0.964pt}{0.400pt}}
\multiput(664.00,306.17)(2.000,1.000){2}{\rule{0.482pt}{0.400pt}}
\put(641.0,307.0){\rule[-0.200pt]{5.541pt}{0.400pt}}
\put(694,307.67){\rule{0.723pt}{0.400pt}}
\multiput(694.00,307.17)(1.500,1.000){2}{\rule{0.361pt}{0.400pt}}
\put(668.0,308.0){\rule[-0.200pt]{6.263pt}{0.400pt}}
\put(734,308.67){\rule{0.723pt}{0.400pt}}
\multiput(734.00,308.17)(1.500,1.000){2}{\rule{0.361pt}{0.400pt}}
\put(697.0,309.0){\rule[-0.200pt]{8.913pt}{0.400pt}}
\put(780,309.67){\rule{0.723pt}{0.400pt}}
\multiput(780.00,309.17)(1.500,1.000){2}{\rule{0.361pt}{0.400pt}}
\put(737.0,310.0){\rule[-0.200pt]{10.359pt}{0.400pt}}
\put(783.0,311.0){\rule[-0.200pt]{12.768pt}{0.400pt}}
\put(706,407){\makebox(0,0)[r]{$g_L,v_F = 0.2$}}
\multiput(728,407)(20.756,0.000){4}{\usebox{\plotpoint}}
\put(794,407){\usebox{\plotpoint}}
\put(176,366){\usebox{\plotpoint}}
\put(176.00,366.00){\usebox{\plotpoint}}
\multiput(179,365)(20.756,0.000){0}{\usebox{\plotpoint}}
\multiput(183,365)(20.756,0.000){0}{\usebox{\plotpoint}}
\multiput(186,365)(20.756,0.000){0}{\usebox{\plotpoint}}
\multiput(189,365)(20.756,0.000){0}{\usebox{\plotpoint}}
\multiput(193,365)(20.756,0.000){0}{\usebox{\plotpoint}}
\put(196.56,364.81){\usebox{\plotpoint}}
\multiput(199,364)(20.756,0.000){0}{\usebox{\plotpoint}}
\multiput(202,364)(20.756,0.000){0}{\usebox{\plotpoint}}
\multiput(206,364)(19.690,-6.563){0}{\usebox{\plotpoint}}
\multiput(209,363)(20.756,0.000){0}{\usebox{\plotpoint}}
\multiput(212,363)(20.136,-5.034){0}{\usebox{\plotpoint}}
\put(216.90,362.00){\usebox{\plotpoint}}
\multiput(219,362)(19.690,-6.563){0}{\usebox{\plotpoint}}
\multiput(222,361)(20.756,0.000){0}{\usebox{\plotpoint}}
\multiput(226,361)(19.690,-6.563){0}{\usebox{\plotpoint}}
\multiput(229,360)(19.690,-6.563){0}{\usebox{\plotpoint}}
\multiput(232,359)(19.690,-6.563){0}{\usebox{\plotpoint}}
\put(237.01,358.00){\usebox{\plotpoint}}
\multiput(239,358)(19.690,-6.563){0}{\usebox{\plotpoint}}
\multiput(242,357)(19.690,-6.563){0}{\usebox{\plotpoint}}
\multiput(245,356)(20.136,-5.034){0}{\usebox{\plotpoint}}
\multiput(249,355)(19.690,-6.563){0}{\usebox{\plotpoint}}
\multiput(252,354)(17.270,-11.513){0}{\usebox{\plotpoint}}
\put(256.50,351.62){\usebox{\plotpoint}}
\multiput(259,351)(19.690,-6.563){0}{\usebox{\plotpoint}}
\multiput(262,350)(19.690,-6.563){0}{\usebox{\plotpoint}}
\multiput(265,349)(17.270,-11.513){0}{\usebox{\plotpoint}}
\multiput(268,347)(20.136,-5.034){0}{\usebox{\plotpoint}}
\multiput(272,346)(17.270,-11.513){0}{\usebox{\plotpoint}}
\put(275.43,343.71){\usebox{\plotpoint}}
\multiput(278,342)(20.136,-5.034){0}{\usebox{\plotpoint}}
\multiput(282,341)(17.270,-11.513){0}{\usebox{\plotpoint}}
\multiput(285,339)(17.270,-11.513){0}{\usebox{\plotpoint}}
\multiput(288,337)(16.604,-12.453){0}{\usebox{\plotpoint}}
\put(293.11,333.26){\usebox{\plotpoint}}
\multiput(295,332)(17.270,-11.513){0}{\usebox{\plotpoint}}
\multiput(298,330)(14.676,-14.676){0}{\usebox{\plotpoint}}
\multiput(301,327)(16.604,-12.453){0}{\usebox{\plotpoint}}
\multiput(305,324)(14.676,-14.676){0}{\usebox{\plotpoint}}
\put(308.99,320.01){\usebox{\plotpoint}}
\multiput(311,318)(14.676,-14.676){0}{\usebox{\plotpoint}}
\multiput(315,314)(10.679,-17.798){0}{\usebox{\plotpoint}}
\multiput(318,309)(12.453,-16.604){0}{\usebox{\plotpoint}}
\put(321.79,303.82){\usebox{\plotpoint}}
\multiput(325,299)(8.176,-19.077){0}{\usebox{\plotpoint}}
\put(330.58,285.12){\usebox{\plotpoint}}
\multiput(331,284)(5.461,-20.024){0}{\usebox{\plotpoint}}
\put(335.68,265.02){\usebox{\plotpoint}}
\multiput(338,254)(0.000,-20.756){7}{\usebox{\plotpoint}}
\put(338,113){\usebox{\plotpoint}}
\multiput(344,113)(0.000,20.756){7}{\usebox{\plotpoint}}
\put(346.25,258.01){\usebox{\plotpoint}}
\multiput(348,265)(7.288,19.434){0}{\usebox{\plotpoint}}
\put(353.58,277.29){\usebox{\plotpoint}}
\multiput(354,278)(12.966,16.207){0}{\usebox{\plotpoint}}
\multiput(358,283)(14.676,14.676){0}{\usebox{\plotpoint}}
\multiput(361,286)(17.270,11.513){0}{\usebox{\plotpoint}}
\multiput(364,288)(14.676,14.676){0}{\usebox{\plotpoint}}
\put(368.29,291.64){\usebox{\plotpoint}}
\multiput(371,293)(19.690,6.563){0}{\usebox{\plotpoint}}
\multiput(374,294)(17.270,11.513){0}{\usebox{\plotpoint}}
\multiput(377,296)(20.136,5.034){0}{\usebox{\plotpoint}}
\multiput(381,297)(19.690,6.563){0}{\usebox{\plotpoint}}
\multiput(384,298)(19.690,6.563){0}{\usebox{\plotpoint}}
\put(387.49,299.12){\usebox{\plotpoint}}
\multiput(391,300)(19.690,6.563){0}{\usebox{\plotpoint}}
\multiput(394,301)(20.756,0.000){0}{\usebox{\plotpoint}}
\multiput(397,301)(19.690,6.563){0}{\usebox{\plotpoint}}
\multiput(400,302)(20.136,5.034){0}{\usebox{\plotpoint}}
\multiput(404,303)(20.756,0.000){0}{\usebox{\plotpoint}}
\put(407.66,303.22){\usebox{\plotpoint}}
\multiput(410,304)(20.756,0.000){0}{\usebox{\plotpoint}}
\multiput(414,304)(19.690,6.563){0}{\usebox{\plotpoint}}
\multiput(417,305)(20.756,0.000){0}{\usebox{\plotpoint}}
\multiput(420,305)(20.756,0.000){0}{\usebox{\plotpoint}}
\multiput(424,305)(19.690,6.563){0}{\usebox{\plotpoint}}
\put(427.96,306.00){\usebox{\plotpoint}}
\multiput(430,306)(19.690,6.563){0}{\usebox{\plotpoint}}
\multiput(433,307)(20.756,0.000){0}{\usebox{\plotpoint}}
\multiput(437,307)(20.756,0.000){0}{\usebox{\plotpoint}}
\multiput(440,307)(20.756,0.000){0}{\usebox{\plotpoint}}
\multiput(443,307)(20.136,5.034){0}{\usebox{\plotpoint}}
\put(448.43,308.00){\usebox{\plotpoint}}
\multiput(450,308)(20.756,0.000){0}{\usebox{\plotpoint}}
\multiput(453,308)(20.756,0.000){0}{\usebox{\plotpoint}}
\multiput(457,308)(19.690,6.563){0}{\usebox{\plotpoint}}
\multiput(460,309)(20.756,0.000){0}{\usebox{\plotpoint}}
\multiput(463,309)(20.756,0.000){0}{\usebox{\plotpoint}}
\put(469.02,309.00){\usebox{\plotpoint}}
\multiput(470,309)(20.756,0.000){0}{\usebox{\plotpoint}}
\multiput(473,309)(19.690,6.563){0}{\usebox{\plotpoint}}
\multiput(476,310)(20.756,0.000){0}{\usebox{\plotpoint}}
\multiput(480,310)(20.756,0.000){0}{\usebox{\plotpoint}}
\multiput(483,310)(20.756,0.000){0}{\usebox{\plotpoint}}
\put(489.62,310.00){\usebox{\plotpoint}}
\multiput(490,310)(20.756,0.000){0}{\usebox{\plotpoint}}
\multiput(493,310)(20.756,0.000){0}{\usebox{\plotpoint}}
\multiput(496,310)(19.690,6.563){0}{\usebox{\plotpoint}}
\multiput(499,311)(20.756,0.000){0}{\usebox{\plotpoint}}
\multiput(503,311)(20.756,0.000){0}{\usebox{\plotpoint}}
\multiput(506,311)(20.756,0.000){0}{\usebox{\plotpoint}}
\put(510.21,311.00){\usebox{\plotpoint}}
\multiput(513,311)(20.756,0.000){0}{\usebox{\plotpoint}}
\multiput(516,311)(20.756,0.000){0}{\usebox{\plotpoint}}
\multiput(519,311)(20.756,0.000){0}{\usebox{\plotpoint}}
\multiput(523,311)(20.756,0.000){0}{\usebox{\plotpoint}}
\multiput(526,311)(19.690,6.563){0}{\usebox{\plotpoint}}
\put(530.80,312.00){\usebox{\plotpoint}}
\multiput(532,312)(20.756,0.000){0}{\usebox{\plotpoint}}
\multiput(536,312)(20.756,0.000){0}{\usebox{\plotpoint}}
\multiput(539,312)(20.756,0.000){0}{\usebox{\plotpoint}}
\multiput(542,312)(20.756,0.000){0}{\usebox{\plotpoint}}
\multiput(546,312)(20.756,0.000){0}{\usebox{\plotpoint}}
\put(551.56,312.00){\usebox{\plotpoint}}
\multiput(552,312)(20.756,0.000){0}{\usebox{\plotpoint}}
\multiput(556,312)(20.756,0.000){0}{\usebox{\plotpoint}}
\multiput(559,312)(20.756,0.000){0}{\usebox{\plotpoint}}
\multiput(562,312)(20.756,0.000){0}{\usebox{\plotpoint}}
\multiput(565,312)(20.756,0.000){0}{\usebox{\plotpoint}}
\multiput(569,312)(20.756,0.000){0}{\usebox{\plotpoint}}
\put(572.30,312.10){\usebox{\plotpoint}}
\multiput(575,313)(20.756,0.000){0}{\usebox{\plotpoint}}
\multiput(579,313)(20.756,0.000){0}{\usebox{\plotpoint}}
\multiput(582,313)(20.756,0.000){0}{\usebox{\plotpoint}}
\multiput(585,313)(20.756,0.000){0}{\usebox{\plotpoint}}
\multiput(589,313)(20.756,0.000){0}{\usebox{\plotpoint}}
\put(592.91,313.00){\usebox{\plotpoint}}
\multiput(595,313)(20.756,0.000){0}{\usebox{\plotpoint}}
\multiput(598,313)(20.756,0.000){0}{\usebox{\plotpoint}}
\multiput(602,313)(20.756,0.000){0}{\usebox{\plotpoint}}
\multiput(605,313)(20.756,0.000){0}{\usebox{\plotpoint}}
\multiput(608,313)(20.756,0.000){0}{\usebox{\plotpoint}}
\put(613.66,313.00){\usebox{\plotpoint}}
\multiput(615,313)(20.756,0.000){0}{\usebox{\plotpoint}}
\multiput(618,313)(20.756,0.000){0}{\usebox{\plotpoint}}
\multiput(622,313)(20.756,0.000){0}{\usebox{\plotpoint}}
\multiput(625,313)(20.756,0.000){0}{\usebox{\plotpoint}}
\multiput(628,313)(20.756,0.000){0}{\usebox{\plotpoint}}
\put(634.42,313.00){\usebox{\plotpoint}}
\multiput(635,313)(20.756,0.000){0}{\usebox{\plotpoint}}
\multiput(638,313)(20.756,0.000){0}{\usebox{\plotpoint}}
\multiput(641,313)(20.756,0.000){0}{\usebox{\plotpoint}}
\multiput(645,313)(19.690,6.563){0}{\usebox{\plotpoint}}
\multiput(648,314)(20.756,0.000){0}{\usebox{\plotpoint}}
\multiput(651,314)(20.756,0.000){0}{\usebox{\plotpoint}}
\put(655.01,314.00){\usebox{\plotpoint}}
\multiput(658,314)(20.756,0.000){0}{\usebox{\plotpoint}}
\multiput(661,314)(20.756,0.000){0}{\usebox{\plotpoint}}
\multiput(664,314)(20.756,0.000){0}{\usebox{\plotpoint}}
\multiput(668,314)(20.756,0.000){0}{\usebox{\plotpoint}}
\multiput(671,314)(20.756,0.000){0}{\usebox{\plotpoint}}
\put(675.77,314.00){\usebox{\plotpoint}}
\multiput(678,314)(20.756,0.000){0}{\usebox{\plotpoint}}
\multiput(681,314)(20.756,0.000){0}{\usebox{\plotpoint}}
\multiput(684,314)(20.756,0.000){0}{\usebox{\plotpoint}}
\multiput(688,314)(20.756,0.000){0}{\usebox{\plotpoint}}
\multiput(691,314)(20.756,0.000){0}{\usebox{\plotpoint}}
\put(696.52,314.00){\usebox{\plotpoint}}
\multiput(697,314)(20.756,0.000){0}{\usebox{\plotpoint}}
\multiput(701,314)(20.756,0.000){0}{\usebox{\plotpoint}}
\multiput(704,314)(20.756,0.000){0}{\usebox{\plotpoint}}
\multiput(707,314)(20.756,0.000){0}{\usebox{\plotpoint}}
\multiput(711,314)(20.756,0.000){0}{\usebox{\plotpoint}}
\multiput(714,314)(20.756,0.000){0}{\usebox{\plotpoint}}
\put(717.28,314.00){\usebox{\plotpoint}}
\multiput(721,314)(20.756,0.000){0}{\usebox{\plotpoint}}
\multiput(724,314)(20.756,0.000){0}{\usebox{\plotpoint}}
\multiput(727,314)(20.756,0.000){0}{\usebox{\plotpoint}}
\multiput(730,314)(20.756,0.000){0}{\usebox{\plotpoint}}
\multiput(734,314)(20.756,0.000){0}{\usebox{\plotpoint}}
\put(738.03,314.00){\usebox{\plotpoint}}
\multiput(740,314)(20.756,0.000){0}{\usebox{\plotpoint}}
\multiput(744,314)(20.756,0.000){0}{\usebox{\plotpoint}}
\multiput(747,314)(20.756,0.000){0}{\usebox{\plotpoint}}
\multiput(750,314)(20.756,0.000){0}{\usebox{\plotpoint}}
\multiput(754,314)(20.756,0.000){0}{\usebox{\plotpoint}}
\put(758.79,314.00){\usebox{\plotpoint}}
\multiput(760,314)(20.756,0.000){0}{\usebox{\plotpoint}}
\multiput(763,314)(20.756,0.000){0}{\usebox{\plotpoint}}
\multiput(767,314)(20.756,0.000){0}{\usebox{\plotpoint}}
\multiput(770,314)(20.756,0.000){0}{\usebox{\plotpoint}}
\multiput(773,314)(20.756,0.000){0}{\usebox{\plotpoint}}
\put(779.54,314.00){\usebox{\plotpoint}}
\multiput(780,314)(20.756,0.000){0}{\usebox{\plotpoint}}
\multiput(783,314)(20.756,0.000){0}{\usebox{\plotpoint}}
\multiput(787,314)(20.756,0.000){0}{\usebox{\plotpoint}}
\multiput(790,314)(20.756,0.000){0}{\usebox{\plotpoint}}
\multiput(793,314)(20.756,0.000){0}{\usebox{\plotpoint}}
\multiput(796,314)(20.756,0.000){0}{\usebox{\plotpoint}}
\put(800.30,314.00){\usebox{\plotpoint}}
\multiput(803,314)(20.756,0.000){0}{\usebox{\plotpoint}}
\multiput(806,314)(20.136,5.034){0}{\usebox{\plotpoint}}
\multiput(810,315)(20.756,0.000){0}{\usebox{\plotpoint}}
\multiput(813,315)(20.756,0.000){0}{\usebox{\plotpoint}}
\multiput(816,315)(20.756,0.000){0}{\usebox{\plotpoint}}
\put(820.93,315.00){\usebox{\plotpoint}}
\multiput(823,315)(20.756,0.000){0}{\usebox{\plotpoint}}
\multiput(826,315)(20.756,0.000){0}{\usebox{\plotpoint}}
\multiput(829,315)(20.756,0.000){0}{\usebox{\plotpoint}}
\multiput(833,315)(20.756,0.000){0}{\usebox{\plotpoint}}
\put(836,315){\usebox{\plotpoint}}
\end{picture}
\caption[Figure of functions $h$ and $g_{T,L}$]{Plot of the functions $h$ 
and $g_{T,L}$ for some representative values of $v_F$.
At $z = 1$, $g_L,h$ become infinitely negative.
For $z\gg 1$, the functions $h,g_T,g_L$ have the asymptotic
values $\frac{1}{3}v_F^2, 2/3, 1/3$, respectively.
\label{fig:gh}}
\end{center}
\end{figure}
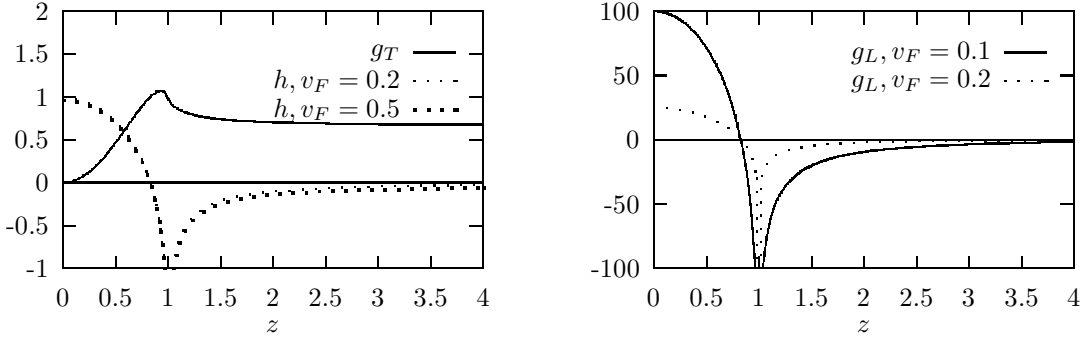

The contribution from the protons can be obtained
straightforwardly by substituting Eqs.\ (\ref{Dqsmall}) and (\ref{ABJdegen}) in 
Eq.\ (\ref{RepiTLfermion}).  In the non-relativistic limit, the result can
be expressed as
\begin{eqnarray}\label{RepiTLdegenprot}
\mbox{Re}\,\pi^{(p)}_T & = &
6e^2\omega_{0p}^2 g_T(z_p) + q^2\chi_0^{(p)}f_T(z_p)
+ q^2\left(\frac{|e|}{m_p\kappa_p}\right)\chi^{(p)}_0 h(z_p)\,,\nonumber\\
\mbox{Re}\,\pi_L^{(p)} & = & 12e^2\omega_{0p}^2 g_L(z_p)
+ q^2\chi_0^{(p)}f_L(z_p)
+ q^2\left(\frac{|e|}{m_p\kappa_p}\right)\chi^{(p)}_0 h(z_p) \,,
\end{eqnarray}
where, in analogy with Eq.\ (\ref{chi0})
\begin{equation}\label{chi0p}
\chi^{(p)}_0 \equiv \frac{\kappa_p^2 m_p{\cal P}_{Fp}}{\pi^2}\,,
\end{equation}
and
\begin{equation}\label{omega0p}
\omega_{0p}^2 = \frac{n_p}{4m_p}\,,
\end{equation}
with $n_p$ denoting the proton number density. The functions $f_{T,L}$
are defined in Eq.\ (\ref{fTL}), while
\begin{eqnarray}\label{gh}
g_T(z) & = & z^2 + \frac{1}{2}z(1 - z^2)
\log\left|\frac{1 + z}{1 - z}\right| \,,\nonumber\\
g_L(z) & = & \left(z^2 - \frac{1}{v_F^2}\right)
\left[-1 + \frac{1}{2}z\log\left|\frac{1 + z}{1 - z}\right|\right]\,,\nonumber\\
h(z) & = & 1 - \frac{1}{6}v_{F}^2 - \left(1 - \frac{1}{2}v_F^2\right)
\frac{1}{2}z\log\left|\frac{1 + z}{1
- z}\right| \,.
\end{eqnarray}
These functions are plotted in Fig.\ \ref{fig:gh}.

Let us consider the dispersion relation for the
transverse modes.  It can be written in the form
\begin{equation}\label{drelelecnucl}
q^2(1 - \chi) = \mbox{Re}\,\pi^{(e)}_T + 6e^2\omega_{0p}^2 g_T(z_p)
\end{equation}
where we have defined
\begin{equation}\label{auxchi}
\chi \equiv \chi^{(n)}_0 f_T(z_n) + \chi^{(p)}_0 f_T(z_p)
+ \chi^{(p)}_0\left(\frac{|e|}{m_p\kappa_p}\right)h(z_p)\,.
\end{equation}
For $\Omega < {\cal Q}$, we have $\pi^{(e)}_T > 0$, as we
also argued before in Section \ref{subsec:neutelectron}, and from
Fig.\ \ref{fig:gh} we can see that $g_T > 0$ also.  Therefore,
the right-hand side of Eq.\ (\ref{drelelecnucl}) is a positive quantity.
On the other hand, since the functions $h$ and $f_T$ never become
larger than unity, then $\chi < 1$ since $\chi^{(n,p)}_0$ are
considerably smaller than unity (they are of order $\alpha v_F$).
Therefore, Eq.\ (\ref{drelelecnucl}) has no solution with $\Omega < {\cal Q}$.
For the opposite regime $\Omega > {\cal Q}$, we have
$z_{n,p}\gg 1$, and $f_T,h,g_T$ go to their asymptotic
values
\begin{eqnarray}\label{fhgasymp}
h & \rightarrow & \frac{1}{3}v_F^2\,,\nonumber\\
f_T & \rightarrow & \frac{1}{3}v_F^2\,,\nonumber\\
g_T & \rightarrow & \frac{2}{3} \,.
\end{eqnarray}
It then follows that the contribution from the $\chi$ terms
in Eq.\ (\ref{drelelecnucl}) is small.  On the right-hand side
of that equation, the term $\omega_{0p}^2 g_T$ from
the protons is competing against the corresponding term
from the electrons.  Since $\omega_{0e}^2 \sim T^2$
and $n_e \sim T^3$, where $T$ is the temperature of the
electron gas, we have $\omega_{0e}^2 \sim n_p/T$,
assuming that $n_e \sim n_p$.  Then from Eq.\ (\ref{omega0p}) it follows
that 
\begin{equation}\label{omega0evsp}
\omega_{0p}^2 \approx \frac{T}{m_p}\omega_{0e}^2 \,,
\end{equation}
so that the proton term on the right-hand side of Eq.\ (\ref{drelelecnucl})
is also small.  The equation for the dispersion relation
is then well approximated by simply
\begin{equation}\label{drelelecnuclapprox}
q^2 = \pi^{(e)}_T \,.
\end{equation}
Thus, the nucleon contribution, represented by the $\chi$
and $g_T$ 
terms in Eq.\ (\ref{drelelecnucl}),  may affect some of the details of 
the transverse dispersion relation for the relativistic electron gas, but it
will not modify the general features in an essential way.

The situation can be different for the longitudinal modes.
The equation for the dispersion relation
can be written in the form
\begin{equation}\label{drelenplong}
q^2(1 - \chi^\prime) = \mbox{Re}\,\pi^{(e)}_L + 12e^2\omega_{0p}^2 g_L(z_p)
\end{equation}
where 
\begin{equation}\label{auxchiprime}
\chi^\prime \equiv \chi^{(n)}_0 v_{Fn}^2 f_L(z_n) + \chi^{(p)}_0 v_{Fp}^2 f_L(z_p)
+ \chi^{(p)}_0\left(\frac{|e|}{m_p\kappa_p}\right)h(z_p)\,.
\end{equation}
The factor $v_F^2$ in the terms proportional to $f_L(z_{n,p})$ makes
those terms unimportant, while the term proportional to $h$ 
need not be negligible, although it is small as discussed above.  
Regarding the terms on the right-hand side
of Eq.\ (\ref{drelenplong}), consider first the region
$\Omega > {\cal Q}$.  In that region, we have $z_p \gg 1$ and hence
$g_L\rightarrow 1/3$.  The same argument as before [Eq.\ (\ref{omega0evsp})]
then implies that the proton contribution is not relevant.
However, in contrast with the equation for the transverse modes,
Eq.\ (\ref{drelenplong}) can have solutions with $\Omega_L < {\cal Q}$
because $g_L$ can be negative and, as we saw in 
Section \ref{subsec:neutelectron},
$\pi^{(e)}_L$ can be negative as well in this range.
Furthermore, for values of $\Omega,{\cal Q}$ such that
$z_p \sim 1$, we have $g_L \sim -\frac{1}{v_F^2}$, and this can
bring the proton term $\omega_{0p}^2 g_L$ to be comparable to
the electron term $\omega_{0e}^2$.
Thus, to summarize the situation for the longitudinal modes,
in Eq.\ (\ref{drelenplong}) the term with $f_L$ is irrelevant, 
the term with $h$ can give a small correction, while
the term with $g_L$ can have an important effect
in the dispersion relation.  The latter effect can occur
in the range of $\Omega,{\cal Q}$ such that $z_p\sim 1$, if 
\begin{equation}\label{omega0epcond}
\frac{\omega_{0p}^2}{v_{Fp}^2} \sim \omega_{0e}^2 \,.
\end{equation}
Using Eq.\ (\ref{omega0evsp}) this can be translated into the
condition $v_{Fp}^2 \sim \frac{T}{m_p}$, where $T$ stands
for the electron gas temperature.

\subsection{Relativistic electron and classical neutron and proton gases}
\label{subsec:classneutprotelectron}

For a classical non-relativistic proton gas, the
contribution to the self-energy is obtained by inserting
the expressions given in Eqs.\ (\ref{Dqsmall}) and
(\ref{Bqsmallclass} - \ref{J0class}) into Eq.\ (\ref{RepiTLfermion}).
For the transverse part we thus obtain
\begin{equation}\label{piprotonTclassnr}
\mbox{Re}\,\pi^{(p)}_T = \frac{|e|^2 n_p}{m_p}G_T({\overline z}_p)
+ \kappa_p^2 n_p \beta_p q^2 F_T({\overline z}_p)
+ \frac{|e|\kappa_p}{m_p}n_p\beta_p q^2 H({\overline z}_p) \,,
\end{equation}
where, to the order that we have calculated,
\begin{eqnarray}\label{FGHT}
G_T({\overline z}_f) & = & \left\{
\begin{array}{ll}
{\overline z}_f^2 & \mbox{(${\overline z}_f \ll 1$)} \\
1 + \frac{1}{{\overline z}_f^2}  & \mbox{(${\overline z}_f \gg 1$)}
\end{array}
\right. \nonumber\\
F_T = H({\overline z}_f) & = & \left\{
\begin{array}{ll}
1 - {\overline z}_f^2& \mbox{(${\overline z}_f \ll 1$)} \\
{\overline v}^2 - \frac{1}{{\overline z}_f^2} & \mbox{(${\overline z}_f \gg 1$)}
\end{array}
\right. \,.
\end{eqnarray}
Similarly, for the longitudinal part,
\begin{equation}\label{piprotonLclassnr}
\mbox{Re}\,\pi^{(p)}_L = \frac{|e|^2n_p}{m_p}G_L({\overline z}_p)
+ \frac{\kappa_p^2 n_p}{m_p} q^2 F_L({\overline z}_p)
+ \frac{|e|\kappa_p}{m_p}n_p\beta_p q^2 H({\overline z}_p) \,,
\end{equation}
where
\begin{eqnarray}\label{FGHL}
G_L({\overline z}_f) & = & \left({\overline z}_f^2 - \frac{1}{{\overline v}_f^2}\right)\times
\left\{
\begin{array}{ll}
-1 + {\overline z}_f^2 & \mbox{(${\overline z}_f \ll 1$)} \\
\frac{1}{{\overline z}_f^2}  + \frac{3}{{\overline z}_f^4} & \mbox{(${\overline z}_f \gg 1$)}
\end{array}
\right. \nonumber\\
F_L({\overline z}_f) & = & \left\{
\begin{array}{ll}
-1 + 2{\overline z}_f^2& \mbox{(${\overline z}_f \ll 1$)} \\
1 + \frac{2}{{\overline z}_f^2} & \mbox{(${\overline z}_f \gg 1$)}
\end{array}
\right. \,.
\end{eqnarray}
The formulas for the neutron contribution correspond to
setting $|e| = 0$ in Eqs.\ (\ref{piprotonTclassnr}) and (\ref{piprotonLclassnr})
and replacing every quantity that refers to the proton by
the analogous quantity for the neutron.

Comparing these results with those given in Eq.\ (\ref{RepiTLdegenprot}) for the
degenerate case, it is easily seen that the qualitative
features between that case and the present one
are similar.  In fact, an analysis along the same lines
to that carried out in Sections \ref{subsec:neutelectron}
and \ref{subsec:neutprotelectron}
reveals similar conclusions.  In particular, the
neutron component does not modify the photon dispersion
relations in a significant way.  The proton contribution
to the transverse component of the self-energy is also
unimportant but it can have a noticeable effect for 
the longitudinal part.  This can occur for values
of $\Omega < {\cal Q}$, but such that $z_p\sim 1$.
In this case, similarly to the situation for the
case of degenerate protons, $G_L\sim 1/{\overline v}_p^2$, and this brings
the proton contribution to $\mbox{Re}\,\pi_L$ to be of the order
of $e^2 n_p\beta_p$.  
Assuming that $n_p\sim n_e$ and $n_e \sim T^3$, where
$T$ stands for the electron gas temperature,
the proton contribution is then of order
$e^2 T^3\beta_p$.  This
can compete with the electron
contribution, which is of order $e^2 T^2$,
if the temperature of both components are comparable.
The proton contribution could in fact dominate
if the proton gas temperature is smaller than that
of the electron gas.

\subsection{Nucleon contribution to the imaginary part}
\label{subsec:impart}

In general, the concept of a propagating mode with a definite
dispersion relation is meaningful provided
that the damping rate $\gamma_{T,L}$ is much smaller
than $\Omega_{T,L}$ in the solution to Eq.\ (\ref{disprel}).
Therefore, in the cases in which the nucleon contribution
to the real part of the photon self-energy may be substantial,
it is imperative to know also the contribution
to the imaginary part.  

In the general case, the calculation of the nucleon 
contribution to the imaginary
part  of the self-energy is cumbersome.
Fortunately, for our purposes, in which the small $q$ formulas
given in Eqs.\ (\ref{Dqsmall}) and (\ref{ABqsmall}) hold, 
we can give a simple recipe to determine it.
It is given by the so-called Landau rule, whose justification
in the present context is provided in Appendix \ref{appendix:impart}.
The rule is simply that, in the small $q$ limit,
the imaginary part of $A_f, B_f$ can be obtained from 
Eq.\ (\ref{ABqsmall}) by making the replacement
$\Omega\rightarrow \Omega + i\epsilon$ in the integrand
of those formulas, and taking the imaginary part of
the resulting expressions.  Thus,
\begin{eqnarray}\label{imABD}
\mbox{Im}\, B_f & = & \frac{\pi\Omega}{2}\int\frac{d^3{\cal P}}{(2\pi)^3}
\delta(\Omega - \vec v_{\cal P}\cdot\vec{\cal Q})
\frac{d}{d{\cal E}}(f_f + f_{\overline f}) \,,\nonumber\\
\mbox{Im}\, A_f & = & \frac{\pi\Omega}{2}\int\frac{d^3{\cal P}}{(2\pi)^3}
(1 - v^2_{\cal P})\delta(\Omega - \vec v_{\cal P}\cdot\vec{\cal Q})
\frac{d}{d{\cal E}}(f_f + f_{\overline f}) \,,\nonumber\\
\mbox{Im}\, D_f & = & \frac{1}{2m_f^2}\mbox{Im}\, A_f \,.
\end{eqnarray}
Further, the imaginary part of $\pi^{(n,p)}_{T,L}$
is given by formulas analagous to Eq.\ (\ref{RepiTLfermion}), 
but with the functions $A_f,B_f,D_f$ replaced by their imaginary
part, given in Eq.\ (\ref{imABD}).

\subsubsection{Degenerate case}

For the particular case of the degenerate gas, these formulas
reduce to
\begin{eqnarray}\label{imABDdegen}
\mbox{Im}\, B_f & = & -\left(\frac{m_f {\cal P}_{Ff}\gamma_f}{8\pi}\right)
z_f\theta(1 - |z_f|) \,,\nonumber\\
\mbox{Im}\, A_f & = & 2m_f^2\,\mbox{Im}\, D_f = (1 - v_{Ff}^2)\mbox{Im}\, B_f \,,
\end{eqnarray}
where $\gamma_f$ and $z_f$ are defined in Eq.\ (\ref{ABJdegenaux}).
These in turn yield, 
\begin{eqnarray}\label{impitldegennucleon}
\mbox{Im}\,\pi^{(f)}_T & = & \pi z_f\theta(1 - |z_f|)\gamma_f\left\{
3e^2\omega_{0p}^2(z_p^2 - 1) 
+ q^2\left(\frac{e_f\chi^{(f)}_0}{2m_f\kappa_p\gamma_f^2}\right)
+ \frac{1}{2}q^2\chi^{(f)}_0 
\left(1 - \frac{1}{2}v_{F}^2 z_f^2\right)\right\}\,,\nonumber\\
\mbox{Im}\,\pi^{(f)}_L & = & \pi z_f\theta(1 - |z_f|)\gamma_f\left\{
-6e_f^2\omega_{0f}^2\left(z_f^2 - \frac{1}{v_{F}^2}\right)
+ q^2\left(\frac{e_f\chi^{(f)}_0}{2m_f\kappa_f}\right)
+ \frac{1}{2}q^2\chi^{(f)}_0 v_{F}^2(z_f^2 - 1)\right\} \,.
\end{eqnarray}
%
%
%
%

\subsubsection{Classical, non-relativistic case}

The formulas in Eq.\ (\ref{imABD}) can be evaluated explicitly also
for the classical non-relativistic case.  With ${\overline z}_f$
as defined in Eq.\ (\ref{zbar}), the results 
can be expressed in the form
\begin{eqnarray}\label{imABDclass}
\mbox{Im}\, B_f & = & -\frac{1}{4}\sqrt{\frac{\pi}{2}}
\beta_f n_f {\overline z}_f e^{-\frac{1}{2}{\overline z}_f^2} \nonumber\\
\mbox{Im}\, A_f & = & 2m_f^2\,\mbox{Im}\, D_f = \mbox{Im}\, B_f[1 - 
{\overline v}_f^2(2 + {\overline z}_f^2)] \,,
\end{eqnarray}
which imply
\begin{eqnarray}\label{impitlnondegennucleon}
\mbox{Im}\,\pi^{(f)}_T & = & \sqrt{\frac{\pi}{2}}
\beta_f n_f {\overline z}_f e^{-\frac{1}{2}{\overline z}_f^2}
\left\{
-e_f^2 {\overline v}_f^2 + \frac{e_f\kappa_f}{m_f}q^2(1 - {\overline v}_f^2 {\overline z}_f^2)
+ \kappa_f^2 q^2 (1 - {\overline v}_f^2 {\overline z}_f^2)\right\}\,,\nonumber\\
\mbox{Im}\,\pi^{(f)}_L & = & \sqrt{\frac{\pi}{2}}
\beta_f n_f {\overline z}_f e^{-\frac{1}{2}{\overline z}_f^2}
\left\{
e_f^2 (1 - {\overline v}_f^2 {\overline z}_f^2) + 
\frac{e_f\kappa_f}{m_f}q^2(1 - {\overline v}_f^2 {\overline z}_f^2)
- 2\kappa_f^2 q^2 {\overline v}_f^2
\right\}\,.
\end{eqnarray}
As usual, these formulas allow us to obtain the results
for the proton and neutron separately, by setting $e_f$ and
$\kappa_f$ to the appropriate value in each case.

\section{Conclusions}\label{sec:conclusions}

In this work we have calculated the 1-loop contribution
to the photon self-energy in a background composed
of electrons and nucleons, taking into account the
anomalous magnetic moment couplings of the nucleons
to the photon.  The generic contribution from a fermion
to the real part of the photon self-energy is given
by the formulas in Eq.\ (\ref{RepiTLfermion}), in terms
of the quantities that we have denoted by $A_f, B_f, D_f$.
The explicit evaluation of these three functions 
is carried out in Section \ref{sec:limitingcases}
for various limiting cases of interest.  The
results obtained there were applied in 
Section \ref{sec:discussion} to consider the effect
of including the nucleon contribution
to the photon dispersion relation in a plasma.
For concreteness, we considered the case of a relativistic
electron gas, superimposed on a nonrelativistic nucleon
background, which we considered to be degenerate or classical
in separate cases.  The results of that exercise
indicate that the magnetic moment couplings of the nucleons do not have
a significant effect on the photon dispersion relations.
Hence the neutrons, which only have these type of coupling, 
do not seem to play a siginificant role in the determination
of the photon dispersion relations in such media.
For the protons, the normal couplings
do not have a significant effect on the transverse modes either,
but they can have an impact on the longitudinal dispersion
relation in some circumstances. As shown in 
Sections \ref{subsec:neutprotelectron} and \ref{subsec:classneutprotelectron},
for values of the photon momentum such that $\Omega \sim {\overline v}_p{\cal Q}$,
where ${\overline v}_p$ is a typical velocity of the protons in the gas,
the proton contribution to the longitudinal component
of the photon self-energy can be of the same order as that
of the electrons.  
Of course we have reached these conclusions by considering
in detail the application of the general formulas
to some particular idealized situations.
In any case, the formulas that we have provided in this
work open the way to similar studies in other possible settings.

This work has been partially supported by the U.S. National
Science Foundation Grant PHY-9600924 (JFN) and by
Grant No. DGAPA-IN100694 at the Universidad Nacional
Aut\'onoma de M\'exico (JCD).
\appendix
\section{Dielectric function}\label{appendix:dielectric}

Although the properties of photons propagating through the medium can be deduced
from the knowledge of the photon self-energy directly,
it is useful to translate those results into the language of macroscopic
electrodynamics.  The transverse and longitudinal 
components of the dielectric constant of the
medium are introduced by writing the induced current, in the rest frame of
the medium, as
\begin{equation}\label{jindclas}
\vec \jmath^{\,(ind)} = i\Omega\left[(1 - \epsilon_l)\vec E_l
+ (1 - \epsilon_t)\vec E_t + i\epsilon_p \hat{\cal Q}\times\vec E\right] \,,
\end{equation}
where the longitudinal and transverse components of the electric field are
defined by  
\begin{eqnarray}\label{Etl}
\vec E_l & = & \hat{\cal Q}(\hat{\cal Q}\cdot\vec E) \,,\nonumber\\
\vec E_t & = & \vec E - \vec E_l \,.
\end{eqnarray}
This is the most general form of the induced current, involving terms that are
linear in the field, and subject only to the assumption of isotropy.  
On the other hand, by comparing Maxwell's
equations with the effective field equation Eq.\ (\ref{claseqmotion}),
it follows that the induced current is given by
\begin{equation}\label{jindft}
j^{(ind)}_\mu = -\pi_{\mu\nu}A^\nu \,,
\end{equation}
which can be written as in Eq.\ (\ref{jindclas}) with
\begin{eqnarray}\label{epsilonpirel}
1 - \epsilon_t & = & \frac{\pi_T}{\Omega^2}\,, \nonumber\\
1 - \epsilon_l & = & \frac{\pi_L}{q^2}\,,\nonumber\\
\epsilon_p & = & \frac{\pi_P}{\Omega^2} \,.
\end{eqnarray}

Alternatively to $\epsilon_{t,l}$, the dielectric and
magnetic permeability functions $\epsilon,\mu$ are
introduced by writing the induced current in the
equivalent form
\begin{equation}\label{jindclasequiv}
\vec \jmath^{(ind)} = i\Omega\left[(1 - \epsilon)\vec E
+ i(1 - \frac{1}{\mu})\vec q\times \vec B + 
i\epsilon_p \hat{\cal Q}\times\vec E\right] \,,
\end{equation}
where 
\begin{equation}\label{BErel}
\vec B = \frac{1}{\Omega}\vec q\times\vec E \,.
\end{equation}
Comparing Eqs.\ (\ref{jindclas}) and (\ref{jindclasequiv}), the relations
\begin{eqnarray}\label{epsilonmu}
\epsilon & = & \epsilon_l\nonumber\\
\frac{1}{\mu} & = & 1 + \frac{\Omega^2}{{\cal Q}^2}(\epsilon_l -
\epsilon_t) 
\end{eqnarray}
then follow, and in particular
\begin{equation}\label{mu}
\frac{1}{\mu} = 1 + \frac{1}{{\cal Q}^2}
\left(\pi_T - \frac{\Omega^2}{q^2}\pi_L\right)\,.
\end{equation}

\section{Imaginary part}\label{appendix:impart}

From Eq.\ (\ref{pieff}), it follows that
\begin{equation}\label{impieff}
\mbox{Im}\,\pi^{(e\!f\!\!f)}_{\mu\nu} = \varepsilon(q\cdot v)\mbox{Im}\,\pi_{\mu\nu} \,,
\end{equation}
where $\mbox{Im}\,\pi_{\mu\nu}$ can be determined in terms of
the components of the self-energy matrix by either
one of the formulas given in Eq.\ (\ref{impi}).  However,
as already mentioned in Section \ref{sec:photonselfenergy}, 
it is much easier in practice to calculate $\pi_{12\mu\nu}$ than it is
to calculate $\mbox{Im}\,\pi_{11\mu\nu}$, although the procedure in both
cases is similar.  As an illustration, we outline
below the main steps of the calculation in terms of
$\pi_{12\mu\nu}$, although we have verified explicitly
that the same result is obtained using 
the alternate form in terms of $\mbox{Im}\,\pi_{11\mu\nu}$.
We start by considering the contribution from the
electron loop, and then generalize the result to the
case of the nucleons afterwards.

From Fig.\ \ref{fig:pimunu},
\begin{equation}\label{pi12}
i\pi_{12\mu\nu}^{(e)} = (-1)(-ie)(ie)\mbox{Tr}\int\frac{d^4p}{(2\pi)^4}
\gamma_\mu iS^{(e)}_{F12}(p^\prime)\gamma\nu iS^{(e)}_{F21}(p) \,,
\end{equation}
where
\begin{equation}\label{pprime}
p^\prime = p + q \,.
\end{equation}
After substituting the propagators given in Eq.\ (\ref{SF}) we use the relation
\begin{eqnarray}\label{identity}
[\eta_e(p^\prime) - \theta(-p^\prime\cdot v)]
[\eta_e(p) - \theta(p\cdot v)] & = &
-n(x)[\eta_e(p)\varepsilon(p^\prime\cdot v)
- \eta_e(p^\prime)\varepsilon(p\cdot v)\nonumber\\
& & \mbox{} + \theta(p\cdot v)\theta(-p^\prime\cdot v)
- \theta(-p\cdot v)\theta(p^\prime\cdot v)]\,,
\end{eqnarray}
which is proven by using the identities
\begin{eqnarray}\label{auxidentities1}
\eta_e(p) - \theta(p\cdot v) & = & -e^{y_e} n_F(y_e)\varepsilon(p\cdot v)
\,,\nonumber\\
\eta_e(p) - \theta(-p\cdot v) & = & n_F(y_e)\varepsilon(p\cdot v) \,,
\end{eqnarray}
together with
\begin{equation}\label{auxidenty}
e^{y_e} n_F(y_e)n_F(y_e^\prime) = n_\gamma(x)(n_F(y_e) - n_F(y_e^\prime)) \,.
\end{equation}
In Eq.\ (\ref{auxidenty}) we have defined
\begin{equation}\label{yeprime}
y_e^\prime = p^\prime - \alpha_e \,,
\end{equation}
while $x$ and $y_e$ are defined in Eqs.\ (\ref{x}) and (\ref{y}), respectively,
and we have used the fact that $y_e^\prime = y_e + x$.
In this way we obtain
\begin{eqnarray}\label{pi12final}
i\pi_{12\mu\nu}^{(e)} & = & 16\pi^2 e^2 n_\gamma(x)\int\frac{d^4p}{(2\pi)^4}
L_{\mu\nu}\delta(p^2 - m_e^2)\delta(p^{\prime 2} - m_e^2)
\nonumber\\
& & \mbox{} \times
\left\{\eta_e(p)\varepsilon(p^\prime\cdot v)
- \eta_e(p^\prime)\varepsilon(p\cdot v)
+ \theta(p\cdot v)\theta(-p^\prime\cdot v)
- \theta(-p\cdot v)\theta(p^\prime\cdot v)
\right\}\,.
\end{eqnarray}
Making the change of variable $p + q\rightarrow p$ in the
second and fourth terms in braces, Eq.\ (\ref{pieff2})
then yields
\begin{eqnarray}\label{impieffaux}
\mbox{Im}\,\pi^{(e\!f\!\!f)}_{e\mu\nu} & = & 4\pi e^2 \int\frac{d^4p}{(2\pi)^4}\left\{
L_{\mu\nu}\delta[(p + q)^2 - m_e^2)]\delta(p^2 - m_e^2)\right.
\nonumber\\
& & \left.\mbox{} \times 
\left[\eta_e(p)\varepsilon((p + q)\cdot v) + 
\theta(p\cdot v)\theta(-(p + q)\cdot v) 
\right]
+ (q\rightarrow -q)\right\} \,,
\end{eqnarray}
for the electron contribution
to the imaginary part of $\pi^{(e\!f\!\!f)}_{\mu\nu}$.
After doing the integral over $p_0$, this reduces to
\begin{eqnarray}\label{impieffexplicit}
\mbox{Im}\,\pi^{(e\!f\!\!f)}_{e\mu\nu} & = & 4\pi e^2\int\frac{d^3{\cal P}}{(2\pi)^3 2{\cal E}}
\left\{L_{\mu\nu}\delta[(p + q)^2 - m_e^2]\right.\nonumber\\
& & \left.\mbox{}\times
\left[\theta((p + q)\cdot v)(f_e + f_{\overline e})
+ \theta(-(p + q)\cdot v)(1 - f_e - f_{\overline e})\right]
- (q\rightarrow -q)
\right\}\,.
\end{eqnarray}
Decomposing $\pi^{(e\!f\!\!f)}_{e\mu\nu}$ according to Eq.\ (\ref{pieff}), Eq.\ (\ref{impieffexplicit})
then implies
\begin{eqnarray}\label{impitl}
\mbox{Im}\,\pi^{(e)}_T & = & -2e^2\left(\mbox{Im}\, A_e + \frac{q^2}{{\cal Q}^2}\mbox{Im}\, B_e\right)
\,,\nonumber\\
\mbox{Im}\,\pi^{(e)}_L & = & 4e^2\frac{q^2}{{\cal Q}^2}\mbox{Im}\, B_e \,,
\end{eqnarray}
where
\begin{eqnarray}\label{imAB}
\mbox{Im}\, A_e & = & \pi\int\frac{d^3{\cal P}}{(2\pi)^3 2{\cal E}}\left\{
\vphantom{\frac{1}{2}}
(2m_e^2 - 2p\cdot q)\delta[(p + q)^2 - m_e^2]\right.
\nonumber\\
& & \mbox{}\times\left.
\vphantom{\frac{1}{2}}
\left[\theta((p + q)\cdot v)(f_e + f_{\overline e})
+ \theta(-(p + q)\cdot v)(1 - f_e - f_{\overline e})\right]
- (q\rightarrow -q)\right\}\,,\nonumber\\
\mbox{Im}\, B_e & = & \pi\int\frac{d^3{\cal P}}{(2\pi)^32{\cal E}}\left\{
\vphantom{\frac{1}{2}}
[2(p\cdot v)^2 + 2(p\cdot v)(q\cdot v) - p\cdot q]
\delta[(p + q)^2 - m_e^2]\right.
\nonumber\\
& & \mbox{}\times\left.
\vphantom{\frac{1}{2}}
\left[\theta((p + q)\cdot v)(f_e + f_{\overline e})
+ \theta(-(p + q)\cdot v)(1 - f_e - f_{\overline e})\right]
- (q\rightarrow -q)\right\}\,.
\end{eqnarray}

The formulas in Eq.\ (\ref{imAB}) complement those in Eq.\ (\ref{ApBp}), which 
hold for the real part.  The counterpart of the approximate
formulas given in Eq.\ (\ref{ABqsmall}) for the small $q$ regime
are obtained as follows.  For $q\ll \langle{\cal E}\rangle$,
we can use 
\begin{equation}\label{zeroterms}
\theta[-(p + q)\cdot v]\delta[(p + q)^2 - m_e^2] = 0
\end{equation}
to discard some terms in Eq.\ (\ref{imAB}).  The terms
discarded in this way are 
the standard vacuum terms as well as 
the temperature corrections to them,
which they are relevant only for $q > 2m_e$
and physically correspond to the creation of electron-positron pairs.
The surviving
terms can be written in the form
\begin{eqnarray}\label{imABqsmallaux}
\mbox{Im}\, B_e & = & \pi\int\frac{d^3{\cal P}}{(2\pi)^3}
\left\{
\frac{(2{\cal E}_{\vec {\cal P}}^2 + {\cal E}_{\vec {\cal P}}\Omega
+ \vec{\cal P}\cdot\vec q)}{4{\cal E}_{\vec {\cal P}}{\cal E}_{\vec{\cal P} +
\vec q}}
\delta(\Omega + {\cal E}_{\vec {\cal P}} - 
{\cal E}_{\vec{\cal P} + \vec q})(f_e + f_{\overline e})
- (q\rightarrow -q)
\right\}\,,\nonumber\\
\mbox{Im}\, A_e & = & \mbox{Im}\, B_e \nonumber\\
& & \mbox{}
+ \pi\int\frac{d^3{\cal P}}{(2\pi)^3}
\left\{
\frac{(2{\vec {\cal P}}^2 - 3{\cal E}_{\vec {\cal P}}\Omega
+ \vec{\cal P}\cdot\vec q)}{4{\cal E}_{\vec {\cal P}}{\cal E}_{\vec{\cal P} +
\vec q}}
\delta(\Omega + {\cal E}_{\vec {\cal P}} - 
{\cal E}_{\vec{\cal P} + \vec q})(f_e + f_{\overline e})
- (q\rightarrow -q)
\right\} \,,
\end{eqnarray}
where we have made explicit the dependence of the
energy on the momentum variable.
Making the change of variable $\vec{\cal P}\rightarrow 
\vec{\cal P} - \frac{1}{2}\vec q$ and expanding the integrand
up to terms that are linear in $q$, we obtain
\begin{eqnarray}\label{imABqsmall}
\mbox{Im}\, B_e & = & -\frac{\pi}{2}\Omega\int\frac{d^3{\cal P}}{(2\pi)^3}
\delta(\Omega - \vec{\cal P}\cdot\vec q)\frac{d}{d{\cal E}}
(f_e + f_{\overline e}) \,,\nonumber\\
\mbox{Im}\, A_e & = & \mbox{Im}\, B_e + \frac{\pi}{2}\Omega\int\frac{d^3{\cal P}}{(2\pi)^3}
v_{\cal P}^2 \delta(\Omega - \vec{\cal P}\cdot\vec q)\frac{d}{d{\cal E}}
(f_e + f_{\overline e})\,.
\end{eqnarray}

The results given in Eq.\ (\ref{imABqsmall}), together with Eq.\ (\ref{impitl}),
provide the justification of the rule stated in Section \ref{subsec:impart},
as can be immediately confirmed by applying it to determine
the imaginary part of
$\pi^{(e)}_{T,L}$ from the formulas in Eq.\ (\ref{Repiqsmall}) for the real part
of the same quantities.  The same arguments given here
also to apply to the nucleon contribution, the only
difference being that in the formula corresponding
to Eq.\ (\ref{impieffexplicit}), the factor $e^2 L_{\mu\nu}$ is replaced by
\begin{equation}\label{replacementfactor}
e_f^2 L_{\mu\nu} + \kappa_f^2 M_{\mu\nu} + 
2e_f\kappa_f m_f q^2\tilde g_{\mu\nu}\,.
\end{equation}
By mimicking the steps that lead to Eq.\ (\ref{RepiTLfermion}), 
we obtain similarly looking formulas for $\mbox{Im}\,\pi^{(f)}_{T,L}$,
but with the functions $A_f,B_f,D_f$ replaced by
their imaginary parts, which in the end yield the results
quoted in Section \ref{subsec:impart}.


\begin{thebibliography}{99}

\bibitem{ftftreview} For a review see, for example, 
N. P. Landsman and Ch. G. van Weert, Phys. Rep. {\bf 145}, 141 (1987).

\bibitem{weldon:cov} H. A. Weldon,  Phys. Rev. D{\bf 26}, 1394 (1982).
See also, E. Braaten and D. Segel, Phys. Rev. D{\bf 48}, 1478 (1993).

\bibitem{weldon:fermions} H. A. Weldon, Phys. Rev. D{\bf 26}, 2789 (1982).

\bibitem{weldon:imag} H. A. Weldon, Phys. Rev. D{\bf 28}, 2007 (1983).

\bibitem{LL:physkin} E. M. Lifshitz and L. P. Pitaevskii,
``Physical Kinetics'', Course of Theoretical Physics Volume 10,
(Pergamon Press, New York 1981), p. 133.

\bibitem{tsytovich} V. N. Tsytovich, 
J. Exptl. Theoret. Phys. (U.S.S.R) {\bf 40}, 1775 (June 1961)
[Sov. Phys. JETP {\bf 13}, 1249 (1961)].

\bibitem{silin} V. P. Silin, Zh. Eksp. Teor. Fiz. {\bf 38}, 1577 (1960)
[Sov. Phys. JETP {\bf 11}, 1136 (1960)].

\bibitem{LL:physkin2} See, for example, Ref.\ \cite{LL:physkin}, p. 130.

\bibitem{fetter} A. L. Fetter and J. D. Walecka,
``Quantum Theory of Many-Particle Systems''
(McGraw-Hill, New York, 1971) p. 49.

\bibitem{mohanty} S. Mohanty and M. K. Samal, Phys. Rev. Lett. {\bf 77}, 806 (1996).

\bibitem{raffelt:critique} Georg G. Raffelt, Phys. Rev. Lett. {\bf 79}, 773 (1997).

\bibitem{canonical} Jos\'e F. Nieves, Phys. Rev. D{\bf 42}, 4123 (1990);
{\em ibid.}{\bf 49}, 3067(E) (1994).

\bibitem{pisubpi} Jos\'e F. Nieves and P. B. Pal,
Phys. Rev. D{\bf 39}, 652 (1989) and {\em ibid.}{\bf 40}, 2148E (1989);
{\em ibid.}{\bf 40}, 1350 (1989).

\bibitem{dnp1} J. C. D'Olivo, Jos\'e F. Nieves and P. B. Pal, 
Phys. Rev. D{\bf 40}, 3679 (1989).

\bibitem{dnnuclmag} J. C. D'Olivo, Jos\'e F. Nieves,
``Nucleon contribution to the neutrino
electromagnetic vertex in matter'', UPR preprint.

\bibitem{LL:physkin3} See, for example, Ref.\ \cite{LL:physkin}, p. 132.

\end{thebibliography}
\end{document}